\documentclass[fleqn,usenatbib]{mnras}

\usepackage[usenames,dvipsnames]{xcolor}
\usepackage{hyperref}
\usepackage[T1]{fontenc}
\usepackage{ae,aecompl}
\usepackage{graphicx}         % Including figure files
\usepackage{amsmath}          % Advanced maths commands
\usepackage{amssymb}          % Extra maths symbols
\usepackage{array}
\usepackage{float}
\usepackage{etoolbox}
\usepackage{CJK}              % for Chinese
\usepackage{newtxtext,newtxmath}
\usepackage{tikz}
\usepackage{comment}
\usetikzlibrary{calc, shapes, arrows, positioning, quotes, chains}

\newcommand {\kpc}            {\,\rm kpc}

\newcommand {\Msun}           {\,\rm{M}_{\sun}}

\newcommand {\gamer}          {\textsc{gamer-2}}
\newcommand {\GALIC}          {\textsc{GALIC}}

\newcommand {\Mh}             {M_{\rm h}}
\newcommand {\Md}             {M_{\rm d}}
\newcommand {\Rd}             {R_{\rm d}}
\newcommand {\md}             {m_{\rm d}}
\newcommand {\Nd}             {N_{\rm d}}

\newcommand {\rc}             {r_{\rm c}}
\newcommand {\rvir}             {r_{\rm vir}}
\newcommand {\ma}             {m_{22}}
\newcommand {\sigmah}         {\sigma_{\rm h}}
\newcommand {\rhoh}           {\rho_{\rm h}}
\newcommand {\rhod}           {\rho_{\rm d}}

\newcommand {\ts}             {t_{\rm sim}}
\newcommand {\tr}             {t_{\rm rel}}

\newcommand {\sref}[1]        {Section~\ref{#1}}
\newcommand {\aref}[1]        {Appendix~\ref{#1}}
\newcommand {\fref}[1]        {Fig.~\ref{#1}}
\newcommand {\tref}[1]        {Table~\ref{#1}}
\newcommand {\eref}[1]        {Eq.~(\ref{#1})}

\newcommand {\be}             {\begin{equation}}
\newcommand {\ee}             {\end{equation}}

\defcitealias{Chiang2023}{C23}

\newcommand{\<}{\langle}
\renewcommand{\>}{\rangle}

\newcommand{\pd}{\partial}
\newcommand{\Mgra}{M_\text{gra}}
\newcommand{\vc}{v_\text{cir}}
\newcommand{\vbf}{\boldsymbol{v}}
\newcommand{\bigM}{\mathcal{M}}
\newcommand{\bigH}{\mathcal{H}}
\newcommand{\bigG}{\mathbb{G}}
\newcommand{\bigD}{\mathcal{D}}

\newcommand{\lambdadB}{\lambda_\text{dB}}
\newcommand{\rhobgeff}{\rho_\text{bg}^\text{eff}}

\newcommand{\Xeff}{X_\text{eff}}
\newcommand{\erf}{\text{erf}}
\newcommand{\zmax}{z_\text{max}}

\tikzstyle{galic} = [rectangle, rounded corners, text width=8em, minimum width=2em, minimum height=2em,text centered, draw=black, fill=red!30]
\tikzstyle{startstop} = [rectangle, rounded corners, text width=8em, minimum width=2em, minimum height=2em,text centered, draw=black, fill=cyan!30]
\tikzstyle{io} = [rectangle, text width=5em, minimum width=2em, minimum height=2em, text centered, draw=black, fill=blue!30]
\tikzstyle{process} = [rectangle, text width=8em, minimum width=2em, minimum height=2em, text centered, draw=black, fill=orange!30]
\tikzstyle{trash} = [rectangle, rounded corners, text width=5em, minimum width=2em, minimum height=2em, text centered, draw=black, fill=gray!30]
\tikzstyle{arrow} = [thick,->,>=stealth]
\tikzstyle{tarrow} = [thick,|->,>=stealth]

\begin{document}
\begin{CJK*}{UTF8}{bkai}

\title[FDM-driven Galactic Disc Heating]{Galactic disc heating by density granulation in fuzzy dark matter simulations}

\author[H. Yang et al.]{Hsun-Yeong Yang$^{1}$,
	Barry T. Chiang$^{1,2,3}$, 
	Guan-Ming Su$^{1}$, 
	Hsi-Yu Schive (薛熙于)$^{1,4,5,6}$\thanks{E-mail: hyschive@phys.ntu.edu.tw},
    \newauthor 
	Tzihong Chiueh (闕志鴻)$^{1,4,5}$, and
	Jeremiah P. Ostriker$^{3,7}$
	\vspace*{8pt}	
	\\
	$^{1}$Institute of Astrophysics, National Taiwan University, Taipei 10617, Taiwan.\\
	$^{2}$Department of Astronomy, Yale University, New Haven, CT 06511, USA.\\
	$^{3}$Department of Astronomy, Columbia University, New York, NY 10027, USA.\\
	$^{4}$Department of Physics, National Taiwan University, Taipei 10617, Taiwan.\\
	$^{5}$Center for Theoretical Physics, National Taiwan University, Taipei 10617, Taiwan.\\
	$^{6}$Physics Division, National Center for Theoretical Sciences, Taipei 10617, Taiwan.\\
	$^{7}$Department of Astrophysical Sciences, Princeton University, 4 Ivy Lane, Princeton, NJ, 08544, USA.}

\date{Accepted XXX. Received YYY; in original form ZZZ}

\pubyear{2023}
\label{firstpage}
\pagerange{\pageref{firstpage}--\pageref{lastpage}}
\maketitle
\end{CJK*}

% abstract
% ----------------------------------------------
\begin{abstract}
Fuzzy dark matter (FDM), an attractive dark matter candidate comprising ultralight bosons (axions) with a particle mass $m_a\sim10^{-22}$~eV, is motivated by the small-scale challenges of cold dark matter and features a kpc-size de Broglie wavelength. Quantum wave interference inside an FDM halo gives rise to stochastically fluctuating density granulation; the resulting gravitational perturbations could drive significant disc thickening, providing a natural explanation for galactic thick discs. Here we present the first self-consistent simulations of FDM haloes and stellar discs, exploring $m_a=0.2$\textendash$1.2\times10^{-22}$~eV and halo masses $\Mh = 0.7$\textendash$2.8\times10^{11}$~M$_\odot$. Disc thickening is observed in all simulated systems. The disc heating rates are approximately constant in time and increase substantially with decreasing $m_a$, reaching $dh/dt \simeq 0.04$ ($0.4$)~kpc\,Gyr$^{-1}$ and $d\sigma_z^2/dt \simeq4$ ($150$)~km$^2$s$^{-2}$Gyr$^{-1}$ for $m_a=1.2$ ($0.2$) $\times10^{-22}$~eV and $\Mh=7\times10^{10} \Msun$, where $h$ is the disc scale height and $\sigma_z$ is the vertical velocity dispersion. These simulated heating rates agree within a factor of two with the theoretical estimates of Chiang et al., confirming that the rough estimate of Church et al. overpredicts the granulation-driven disc heating rate by two orders of magnitude. However, the simulation-inferred heating rates scale less steeply than the theoretically predicted relation $d\sigma^2_z/dt \propto m_a^{-3}$. Finally, we examine the applicability of the Fokker\textendash Planck approximation in FDM granulation modelling and the robustness of the $m_a$ exclusion bound derived from the Galactic disc kinematics.
\end{abstract}


% section 1
% ----------------------------------------------
\section{Introduction}
\label{sec:intro}

The rich complexity in dynamical evolution of galactic discs has become observationally evident in the past decades. Locally, large-scale spectroscopic surveys including \textit{Gaia} \citep{GaiaCollaboration2016A&A595A}, APOGEE \citep{Majewski2017AJ154}, \textit{Kepler} \citep{Borucki2010Sci327}, LAMOST \citep{Zhao2012LAMOST}, RAVE \citep{RAVE2006AJ1321645S}, and GALAH \citep{Martell2017MNRAS465} have revealed many disequilibrium structures \citep[e.g.][]{Antoja2018Natur561360A, Schonrich2018MNRAS4783809S, Bennett2019MNRAS4821417B} and made possible detailed inferences of the secular thickening process \citep[e.g.][]{Mackereth2019MNRAS, Sharma2021MNRAS506} in the Milky Way (MW) stellar disc. To better understand these out-of-equilibrium features, there are renewed theoretical interests in the response of stellar discs to generic perturbations \citep*[e.g.][]{Banik2022ApJ935135B, Dootson2022arXiv220515725D} and a plethora of proposed physical interpretations \citep[e.g.][]{Widrow2012ApJ750L41W, Binney2018MNRAS4811501B, Michtchenko2019ApJ87636M, Khoperskov2019A&A622L6K}. Recently, \citet*{Tremaine2023MNRAS521114T} investigated an intriguing possibility that Gaia phase spiral could be generated by gravitational perturbations caused by giant molecular clouds or dark substructures of similar masses.

Persistent local perturbations in the background gravitational potential, agnostic to the sources, can lead to significant stellar heating over a Hubble time \citep[e.g.][]{Chandrasekhar1942psdbookC, Chavanis2013A&A556A93C}. The Galactic thick disc populations, with stellar ages $\gtrsim 8$~Gyr, exhibit substantially higher velocity dispersion than the kinematically cold and young thin disc populations \citep[e.g.][]{Bland-Hawthorn2016ARA&A54, Miglio2021A&A645A}. This morphological evolution of disc thickening is increasingly well constrained \citep[e.g.][]{Ting2019ApJ878, Mackereth2019MNRAS, Sharma2021MNRAS506}, albeit uncertain in the underlying physical mechanisms. Beyond the MW, the formation of old and kinematically hot thick disc components \citep{Yoachim2005ApJ624701Y, Yoachim2008ApJ683707Y} are also observed in external galaxies across wide ranges of Hubble types, stellar masses, stellar mass-to-light ratios, and thick-to-thin disc mass ratios \citep{Tsikoudi1979ApJ234842T, Burstein1979ApJ234829B, vanderKruit1981A&A95105V, Dalcanton2002AJ1241328D, Yoachim2006AJ131226Y, Comeron2014A&A571A58C}. 

The exact thick disc formation pathway is still debated in the literature, and various qualitatively distinct models have been proposed and can be largely grouped into two classes. In the first category, disc stars are kinematically cold at birth, and the morphologically thin disc is dynamically heated over time to form the thick disc. Some possible sources of stellar heating include the gravitational interactions of disc stars with infalling satellites \citep{Toth1992ApJ3895T, Quinn1993ApJ40374Q}, giant molecular clouds \citep{Spitzer1951ApJ114385S, Lacey1984MNRAS208687L}, spiral arms \citep{Barbanis1967ApJ150461B, Carlberg1985ApJ29279C}, or the Galactic bar \citep{Minchev2010ApJ722112M}. These secular processes could also drive radial migration of disc stars \citep{Schonrich2009MNRAS396203S, Loebman2011ApJ7378L}. The second class of models advocates that, via violent and rapid gravitational-instability-driven heating, disc stars are already kinematically hot during the initial assembly process to produce a morphologically thick disc. This scenario could be achieved via gas-rich major mergers \citep{Brook2004ApJ612894B}, early turbulent and clumpy phases of disc evolution \citep{Kroupa2002MNRAS330707K, Bournaud2009ApJ707L1B}, clustered star formation \citep{Kroupa2002MNRAS330707K}, and/or direct accretion of tidally stripped satellite galaxies \citep{Abadi2003ApJ59721A, Martin2004MNRAS34812M}.

\citet*{Chiang2023}, henceforth \citetalias{Chiang2023}, recently explored another possible Galactic thick disc formation scenario in a Fuzzy Dark Matter (FDM) universe. As a promising alternative to cold dark matter (CDM), the FDM paradigm reproduces the large-scale successes of the $\Lambda$CDM cosmological model \citep*{Schive2014a}. The ultralight mass scale $m_a\sim 10^{-22}$~eV yields a kpc/sub-kpc-size de Broglie wavelength and could address several small-scale discrepancies of $\Lambda$CDM \citep[e.g.][]{Schive2014PhRvL, Hui:2016ltb,Marsh:2015wka,Calabrese:2016hmp, Chen2017MNRAS4681338C, Leung2018ApJ862156L,Wasserman2019ApJ885, Amruth2023NatAstmp104A}; on the flip side, this mass range is disfavoured in some other analyses (e.g. \citealp{Bernal2018MNRAS4751447B, Schutz2020PhRvD101l3026S, Hayashi2021ApJ912L3H, Rogers2021PhRvL126g1302R, Nadler2021PhRvL126i1101N, Powell2023arXiv230210941P}; see however \citetalias{Chiang2023}). 

Quantum wave interference inside an FDM halo gives rise to ubiquitous de-Broglie-scale density granulation that are stochastically distributed and undamped on cosmological timescales (e.g. \citealp{Schive2014a}; \citealp*{Veltmaat:2018dfz}). In an MW-sized host halo, these locally fluctuating granules weight comparably to giant molecular clouds $\sim 10^{6}$~M$_\odot$. By incorporating the full baryon and dark matter distributions of the MW together with stellar disc kinematics inferred from \textit{Gaia}, APOGEE, and LAMOST surveys, \citetalias{Chiang2023} provided a detailed estimate of the Galactic disc heating rate based on the Fokker\textendash Planck approximation formalism first developed in \cite*{Bar-Or2019ApJ}. We found that granulation- and subhalo-induced stellar heating can quantitatively reproduce both the observed `U-shaped' disc vertical velocity dispersion profile \citep{Sanders2018MNRAS4814093S} and the age\textendash velocity dispersion relation in the solar neighbourhood \citep{Sharma2021MNRAS506}. This particular observable places a conservative exclusion bound $m_a \gtrsim 0.4\times 10^{-22}$~eV and favours $m_a \simeq0.5$\textendash$0.7\times 10^{-22}$~eV. 

The main aim of this paper is to numerically verify the analytical estimates of FDM-granulation-driven stellar heating rate derived in \citetalias{Chiang2023}. We present the first self-consistent simulations of FDM haloes with embedded $N$-body stellar discs, covering a range of FDM particle masses $m_a = 0.2$\textendash$1.2\times10^{-22}$~eV and host halo masses $\Mh = 0.7$\textendash$2.8\times10^{11}$~M$_\odot$. The granulation-driven stellar disc thickening is present in all simulated systems and could provide a natural mechanism for the thick disc formation observed in external galaxies with assembly histories likely dissimilar from that of the MW.

We begin in \sref{sec:theo} with a brief review of the analytical framework for estimating FDM granulation-driven disc heating rates presented in \citetalias{Chiang2023}. An overview of the simulation setup is presented in \sref{sec:numer}, where the suitable initial condition construction methods for FDM haloes and galactic discs are detailed. \sref{subsec:disc_scale_height} presents the simulation results, with a particular focus on the galactic disc morphological evolution in FDM haloes with different $m_a$ and $\Mh$. In \sref{subsec:Simulations_vs_Theory}, the halo-granulation-induced stellar heating rate measured in self-consistent FDM simulations is compared directly against the analytical predictions of \citetalias{Chiang2023}. In \sref{sec:implications}, the relevant implications on the intrinsic uncertainty of hydrostatic-equilibrium-based $\sigma_z$ in observed external disc galaxies are discussed. We conclude in \sref{sec:concl}. Technical details are organised as follows: FDM halo-disc co-relaxation (\aref{app:RelaxationOfInitialConditions}), numerical convergence tests (\aref{app:ConvergenceTest}), and heating in CDM halo-disc simulations (\aref{app:CDM}).

In this paper, cylindrical coordinates $(R,z)$ denote the axial distance along and vertical height from the mid-plane in a galactocentric frame; $r \equiv \sqrt{R^2+z^2}$ gives the radial distance from the instantaneous galactic centre. For notational convenience, we introduce a dimensionless FDM particle mass $\ma \equiv m_a/10^{-22}\,{\rm eV}$.

% section 2
% ----------------------------------------------
\section{Theory review}\label{sec:theo}

In this section, we briefly review the analytical formalism of \citetalias{Chiang2023} for estimating FDM-granulation-driven galactic disc heating rates. Consider an axisymmetric, locally isothermal galactic disc embedded in a spherically symmetric background halo potential $\Phi_\text{bg}$. The disc three-dimensional density profile $\rhod(R,z)$ and scale height $h(R)$ are determined by solving the appropriate Poisson equation and vertical hydrostatic equilibrium condition \citep[e.g.][]{Lacey1985ApJ}
\be\label{eqn:Hydrostatic_Poisson}
\begin{cases}
	\frac{\pd^2\Phi_\text{tot}}{\pd z^2} =  4\pi G (\rhod+\rhobgeff) ,\\
	-\frac{\sigma_z^2}{\rhod}\scalebox{0.8}{\Big(}\frac{\pd \rhod}{\pd z}\scalebox{0.8}{\Big)} = \frac{\pd\Phi_\text{tot}}{\pd z},
\end{cases}
\ee
where $\Phi_\text{tot} \equiv \Phi_\text{d}+\Phi_\text{bg}$, and the effective background density reads
\be\label{eqn:rho_bg_eff}
\rhobgeff(R) \equiv  \rho_\text{bg}(R) - \frac{\pd \vc^2/\pd R}{4\pi G R},
\ee
where $\vc$ denotes the total circular velocity. Exact closed-form solutions to \eref{eqn:Hydrostatic_Poisson} exist only in either the self-gravity-dominated (SGD) limit $\rhod\gg \rhobgeff$ or the background-dominated (BGD) limit $\rhod \ll \rhobgeff$
\be\label{eqn:Disc_Den_Profiles}
\rhod(R,z)=
\begin{cases}
	\frac{\Sigma(R)}{2 h(R)} \text{sech}^2[z/h(R)]\;\;\;\;\;\;\;\;\;\;\:\text{(SGD)},\\
	\frac{\Sigma(R)}{2 h(R)} e^{-\pi z^2/[4 h^2(R)]}\;\;\;\;\;\;\;\;\;\:\text{(BGD)},
\end{cases}
\ee
with corresponding disc scale heights (\citetalias{Chiang2023})
\be\label{eqn:Scale_Height_Limits}
h(R) = 
\begin{cases}
	\frac{\sigma_z^2(R)}{\pi G \Sigma(R)}\;\;\;\;\;\;\;\;\;\;\;\;\;\;\;\;\: \text{(SGD)},\\
	\frac{\sigma_z(R)}{\sqrt{8 G \rhobgeff(R)}} \;\;\;\;\;\;\;\;\;\;\;\:\: \text{(BGD)}.
\end{cases}
\ee
The transition between these two limits can be estimated by equating the two expressions of \eref{eqn:Scale_Height_Limits}
\be\label{eqn:sigma_z_transition}
\kappa \equiv \frac{\sigma_z(R)\sqrt{8 G \rhobgeff(R)}}{\pi G \Sigma(R)},
\ee
where $\kappa \ll 1$ ($\gg1$) corresponds to the SGD (BGD) limit. For individual disc stars, the vertical oscillation period reaching the maximum vertical displacement $\pm\zmax$ away from the mid-plane $z = 0$ takes the form
\be\label{eqn:P_Definition}
P(R,\zmax) = 2\int_{-\zmax}^{\zmax} \frac{dz}{\sqrt{2[\Phi_\text{tot}(R,\zmax)-\Phi_\text{tot}(R,z)]}}.
\ee

Quantum-wave-interference-induced stochastic density fluctuations inside an FDM halo, under a suitable set of assumptions listed below, can be treated as classical quasi-particles with a typical effective mass of \citep{Bar-Or2019ApJ}
\be\label{eqn:M_gra}
\Mgra \equiv \rhoh\scalebox{0.9}{\bigg(}\frac{\lambdadB}{2\sqrt{\pi}}\scalebox{0.90}{\bigg)}^3,
\ee
where $\rhoh$ denotes the local halo density. The de Broglie wavelength $\lambdadB$ is defined as \footnote{\label{fn:Halo_Velocity_Dispersion} It is worth stressing that the \textit{genuine} halo one-dimensional velocity dispersion $\sigmah$ (i.e. kinetic energy sourced by the turbulent motion) differs by definition from the \textit{total effective} velocity dispersion $\sigma_\text{Jeans}$ (i.e. kinetic energy + quantum internal energy) obtained from solving the Jeans equation. One must take equipartition of energy $\sigmah = \sigma_\text{Jeans}/\sqrt{2}$ into account for an FDM halo, unlike the CDM counterparts; see \citet{Dutta_Chowdhury_21} and \citetalias{Chiang2023}. This physical distinction is crucial in accurately estimating the FDM granulation-driven disc heating rate that scales as $\propto \sigma_h^{-6}$; see \eref{eqn:Heating_Time_Scale}.}
\be 
\lambdadB=\frac{2\pi \hbar}{m_a(\sqrt{2}\sigmah)},
\label{eq:lambda}
\ee
where $\hbar$ is the reduced Planck constant. The granule effective radius then reads $R_\text{gra} \equiv \big(\frac{3}{4\pi}\big)^{1/3}\frac{\lambdadB}{2\sqrt{\pi}} \simeq 0.175 \lambdadB$. Under the \textit{local} assumption that these self-similar granules (1) all have identical masses and physical cutoff scales $b_\text{min}$, (2) are homogeneously distributed in space, (3) are statistically uncorrelated over the stellar oscillation timescale $P(R,\zmax)$, (4) cause net energy transfer dominantly sourced by weak encounters $b_\text{max} \equiv P\sigmah/2 \gg b_\text{min} \equiv (\lambdadB/2\pi)/2$, and (5)~exhibit a Maxwellian velocity dispersion $\sigmah$, the granulation-induced stellar heating rate $\bigH$ can be  analytically estimated via the Fokker\textendash Planck approximation. The validity thereof and predicted $\bigH$ are examined by comparing with self-consistent simulation results in \sref{subsec:Simulations_vs_Theory}. 

Following the derivation of \citetalias{Chiang2023}, we first reparametrise the one-dimensional velocity of disc stars as
\be\label{eqn:Definition_nu}
\sqrt{\frac{\vc^2 + (\sigma_\phi^2+\sigma_z^2+\sigma_R^2)}{3}} \equiv \mu(R)\sigma_z(R),
\ee
where the dimensionless factor $\mu(R)$ absorbs the radius dependence of all the non-$\sigma_z$ velocity components appearing in the Fokker\textendash Planck formulation. For the MW stellar disc, $\mu(R)\simeq 3$\textendash7 varies by about a factor of two across $R=2$\textendash20~kpc (\citetalias{Chiang2023}). The first-order and second-order diffusion coefficients describing the dynamical evolution of a test star $m$ orbiting in a granular FDM halo are \citep{Bar-Or2019ApJ}
\be\label{eqn:FDM_Diffusion_Coeff_1}
\begin{cases}
	vD[\Delta v_\parallel] = -\bigD\Xeff\scalebox{0.9}{\Big[}\bigG(\Xeff)+\frac{m}{2\Mgra}\bigG\scalebox{0.9}{\Big(}\frac{\Xeff}{\sqrt{2}}\scalebox{0.9}{\Big)}\scalebox{0.9}{\Big]},\\
	D[(\Delta v_\parallel)^2] = \bigD\frac{\bigG(\Xeff)}{\Xeff},\\
	D[(\Delta \vbf_\bot)^2] = \bigD\frac{\erf(\Xeff)-\bigG(\Xeff)}{\Xeff},
\end{cases}
\ee
where $\Xeff = \frac{\sqrt{3}\mu\sigma_z}{\sqrt{2}\sigmah}$, $\bigG(x) \equiv \frac{1}{2x^2}\big[\erf(x)-\frac{2x}{\sqrt{\pi}}e^{-x^2}\big]$, and
\be\label{eqn:FDM_Diffusion_Coeff_2}
\begin{cases}
	\bigD= \frac{4\sqrt{2}\pi G^2 \rhoh \Mgra }{\sigmah}\ln\Lambda,\\
	\Lambda=\frac{b_\text{max}}{b_\text{min}} = \frac{P\sigmah/2}{(\lambdadB/2\pi)/2} \equiv \frac{P}{\tau},
\end{cases}
\ee
and therefore $\tau = \hbar/(\sqrt{2} m_a\sigmah^2)$. In the limit $\Mgra \gg m$, the ensemble-averaged change rates in the disc vertical velocity dispersion squared in the SGD and BGD limits are (\citetalias{Chiang2023})
\be\label{eqn:Heating_Eq_SGD_BGD_Limits}
\frac{d\sigma_z^2}{dt} = 
\begin{cases}
	\frac{2\sigma_z^2}{3}\frac{d\ln\Sigma}{dt}+ 0.526\<\bigH\>_1 \;\;\;\;\;\;\;\;\:\:\:\text{(SGD)},\\
	\<\bigH\>_2 \;\;\;\;\;\;\;\;\;\;\;\;\;\;\;\;\;\;\;\;\;\;\;\;\;\;\;\;\;\;\;\;\;\;\;\:\:\text{(BGD)},
\end{cases}
\ee
where the FDM-granulation-driven heating rates $\<H\>_i$ read
\be\label{eqn:Heating_Eq_SGD_BGD_Limits_2}
\begin{cases}
	\<\bigH\>_i  \equiv \bigM\ln\Big(\frac{\<P\>_i}{\tau}\Big),\\
	\bigM \equiv \frac{\pi^{5/2} G^2 \rhoh^2\hbar^3}{2m_a^3\sigmah^4} \bigg\{\frac{4e^{-\Xeff^2}}{3\mu^2\sqrt{\pi}}+\frac{(1-\mu^{-2})[\erf(\Xeff)-\bigG(\Xeff)]}{\Xeff}\bigg\},\\
	\<P\>_1 = \frac{4.82 \sigma_z }{\pi G \Sigma},\\
	\<P\>_2 = 2\int_{-\zmax}^{\zmax} \frac{dz}{\sqrt{2[\Phi_\text{bg}(R,\zmax)-\Phi_\text{bg}(R,z)]}}.
\end{cases}
\ee
We directly apply \eref{eqn:Heating_Eq_SGD_BGD_Limits} only when $\kappa \leq 0.5$ ($\geq 1.5$) in the SGD (BGD) limit, corresponding respectively to $\frac{\Phi_\text{d}(R,\zmax)-\Phi_\text{d}(R,0)}{\Phi_\text{bg}(R,\zmax)-\Phi_\text{bg}(R,0)} \gtrsim 5$ ($\lesssim 0.2$). In the intermediate regime $0.5 < \kappa < 1.5$, we compute and linearly interpolate the heating rates in both limits \eref{eqn:Heating_Eq_SGD_BGD_Limits}.

For a sufficient large Coulomb factor $\<P\>_i/\tau \gg 1$, the stellar heating rate $\<\bigH\>_i$ is comparatively insensitive to $\<P\>_i$ and roughly scales with $\bigM$. The associated characteristic heating timescale driven by FDM halo density granulation is to leading order
\be\label{eqn:Heating_Time_Scale}
T_\text{heat} \equiv \frac{\sigmah^2}{d\sigma_z^2/dt} \propto m_a^3\sigmah^6\rhoh^{-2},
\ee
identical in both the SGD and BGD limits. The granulation-driven disc heating rate $\bigH \propto T_\text{heat}^{-1}$ is sensitive to the halo attributes $\sigmah$ and $\rhoh$. In contrast, the $\sigma_z$-dependence enters $\bigH$ only through the vertical oscillation period in the SGD limit and is logarithmically suppressed. 
The disc vertical heating rate is thus independent of $\sigma_z(t)$ to zeroth order. For a given FDM halo (i.e. fixing $m_a, \sigmah,$ and $\rhoh$), the approximately time-independent $\bigH$ implies that $\sigma_z^2$ and the disc scale height $h\propto \sigma_z^2$ in the SGD limit \eref{eqn:Scale_Height_Limits} grow linearly with time, a prediction to be compared with simulation results in \sref{subsec:Simulations_vs_Theory}.

% section 3
% ----------------------------------------------
\section{Simulation methods}
\label{sec:numer}
We describe the initial condition construction of FDM haloes in \sref{subsec:init_halo} and galactic discs in \sref{subsec:init_disc}. The adaptive mesh refinement (AMR) criteria suitable for simulating dynamical co-evolution of FDM haloes and $N$-body galactic discs are detailed in \sref{subsec:amr}. We outline the analysis of disc properties in  \sref{subsec:disc_property}.

\subsection{Initial conditions of FDM haloes}
\label{subsec:init_halo}

The FDM halo initial conditions are constructed using the algorithm presented in \citet{Lin2018} that, provided an input density profile, iteratively solves for self-consistent wave distribution function from the Schr\"{o}dinger\textendash Poisson equations. The distribution function for each wave function eigen-state follows the King model \citep{King1966AJ.....71...64K}
\be\label{eqn:Fermionic model}
    f_{\mathrm{King}} = 
    \begin{cases}
    A\exp(-\beta(E-E_c))-1 \;\quad &\;\text{(if } E\le E_c\text{)},\\
    0 &\;\text{(otherwise),}
    \end{cases}
\ee
which dependents merely on the eigen-energy $E$. The parameters $A$, $\beta$, and $E_c$ are chosen properly for each halo, such that the final density profile of the outer halo can roughly follow the assigned NFW profile \citep{NFW1996ApJ} with halo concentration parameter $c_{\rm h}$ and halo virial mass $\Mh$ given as input, meanwhile the total FDM halo mass inside the virial radius $\rvir$ can be kept as close to the desired $\Mh$ as possible. We also confirm that the selected parameters can render a convergent final density profiles after sufficient iterations. Alternatively, one can also use other distribution functions to construct the FDM halo, such as the distribution function obtained from the Eddington formula \citep[see][]{Dalal2021}. 

In this work, $\rvir$ is defined as the radius within which the mean enclosed density is 347 times the critical density.
We construct four FDM haloes of identical virial masses $\Mh=7\times10^{10}\Msun$ with different FDM particle masses $\ma=0.2$, 0.4, 0.8, and 1.2, all yielding $\rvir \simeq 102.6$~kpc. Fixing $\ma=0.4$, we add two additional cases with heavier halo masses $\Mh=1.4\times10^{11}\Msun$ and $2.8\times10^{11}\Msun$ that have $\rvir$ 129.3~kpc and 162.9~kpc, respectively. The adopted halo properties are listed in \tref{tab:SimulationSetup}.
\begin{table}
	\centering
	\begin{tabular}{lcccc}
		\hline
		Case     &$\ma$            & $\Mh$                     & $\Nd$              & Maximum resolution\\ \hline
		1        &0.2             & $7.0\times10^{10}\Msun$   & $1.6\times10^8$    & $0.062 \kpc$\\
		1(a)     &0.2             & $7.0\times10^{10}\Msun$   & $8.0\times10^7$    & $0.062 \kpc$\\			
		1(b)     &0.2             & $7.0\times10^{10}\Msun$   & $2.0\times10^7$    & $0.062 \kpc$\\
        1(c)     &0.2             & $7.0\times10^{10}\Msun$   & $1.6\times10^8$    & $0.124 \kpc$\\
		2        &0.4             & $7.0\times10^{10}\Msun$   & $8.0\times10^7$    & $0.060 \kpc$\\		
		2(a)     &0.4             & $7.0\times10^{10}\Msun$   & $2.0\times10^7$    & $0.060 \kpc$\\		
		2(b)     &0.4             & $7.0\times10^{10}\Msun$   & $8.0\times10^7$    & $0.120 \kpc$\\
		3        &0.8             & $7.0\times10^{10}\Msun$   & $1.6\times10^8$    & $0.056 \kpc$\\
		3(a)     &0.8             & $7.0\times10^{10}\Msun$   & $8.0\times10^7$    & $0.056 \kpc$\\
		3(b)     &0.8             & $7.0\times10^{10}\Msun$   & $2.0\times10^7$    & $0.056 \kpc$\\
        3(c)     &0.8             & $7.0\times10^{10}\Msun$   & $1.6\times10^8$    & $0.111 \kpc$\\
		4        &1.2             & $7.0\times10^{10}\Msun$   & $8.0\times10^7$    & $0.019 \kpc$\\
		4(a)     &1.2             & $7.0\times10^{10}\Msun$   & $8.0\times10^7$    & $0.038 \kpc$\\
		4(b)     &1.2             & $7.0\times10^{10}\Msun$   & $2.0\times10^7$    & $0.038 \kpc$\\
        4(c)     &1.2             & $7.0\times10^{10}\Msun$   & $8.0\times10^7$    & $0.076 \kpc$\\
		5        &0.4             & $1.4\times10^{11}\Msun$   & $8.0\times10^7$    & $0.045 \kpc$\\
		5(a)     &0.4             & $1.4\times10^{11}\Msun$   & $2.0\times10^7$    & $0.045 \kpc$\\
        5(b)     &0.4             & $1.4\times10^{11}\Msun$   & $8.0\times10^7$    & $0.090 \kpc$\\
		6        &0.4             & $2.8\times10^{11}\Msun$   & $8.0\times10^7$    & $0.036 \kpc$\\
		6(a)     &0.4             & $2.8\times10^{11}\Msun$   & $2.0\times10^7$    & $0.036 \kpc$\\
        6(b)     &0.4             & $2.8\times10^{11}\Msun$   & $8.0\times10^7$    & $0.071 \kpc$\\
	\end{tabular}
	\caption{The FDM particle mass $\ma$ and virial halo mass $\Mh$ for all six cases simulated in this work. In each setup, the production run adopts the highest disc particle number $\Nd$ and spatial resolution. Simulations with lower particle resolutions (denoted with lowercase letters) are carried out to verify numerical convergence, as discussed in \aref{app:ConvergenceTest}.}
	\label{tab:SimulationSetup}
\end{table}

\begin{figure}
	\centering
	\includegraphics[width=\columnwidth]{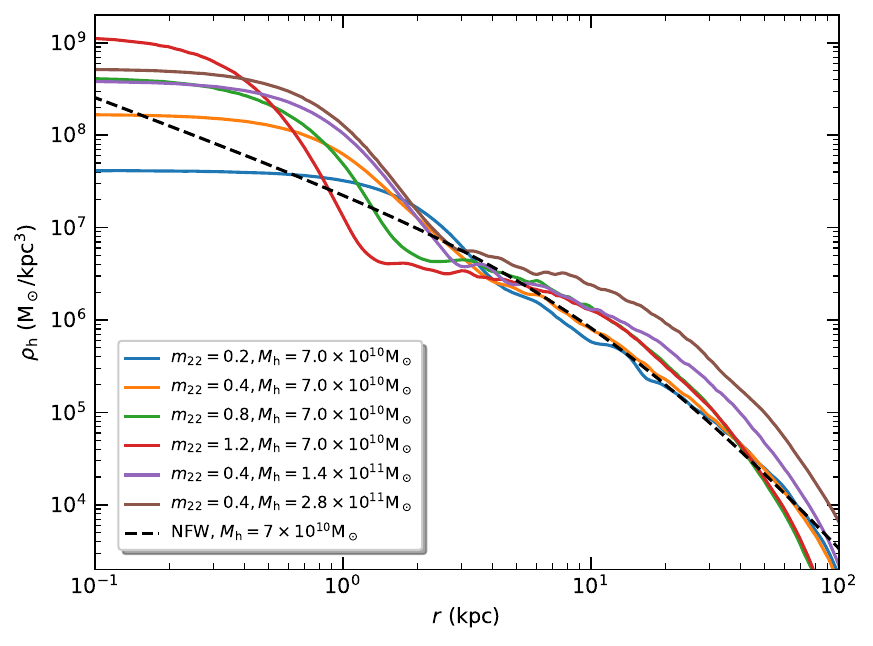}
	\caption{
		Initial shell-averaged density profiles of all six FDM haloes listed in \tref{tab:SimulationSetup} (solid), compared with an NFW profile with $\Mh=7.0\times10^{10}\Msun$ and $c_\text{h} = 8$ (dashed). For a fixed halo mass $\Mh$, the central soliton core exhibits higher peak density and smaller core radius $\rc$ for heavier FDM particle mass $\ma$. Beyond the characteristic density profile transition at $\simeq 3.3\rc$, the halo density distribution roughly follows the NFW profile.}
	\label{fig:halo_dens}
\end{figure}

The physical properties of the central soliton core follow the soliton core-halo relation of \citet{Schive2014PhRvL}
\be
\rc=1.6\ma^{-1}\left(\frac{\Mh}{10^{9}\Msun}\right)^{-1/3}\kpc
\label{eq:core-halo}
\ee 
and the soliton scaling relation
\be
\rho_0=0.019\ma^{-2}\left(\frac{\rc}{\kpc}\right)^{-4}\Msun{\rm pc}^{-3},
\label{eq:core-scale}
\ee
where $\rho_0$ is the time-averaged peak density of the soliton and $\rc$ is the half-density radius $\rho_{\rm soliton}(\rc)=\rho_0/2$. The soliton core extends out to $\simeq 3.3\rc$ \citep{Chiang:2021uvt}, outside of which the halo density profiles listed in  \tref{tab:SimulationSetup} can all be well fitted by the NFW profile with $c_{\rm h}=8$.
The shell-averaged initial halo density profiles simulated in this work are shown in \fref{fig:halo_dens}. Within $3.3\rc \leq r \leq15$~kpc, the one-dimensional halo velocity dispersion is roughly radius-independent and ranges from $\sigmah \simeq 35$km s$^{-1}$ for the lightest haloes simulated in this work $\Mh=7.0\times10^{10}\Msun$ to $\sigmah \simeq 60$km s$^{-1}$ for the most massive halo $\Mh=2.8\times10^{11}\Msun$.

\begin{figure*}
\centering
\includegraphics[width=\textwidth]{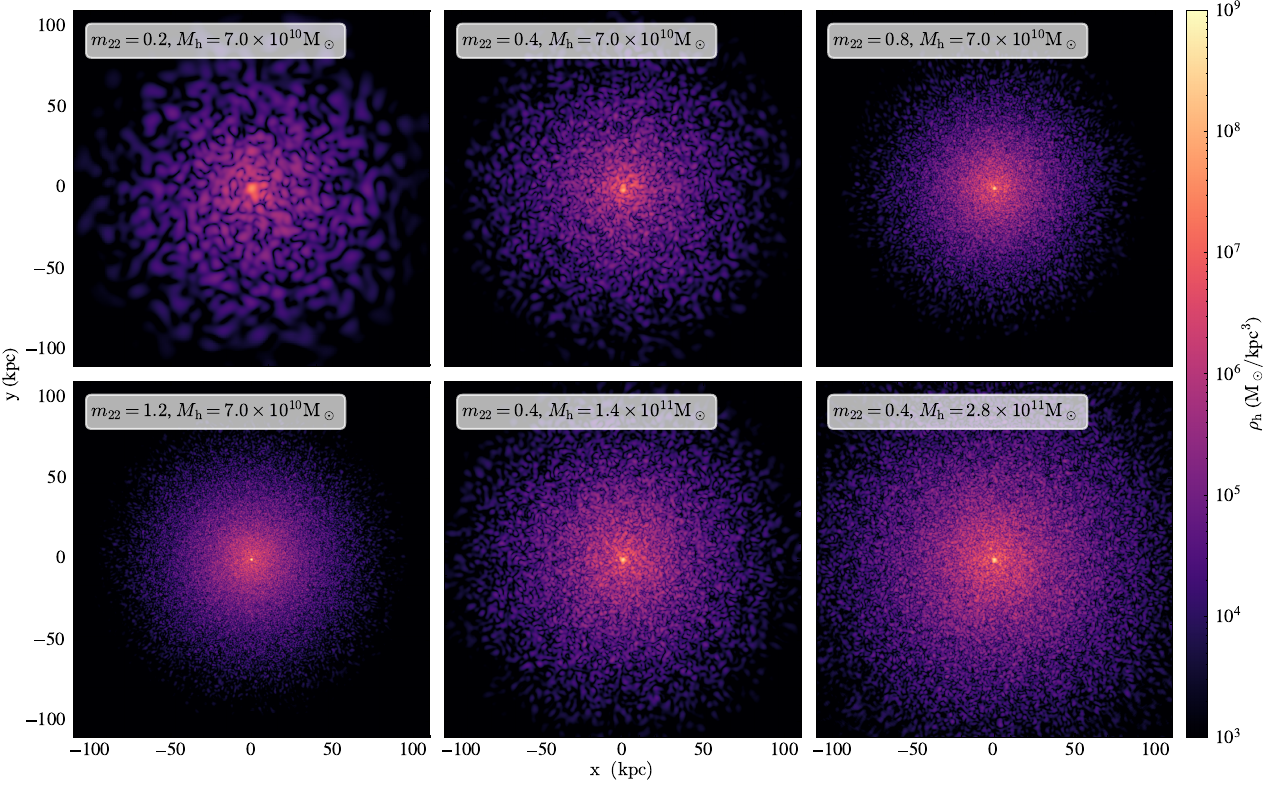}
\caption{
Density slices of relaxed haloes at $\tr = 0$~Gyr for all six cases listed in \tref{tab:SimulationSetup}. The de Broglie wavelength $\lambdadB$, \eref{eq:lambda}, and consequently the FDM density granulation length scale decrease for higher FDM particle mass $\ma$ and larger virial halo mass $\Mh$.}
\label{fig:halo_slice}
\end{figure*}

The free-fall times of these six FDM haloes
\be
T_{\rm ff}=\sqrt{\frac{\pi^2 \rvir^3}{8\Mh G}}.
\label{eq:free_fall}
\ee
are all comparable, giving $T_{\rm ff} \simeq 2.1$~Gyr. Since the constructed FDM haloes generally reach a state of dynamical quasi-equilibrium within $\lesssim 0.5T_{\rm ff}$, we first dynamically relax each newly constructed FDM halo in isolation for $2.1$~Gyr.
The density slices of the relaxed haloes are presented in \fref{fig:halo_slice}. The ubiquitous granular structures in each halo density slice arise from the quantum wave interference.
The granules have characteristic length scales comparable to the local de Broglie wavelength \eref{eq:lambda}. The typical physical sizes of density granulation increase for smaller FDM particle masses $\ma$.
In this work, we quantify the disc heating effect caused by these fluctuating granular structures.

\subsection{Initial conditions of galactic discs}
\label{subsec:init_disc}

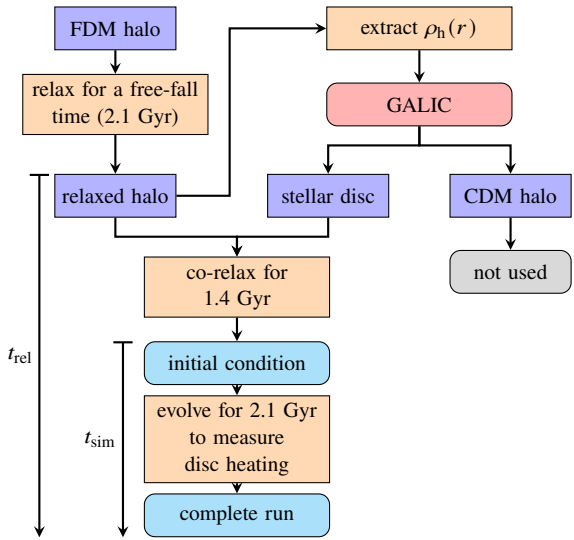
\begin{figure}
	\centering
	\begin{tikzpicture}[node distance=1cm]
		\node (halo)     [io] {FDM halo};
		\node (rel_ff)   [process, below of=halo] {relax for a free-fall time (2.1 Gyr)};
		\node (rel_halo) [io, below of=rel_ff, node distance=1.2cm] {relaxed halo};
		\node (get_rho)  [process, right of=halo, xshift=3cm] {extract $\rho_{\rm h}(r)$};
		\node (GALIC)    [galic, below of=get_rho] {GALIC};
		\node (disc)     [io, below of=GALIC, node distance=1.2cm, xshift=-1.2cm] {stellar disc};
		\node (cdm)      [io, right of=disc, node distance=1.2cm, xshift=1.2cm] {CDM halo};
		\node (no_use)   [trash, below of=cdm] {not used};
		\node (co_rel)   [process, below of=disc, node distance=1.2cm, xshift=-1.2cm] {co-relax for 1.4 Gyr};
		\node (IC)       [startstop, below of=co_rel] {initial condition};
		\node (sim)      [process, below of=IC] {evolve for 2.1~Gyr to measure disc heating};
		\node (end)      [startstop, below of=sim] {complete run};
		\coordinate [left=1cm of rel_halo.north] (tr) ;
		\coordinate [left=1.5cm of IC.north] (ts) ;
		\draw[arrow] (halo) -- (rel_ff);
		\draw[arrow] (rel_ff) -- (rel_halo);
		\draw[arrow] (rel_halo) -|([xshift=0.7cm, yshift=0cm]rel_halo.east)|- (get_rho);
		\draw[arrow] (get_rho) -- (GALIC);
		\draw[arrow] (GALIC.south) - ++ (0, -0.25cm) -| (disc);
		\draw[arrow] (GALIC.south) - ++ (0, -0.25cm) -| (cdm);
		\draw[arrow] (cdm) -- (no_use);
		\draw[arrow] (disc.south) -- ++(0,-0.25cm) -| (co_rel);
		\draw[arrow] (rel_halo.south) -- ++(0,-0.25cm) -| (co_rel);
		\draw[arrow] (co_rel) -- (IC);
		\draw[arrow] (IC) -- (sim);
		\draw[arrow] (sim) -- (end);
		\draw[tarrow] (tr) -- node[xshift=-0.25cm]{$\tr$} (end.south -|tr) ;
		\draw[tarrow] (ts) -- node[xshift=-0.3cm]{$\ts$} (end.south -|ts) ;
	\end{tikzpicture}
	\caption{
		Flowchart of the disc-halo simulation setup.
			In each of the six cases listed in \tref{tab:SimulationSetup}, we first dynamically co-evolve the $\GALIC$-generated stellar disc and the relaxed FDM halo over $\tr = 0$\textendash1.4~Gyr. At the end of this co-relaxation phase that we define as $\ts = 0$~Gyr, the disc configuration reaches a quasi-equilibrium state; we continue each simulation for another 2.1~Gyr and quantify the granulation-driven disc heating rate. Hence $\tr\equiv\ts+1.4$~Gyr for all cases.
	}
	\label{fig:flowchart}
\end{figure}

The stellar discs are constructed using $\GALIC$ \citep{Yurin2014}, a code that generates $N$-body halo-disc initial conditions by iteratively optimizing the velocities of populated particles to reach an approximate equilibrium solution to the collisionless Boltzmann equation. However, the predefined analytic density distribution functions implemented in $\GALIC$ only support CDM haloes parameterised by the Hernquist profile. We have thus modified the source code to take any shell-averaged FDM halo density profiles constructed in \sref{subsec:init_halo} as inputs, and output initial conditions of galactic discs and $N$-body haloes that match the desired density profiles. After discarding the halo particles, we introduce the disc component to the dynamically relaxed FDM halo and perform self-consistent halo-disc simulations as outlined in the flowchart \fref{fig:flowchart}.

The initial axisymmetric stellar discs follow the canonical exponential surface density profile
\be\label{eqn:Expo_Surface_Den}
\Sigma(R) = \int_{-\infty}^\infty dz \rhod(R,z) = \Sigma_0 e^{-R/\Rd},
\ee
where $\Rd$ denotes the disc scale radius, and the disc total mass is $\Md = \int_{0}^{\infty} 2\pi R \Sigma(R) dR = 2\pi \Sigma_0\Rd^2.$ The initial three-dimensional disc density profile is taken to be (cf. \eref{eqn:Disc_Den_Profiles})
\be
\rhod(R,z)=\frac{\Sigma_0}{2 h(R)}{\rm {sech}}^2\left(\frac{z}{h(R)}\right){\rm {exp}}\left(\frac{-R}{\Rd}\right),
\label{eq:disc}
\ee
where $h(R)$ is the disc scale height profile.
In all the six simulated FDM haloes, the discs have the same total mass $\Md=3.16\times 10^{9} \Msun$, scale radius $\Rd=3.0 \kpc$, and initially radius-independent scale 
height $h=0.15 \kpc$. Each $N$-body galactic disc is constructed with a total number of $\Nd$ identical particles; each disc star particle hence weights $\md=\Md/\Nd$. The adopted $\Nd$ in each production run listed in \tref{tab:SimulationSetup} has been numerically verified to have sufficiently high mass resolution for numerical convergence (see Appendix~\ref{app:ConvergenceTest} for details).

The initial disc velocity dispersion profile is obtained by solving the cylindrically symmetric Jeans equations
\be
\begin{cases}\label{eqn:Jeans_Disc}
\frac{\pd\left(\rhod\sigma_z^2\right)}{\pd z}+\rhod\frac{\pd\Phi_\text{tot}}{\pd z}=0,\\
\left<v_\phi^2\right>=\sigma_R^2+\frac{R}{\rhod}\frac{\pd\left(\rhod\sigma_R^2\right)}{\pd R}+R\frac{\pd\Phi_\text{tot}}{\pd R},
\end{cases}
\ee
where $\Phi_\text{tot} \equiv \Phi_\text{d}+\Phi_\text{bg}$ denotes the total gravitational potential sourced by both the stellar disc and the background halo. $v_\phi$ is the azimuthal velocity of disc stars and related to the total circular velocity defined in \eref{eqn:rho_bg_eff} via $\vc^2 = \left<v_\phi^2\right> - \sigma_\phi^2$; $\left<\right>$ is the ensemble average. We let $\sigma_z=\sigma_R=\sigma_\phi$ when the disc is generated, where $\sigma_z$, $\sigma_R$, and $\sigma_\phi$ are the vertical, radial, and 
azimuthal stellar velocity dispersion. The velocity distribution is set to be Gaussian.

Although each constructed disc is initially in equilibrium with the spherical and smoothed background halo potential, the disc configuration can still significantly evolve during the initial disc-halo co-evolution due to following three sources of perturbations: (1) The additional of a disc component causes an oscillatory quadrupolar distortion in the initially spherical host halo potential. (2) At a fixed time slice, FDM haloes are filled with local granular fluctuations that deviate from the assumption of spherical symmetry used in $\GALIC$. (3) These local density fluctuations are time-varying. The resulting potential perturbations can temporarily destabilise the initial disc configuration.
We therefore need to relax the system further before any disc properties can be reliably measured (see \aref{app:RelaxationOfInitialConditions} for details). The disc stability can be quantified by the Toomre $Q$ parameter \citep{Toomre1964ApJ1391217T}
\be
	Q \equiv \frac{\sigma_R k}{\pi G \Sigma},
\ee
where $k$ denotes the disc epicyclic frequency. The disc is said to be Toomre stable if $Q>1$ (or unstable if $Q<1$).

Dynamical evolution of FDM haloes and stellar discs are carried out using $\gamer \, $\citep{Schive2018}, a GPU-accelerated grid-based simulation code with adaptive mesh refinement (AMR). The flowchart of the entire simulation setup is presented in \fref{fig:flowchart}. We define the relaxation time $\tr$, where $\tr=0$~Gyr corresponds to the start of each self-consistent simulation. During this phase of initial relaxation where $Q(R\leq15\text{~kpc})\lesssim1$, disc spiral arm structures and local density clumps quickly develop (see \fref{fig:disc_relax}), causing the disc rotation curve to fluctuate considerably with time. Due to the aforesaid sources of potential perturbations, the disc heating rate in this period is significantly higher than that in the post-relaxation phase, making the robust inference of the granulation-driven disc heating rate difficult. This rapid increase in $\sigma_z(R)$ in turn raises the values of $Q(R)$ across the entire disc.

\begin{figure}
	\centering
	\includegraphics[width=\columnwidth]{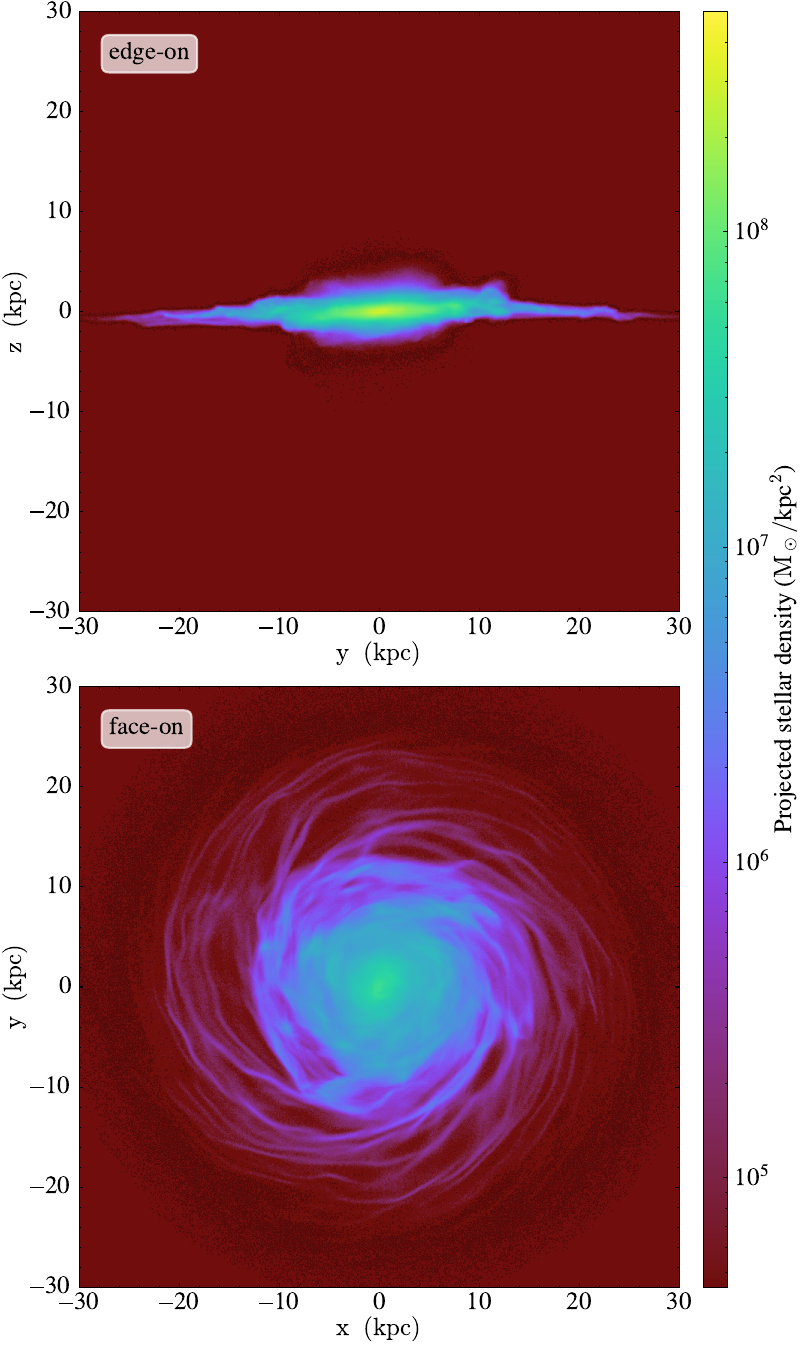}
	\caption{
		Edge-on (top) and face-on (bottom) projections of the dynamically co-relaxed stellar disc density at $\ts = 0$~Gyr for the case $\ma=0.4$ and $\Mh=7\times10^{10}\Msun$. At this stage, the disc rotation curve and surface density profiles have largely stabilised against transient structures (see also \aref{app:RelaxationOfInitialConditions} and \fref{fig:disc_relax}). For all six cases simulated in this work, the relaxed disc configuration is taken as the suitable initial condition after which granulation-driven disc heating rate can be reliably measured.
	}
	\label{fig:disc_views}
\end{figure}

\begin{figure*}
	\centering
	\includegraphics[width=\textwidth]{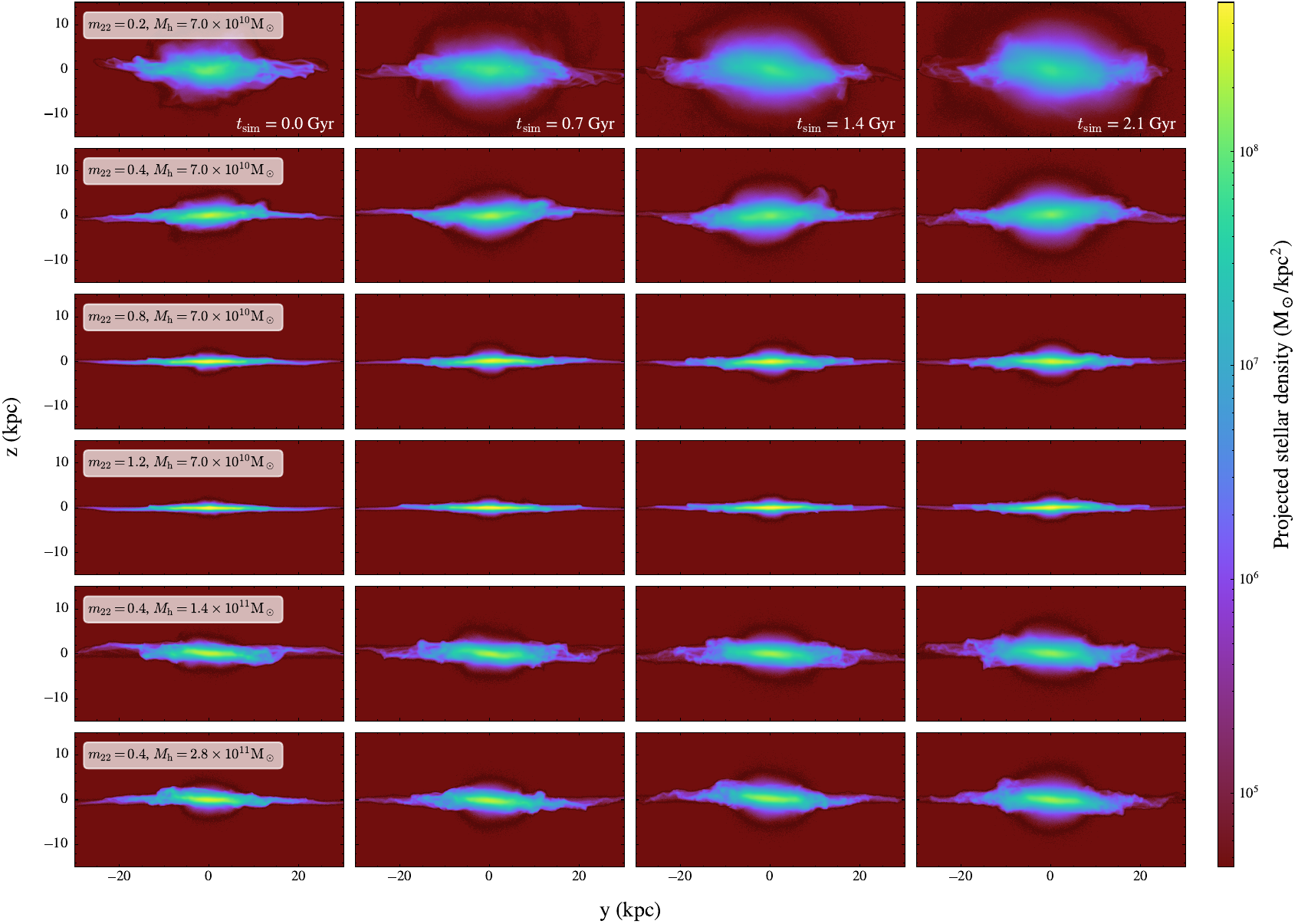}
	\caption{
		Edge-on disc density projections of all six cases listed in \tref{tab:SimulationSetup} (top to bottom row) at $\ts = 0.0, 0.7, 1.4,$~and~$2.1$~Gyr (first to fourth column). Although the relaxed initial conditions at $\ts = 0$~Gyr can non-trivially differ, all galactic discs experience granulation-driven stellar heating. For a fixed halo mass $\Mh$, the rate of disc thickening is higher for smaller $\ma$. For a fixed $\ma=0.4$, thickening rates decrease slightly with increasing $\Mh$.}
	\label{fig:disc_edge_on}
\end{figure*}

After about half a free-fall time $0.5 T_\text{ff}\simeq 1.05$~Gyr, the spiral arms and local density clumps become less pronounced; furthermore, both the disc rotation curve and halo density profile stabilise. We thus measure the granulation-driven disc heating only after $\tr \geq 1.4$~Gyr when $Q(R\leq15\text{~kpc})>1$ is always satisfied, and define $\ts=\tr-1.4$~Gyr for all cases. Each system at $\ts = 0$~Gyr is regarded to have achieved dynamical quasi-equilibrium, where non-granulation-driven disc heating sources have been minimised. \fref{fig:disc_views} shows the edge-on (top) and face-on (bottom) views of the relaxed initial condition of the disc (i.e. at $\ts = 0$~Gyr) hosted by the FDM halo with $\ma=0.4$ and $\Mh=7\times10^{10}\Msun$. The disc-relaxed initial conditions of all six cases now look very distinguishable, as shown in the leftmost column of \fref{fig:disc_edge_on}. This post-relaxation disc-halo configuration is adopted as the suitable initial condition in all our simulation runs, and further details on the relaxation process can be found in \aref{app:RelaxationOfInitialConditions}. For comparison, in \aref{app:CDM}, we have also simulated and examined a CDM halo-galactic disc system constructed by $\GALIC$, following the same numerical setup (\fref{fig:flowchart}).

\subsection{AMR criteria for halo-disc simulations}
\label{subsec:amr}

The dynamical evolution of disc-halo systems requires sufficient spatial and temporal resolution over the entire region of interest to prevent artificial disc thickening and outer mass infall of the FDM halo caused by numerical errors. In each simulation run, we adopt separate AMR schemes for the FDM halo and the galactic disc to ensure both components are always adequately resolved. If at a given location two schemes yield different refinement requirements, the more stringent (i.e. more finely resolved) condition is applied.

\begin{figure}
	\centering
	\includegraphics[width=\columnwidth]{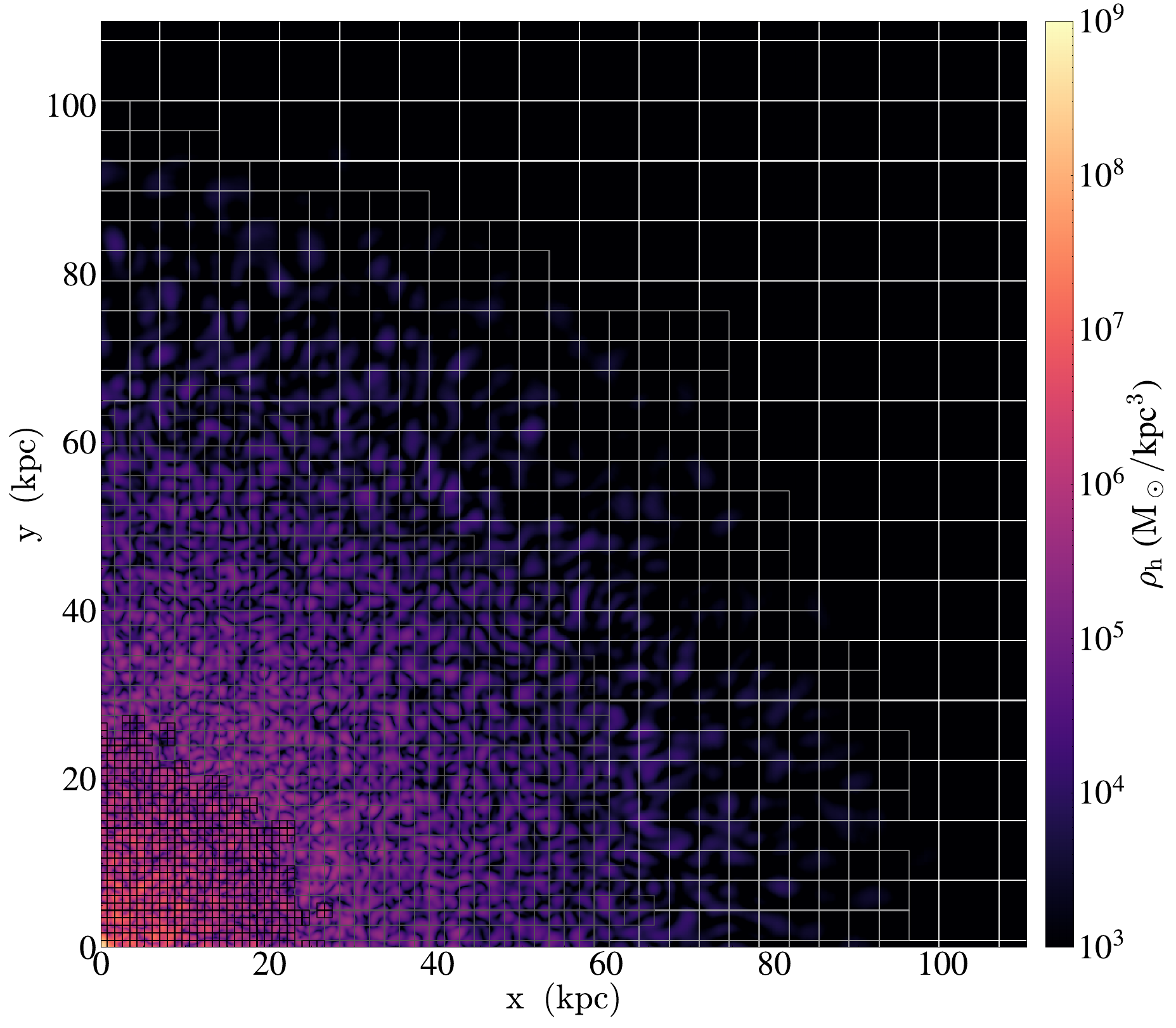}
	\caption{
		One quarter of a density slice for the halo with $\ma=0.8$, $\Mh=7\times10^{10}\Msun$ (upper right panel of \fref{fig:halo_slice}) with AMR grid overlaid. Under the disc-halo AMR scheme described in \sref{subsec:amr}, the highest refinement level should resolve both the pre-relaxed stellar disc vertically by at least five cells (0.05\textendash0.06~kpc in cell size) and the minimum granulation length scale by at least 20 cells. Here, the innermost AMR boundary occurs at the radius  $r\simeq40\rc\simeq20$~kpc. The second (third) AMR boundary transition, where the granulation length scale increases twofold (fourfold) compared to the global minimum, is located at $r\simeq60$~kpc ($\simeq100$ kpc).}
	\label{fig:halo_amr}
\end{figure}

Since the local de Broglie wavelength $\lambdadB$ \eref{eq:lambda} and hence the granulation length scale decrease with increasing FDM particle masses $\ma$ and halo velocity dispersion $\sigmah$, FDM haloes with heavier $\ma$ and/or at smaller radii require higher spatial resolution. The finest grid size 
is set to resolve the smallest density granulation size by at least 20 cells, and this finest refinement level covers the central halo region. The transition from the finest to the second finest refinement levels is set to occur roughly at $40\rc$, where $\rc$ is the soliton core radius defined in \eref{eq:core-halo}. The boundary between the second highest and the third highest refinement levels corresponds to the radius at which the granulation size doubles compared to the smallest one across the entire simulation box, which occurs at $r\simeq$ 60\textendash80~kpc for $\Mh=7\times10^{10}\Msun$ haloes.
Lastly, the transition from the third to the fourth refinement level marks the radius at which the granulation size increases fourfold compared to the smallest one. As an example, \fref{fig:halo_amr} shows the FDM halo density slice together with the grid refinement structures for the case $\ma=0.8$. The innermost AMR boundary is located at $r\simeq40\rc\simeq20$~kpc, and the transition from the second highest to the third highest refinement levels lies at $r\simeq60$~kpc.

For the stellar discs, the finest refinement level is set to resolve the initial disc scale height vertically by a total of five to six cells, giving a cell size of 0.05\textendash0.06~kpc. In practice, the refinement criteria are determined by the particle number in each cell and dynamically updated to ensure that $\geq 97$\% of all disc particles are always resolved to this highest level throughout each simulation run. In general, the cell sizes at the highest refinement level differ under the halo and disc refinement criteria. As mentioned earlier, we always adopt the more stringent refinement criteria (smaller cell size). For the six cases listed in \tref{tab:SimulationSetup}, the finest cell size is determined by the disc (halo) refinement scheme for the cases $\ma=0.2$ and $0.4$ ($\ma=0.8$ and $1.2$). These AMR criteria are tailored towards the finite difference numerical scheme for FDM simulations (see \citealp{Schive2014a} for details).

\subsection{Disc property analysis}
\label{subsec:disc_property}

The prime focus of this work is on the kinematic and morphological evolution of galactic discs with time. During the disc-halo co-evolution, the disc dynamical centre traces the trajectory of the central soliton core (\fref{fig:halo_dens}) and can differ from the centre of mass of the entire disc. Hence to reliably infer the stellar disc properties at a given time slice, we adopt the instantaneous location of the soliton peak density as the disc centre. With respect to this reference centre, the $z$-axis is taken to be the instantaneous direction of total disc angular momentum to account for any possible halo-disc misalignment. The disc properties are then computed in this cylindrical coordinate system, with $z=0$ as the disc mid-plane.

To compute the ensemble-averaged disc scale height $h(R)$ and vertical velocity dispersion $\sigma_z$, we group all disc particles into distinct annuli evenly spaced in radius $R$. In each radial bin, the scale height $h(R)$ is defined as the vertical distance such that the ratio of mass enclosed within $|z| \leq h(R)$ and the total mass equals $\tanh1\simeq0.762$, a definition consistent with \eref{eq:disc}. After projecting all disc particles onto the mid-plane, the vertical velocity dispersion $\sigma_z$ can be obtained by
\be
\sigma_z=\sqrt{\left<(v_z-\left<v_z\right>)^2\right>}
\label{eq:sigma_z}
\ee
where $v_z$ denotes the vertical velocity of individual star particles.

% section 4
% ----------------------------------------------
\section{DISC HEATING: Simulations vs. Theory}
\label{sec:result}

We quantify in \sref{subsec:disc_scale_height} the rate of galactic disc thickening observed in the self-consistent FDM simulations. Section~\ref{subsec:Simulations_vs_Theory} examines the level of consistency between simulated and predicted granulation-driven disc heating rates. We discuss in \sref{subsec:disc_heating_rate_Fokker_Planck} the applicability of the
Fokker\textendash Planck approximation in deriving the disc heating rate by \citetalias{Chiang2023}.

\subsection{Simulation results}
\label{subsec:disc_scale_height}

We first examine the disc morphological evolution in self-consistent FDM simulations. Starting from the suitably relaxed disc-halo initial conditions at $\ts = 0$~Gyr (see \sref{subsec:init_disc} and Fig.~\ref{fig:flowchart}), we dynamically co-evolve the disc-halo systems further to $\ts = 2.1$~Gyr. \fref{fig:disc_edge_on} shows the edge-on disc density projections of the six cases listed in \tref{tab:SimulationSetup} (top to bottom row) at $\ts = 0.0, 0.7, 1.4,$~and~$2.1$~Gyr (first to fourth column). Disc thickening is observed in all simulation runs. For the same FDM halo mass $\Mh$, the level of disc thickening increases for smaller $\ma$, as larger de Broglie wavelengths \eref{eq:lambda} give rise to more massive granular structures and thus stronger disc heating (see \sref{sec:theo} for the precise theoretical formulation). Similarly for a fixed $\ma = 0.4$, the disc heating rate increases for lighter halo masses $\Mh$ that exhibit lower $\sigmah(R)$ and consequently larger de Broglie wavelengths. These overall qualitative trends are carefully assessed below.

\begin{figure}
	\centering
	\includegraphics[width=\columnwidth]{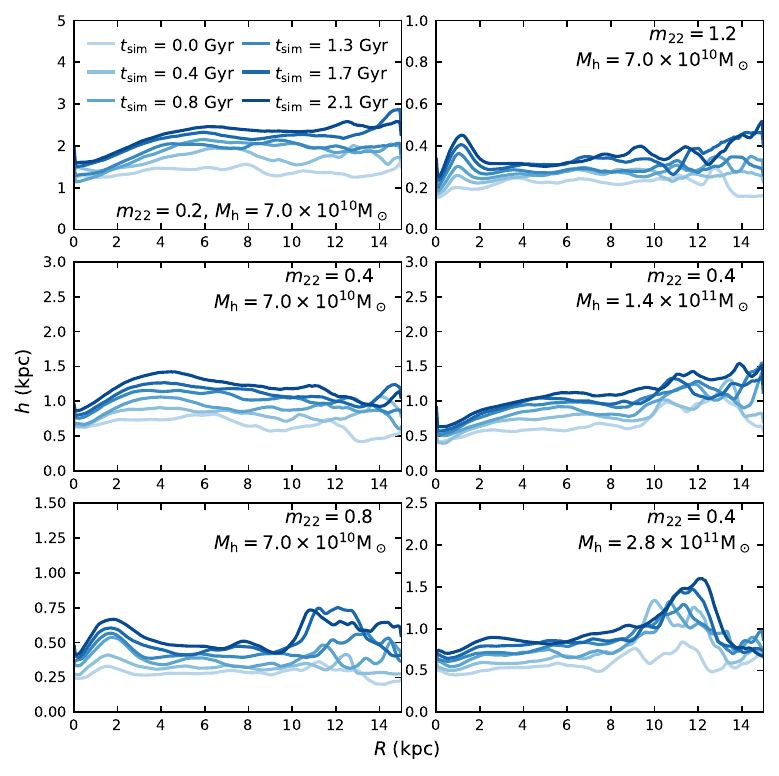}
	\caption{
		Disc scale height profiles from $\ts = 0$~Gyr (light blue) to 2.1~Gyr (dark blue). Granulation-driven stellar disc thickening occurs at all radii of interest. The disc thickening rates are noticeably higher in FDM haloes with smaller $\ma$, fixing $\Mh = 7\times10^{10} \Msun$. For the cases with $\ma=0.4$, the thickening rates are slightly higher in less massive haloes (see also \fref{fig:height_t}).}
	\label{fig:height_r}
\end{figure}

\begin{figure}
	\centering
	\includegraphics[width=\columnwidth]{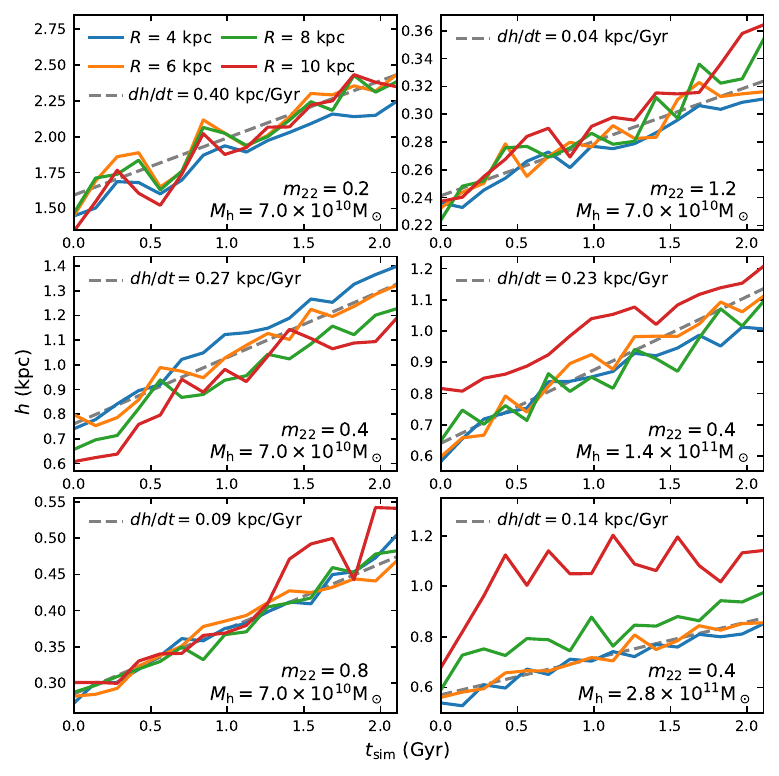}
	\caption{
		Ensemble-averaged disc scale heights in $2$-kpc-wide radial bins centred on $R = 4$~kpc~(blue), 6~(orange), 8~(green), and 10~kpc~(red) over $\ts = 0$\textendash2.1~Gyr. The dashed lines represent the linear best-fit growth curves of $R=6$~kpc bin data. The scale height grows roughly linear with time, and the thickening rate depends strongly on the value of $\ma$. In particular, the scale height increases by about 0.8~kpc (0.08~kpc) within 2~Gyr for the case $\ma=0.2$ ($\ma=1.2$), fixing $\Mh = 7\times10^{10}$~M$_\odot$. For the cases with $\ma=0.4$, smaller $\Mh$ yields slightly higher time-averaged scale height growth rates.}
	\label{fig:height_t}
\end{figure}

The time evolution of azimuthally averaged disc scale heights $h(R, \ts)$ is compared in \fref{fig:height_r}, showing a systematic increase in $h(R, \ts)$ with time. The wave-like fluctuating feature in all scale height profiles in the outer disc region $R \gtrsim 10\text{ kpc}$ is caused by the persistent spiral structures and local density clumps. For haloes with more compact solitons ($\ma=0.8, 1.2$), the scale height profiles also form sharp peaks around the transition between solitons and haloes, which occurs at $\simeq3.3r_{\rm c}=1.6$ (1.1) kpc for $\ma=0.8$ (1.2). To further quantify the disc thickening rate, \fref{fig:height_t} plots the ensemble-averaged scale height of four radial bins with $\Delta R = 2$~kpc in width centred on $R=$ 4~kpc~(blue), 6~(orange), 8~(green), and 10~kpc~(red). In all six cases, the disc scale heights increase approximately linearly with time during $\ts = 0$\textendash$2.1$~Gyr, which can be easily seen by comparing to the linear best-fit growth curves of the respective $R=6$~kpc bin data (dashed lines). Furthermore, the level of disc thickening is sensitive to $\ma$. For FDM haloes with the same virial mass $\Mh = 7.0\times10^{10}$~M$_\odot$, the averaged disc scale height increases by $\simeq 0.8$~kpc within $2$~Gyr for $\ma=0.2$, and only by $\simeq 0.08$~kpc for $\ma=1.2$. For the three haloes with $\ma=0.4$ and varying $\Mh$, the disc thickening rate is slightly higher for smaller $\Mh$. 

\begin{figure}
	\centering
	\includegraphics[width=\columnwidth]{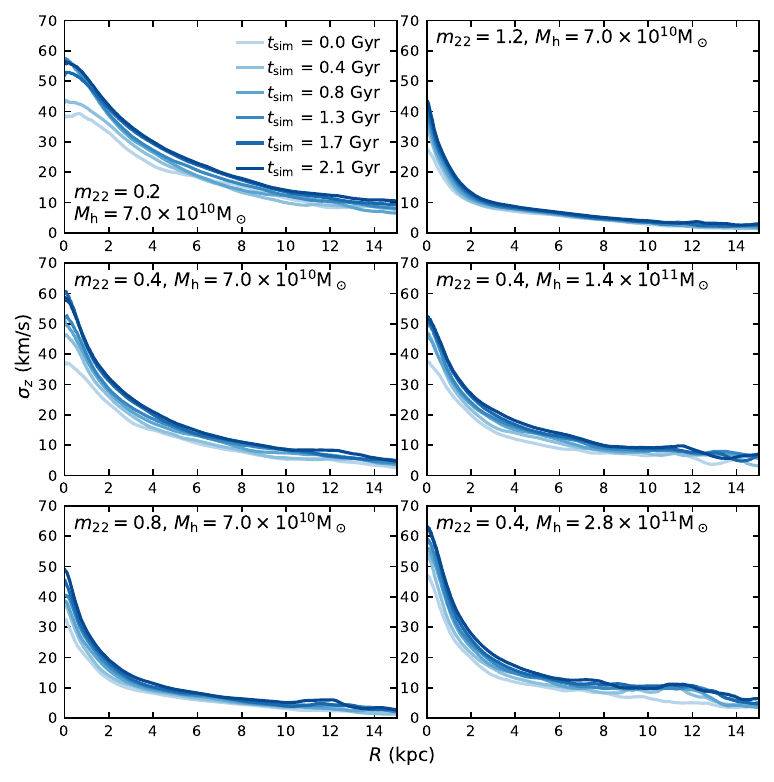}
	\caption{
		Disc vertical velocity dispersion profiles $\sigma_z(R)$ from $\ts = 0$~Gyr (light blue) to $2.1$~Gyr (dark blue). Granulation-driven heating smoothly increases $\sigma_z$ at all radii of interest.}
	\label{fig:sigma_z}
\end{figure}

The granulation-driven heating can be most directly quantified by measuring the disc vertical velocity dispersion $\sigma_z(R, \ts)$. \fref{fig:sigma_z} shows the time evolution of $\sigma_z(R\leq 15\text{ kpc}, \ts)$ from $\ts = 0$~Gyr (light blue) to $2.1$~Gyr (dark blue) for all six simulated cases listed in \tref{tab:SimulationSetup}. As expected from the universal disc thickening seen in Figs.~\ref{fig:disc_edge_on}~and~\ref{fig:height_r}, we observe that disc particles all become kinematically hotter over time. For each individual profile, the disc vertical velocity dispersion increases more rapidly for smaller $R$. Since the halo velocity dispersion $\sigmah$ between $3.3\rc$ and $15$~kpc is approximately constant for all the FDM haloes simulated in this work, the radial dependence of the disc heating rate is primarily sourced by $\rhoh(R)$. Indeed with decreasing $R$, other things being equal, the granulation-driven potential perturbations increase in magnitude due to higher local halo density $\rhoh(R)$ (\fref{fig:halo_dens}).

\begin{figure}
	\centering
	\includegraphics[width=\columnwidth]{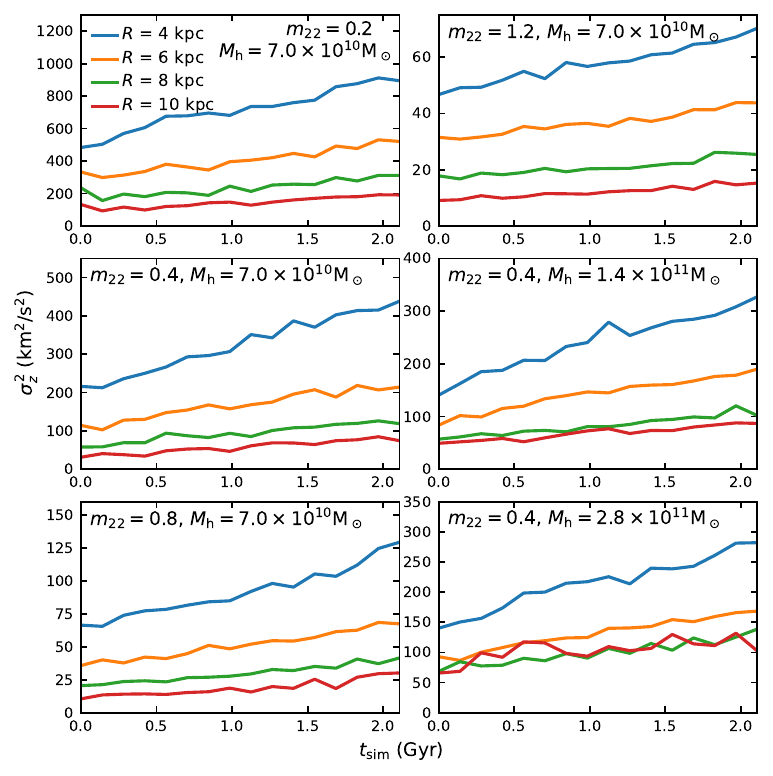}
	\caption{
		Ensemble-averaged $\sigma_z^2$ in $2$-kpc-wide radial bins centred on $R = 4$~kpc~(blue), 6~(orange), 8~(green), and 10~kpc~(red) over $\ts = 0$\textendash2.1~Gyr. We observe that $\sigma_z^2$ increases roughly linear with time, and the granulation-driven disc heating rate is generally higher for smaller $\ma$ and $R$. In contrast, the disc evolved in a CDM halo shows negligible heating $d\sigma_z^2/dt \lesssim1$~km$^2$s$^{-2}$Gyr$^{-1}$ (\fref{fig:cdm}).
	}
	\label{fig:sigma_z_sqr}
\end{figure}

We next examine the time evolution of vertical velocity squared $\sigma_z^2$ of disc particles grouped in four $2$-kpc-wide radial bins centred on $R=$~4~kpc~(blue), 6~(orange), 8~(green), and 10~kpc~(red), as shown in \fref{fig:sigma_z_sqr}. The simulation data outputs are evenly spaced in time across $\ts=0$\textendash$2.1$~Gyr, and the slope between any two adjacent data points in $\sigma_z^2$ yields the instantaneous stellar heating rate $d\sigma_z^2/dt$. For a fixed value of $\Mh$, stellar discs in FDM haloes with smaller $\ma$ experience stronger heating due to relatively enhanced granulation-driven gravitational perturbations. For the three cases with $\ma=0.4$, we observe a gentle increase in the overall disc heating rates with decreasing halo masses $\Mh = 0.7$\textendash$2.8\times10^{11}$~M$_\odot$ probed here. These trends are consistent with the disc scale height evolution shown in Figs.~\ref{fig:height_r}~and~\ref{fig:height_t}. Across all six cases, the $\sigma_z^2$ profiles generally exhibit an approximately linear growth. The only exception is the $R=10$~kpc bin in the $\ma=0.4$ and $\Mh=2.8\times10^{11}\Msun$ host halo, which has a small bump around $\ts=0.6$~Gyr, and its scale height also grows more rapidly compared to other radii as shown in \fref{fig:height_t}.


\subsection{Disc heating rates: Simulations vs. Theory}\label{subsec:Simulations_vs_Theory}

\begin{figure}
	\centering
	\includegraphics[width=\columnwidth]{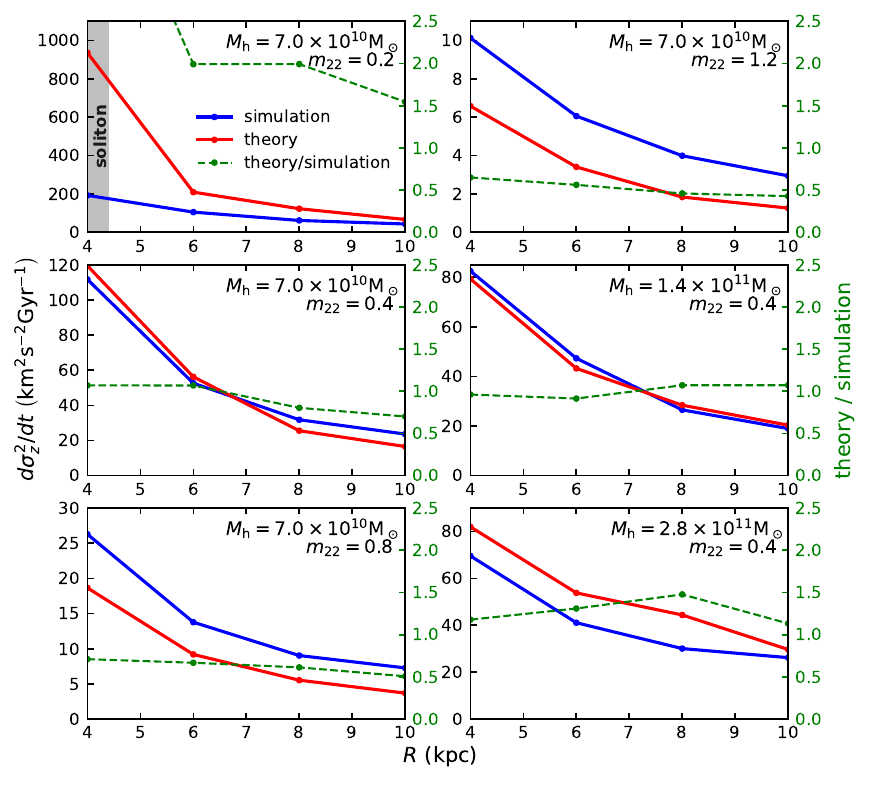}
	\caption{
		Ensemble- and time-averaged disc heating rates $d\sigma_z^2/dt$ from self-consistent simulations (blue) and theoretical predictions (red) at $R = 4, 6, 8,$~and~10~kpc. The ratio of theoretical to simulated heating rates (green) is shown on the right $y$-axes.
		Outside of the soliton core $R\lesssim3.3\rc$ (gray shaded), theoretical estimates agree within about a factor of two to simulations at all radii for all six cases. In the three FDM haloes with $\ma=0.4$, analytical estimates are highly consistent with the measured heating rates, agreeing within 10\textendash50$\%$ at all radii. Interestingly, the theory overpredicts (underestimates) the heating rate for the case $\ma=0.2$ (cases $\ma = 0.8$ and $1.2$). The Fokker\textendash Planck approximation breaks down and results in significant overestimation in the soliton-occupied region. We discuss the possible factors that contribute to the heating rate discrepancies in \sref{subsec:disc_heating_rate_Fokker_Planck}.
	}
	\label{fig:heating}
\end{figure}

Having discussed the simulation results in \sref{subsec:disc_scale_height} and reviewed the corresponding theoretical framework in \sref{sec:theo}, we present in this subsection the comparative analysis of simulated vs. predicted disc heating rates. To compare directly with the \textit{orbit-averaged} theoretical predictions of disc vertical heating rate $d\sigma_z^2/dt$ \eref{eqn:Heating_Eq_SGD_BGD_Limits} in each radial bin, we first compute the linear slope $\Delta \sigma_z^2/\Delta \ts$ of all adjacent data point pairs. Since the heating rate is approximately time-independent over $\Delta \ts = 2.1\text{ Gyr} \gg P(R\leq 10~\text{kpc}, \zmax) \simeq 0.05\text{\textendash}1.0~\text{Gyr}$, the time-averaged slope then gives the orbit-averaged disc heating rate in each radial bin. \fref{fig:heating} compares the ensemble-averaged disc heating rates between numerical simulations (blue solid) and analytical estimates (red solid) \footnote{Since our simulation setup does not include external mass infall (cf. the continuous disc accretion scenario considered in \citetalias{Chiang2023}), the disc-mass-accretion-driven heating term $\frac{d\ln\Sigma}{dt}$ in the SGD limit, \eref{eqn:Heating_Eq_SGD_BGD_Limits}, is not considered when computing the theoretical heating rates here. The disc surface density can still fluctuate over $\ts=0$\textendash$2.1$~Gyr due to radial migration, as shown in \fref{fig:surface_dens}. We have explicitly verified that this time-averaged change in $\Sigma$ could have altered the theoretical ensemble-averaged disc heating rates only by at most $\simeq10$\% in the outer radial bins for the case $\ma = 1.2$; this heating source becomes negligible for smaller $R$ and in lighter $\ma$ cases.} on the left $y$-axis; the ratio of theoretical over simulated heating rates (green dashed) is labelled on the right $y$-axis. 

The simulation results generally agree well with the theoretical estimates. For $\ma=0.4$, analytical estimates agree within 10\textendash50$\%$ to the simulation measurements across all radial bins with different $\Mh$. However, the theory overpredicts (underestimates) the heating rate for $\ma = 0.2$ ($\ma \geq 0,8$) by a factor of $\simeq1.5$\textendash$2$ at all radii (except for the soliton-occupied region for $\ma=0.2$, at which the theory overpredicts the heating rate by more than a factor of five). We discuss in \sref{subsec:disc_heating_rate_Fokker_Planck} some possible factors contributing to the heating rate discrepancies between theory and simulations.

We furthermore observe that the predicted $\ma$- and $\Mh$-dependence of the disc heating rate $\bigH \propto T_\text{heat}^{-1}\propto \ma^{-3}\sigmah^{-6}\rhoh^2$ in \eref{eqn:Heating_Time_Scale} is qualitatively consistent with simulations. In essence, smaller $\ma$ with fixed $\Mh$ corresponds to a larger de Broglie wavelength \eref{eq:lambda}, which gives rise to more massive granular structures \eref{eqn:M_gra} and hence higher disc heating rates. Similarly for a fixed  $\ma$, since lighter halo masses $\Mh$ correspond to smaller local halo density $\rhoh$ (\fref{fig:halo_dens}) and velocity dispersion $\sigmah$, the sharper $\sigmah$ dependence compared to that of $\rhoh$ in the disc heating rate $\bigH \propto \sigmah^{-6}\rhoh^2$, implies stronger disc heating in less massive haloes. However, the quantitative heating rate scaling with respect to $\ma$ found in simulations appears to be `shallower' than the analytical prediction $\bigH\propto \ma^{-3}$. 
 
Next concerning the $R$-dependence of $d\sigma_z^2/dt$, \fref{fig:heating} clearly shows that the disc heating rate increases with decreasing $R$ in all six cases. Since the halo velocity dispersion $\sigmah$ between $3.3\rc$ and $15$~kpc is approximately constant for all the FDM haloes simulated in this work, the inner and outer disc regions in a given halo differ mainly in $\rhoh$, varying by up to a factor of $\simeq4$ between $R = 4$~kpc and $10$~kpc (\fref{fig:halo_dens}).
Given the scaling relation $\bigH\propto \rhoh^2$, the larger $\rhoh(R)$ with decreasing radius is expected to yield higher local disc heating rates, consistent with the general trend in \fref{fig:heating} (see also Figs.~\ref{fig:sigma_z}~and~\ref{fig:sigma_z_sqr}). 

Having uncovered great simulation\textendash theory consistency in the radial dependence, we now focus on the time dependence of $d\sigma_z^2/dt$. Across all six halo-disc systems, the approximate linear growth in $\sigma_z^2$ (\fref{fig:sigma_z_sqr}) corresponds to a roughly time-invariant $d\sigma_z^2/dt$. As the relaxed halo profiles $\sigmah(R)$ and $\rhoh(R)$ remain stable throughout $\ts = 0$\textendash$2.1$~Gyr with negligible secular evolution (see \fref{fig:halo_dens_evo}), the above observation implies a $\sigma_z$-independent disc heating rate, consistent with the analytical prediction \eref{eqn:Heating_Time_Scale}. To interpret this result in relation to the disc scale height growth via \eref{eqn:Scale_Height_Limits}, we first need to determine when either of the analytical expression $h(R)$ can be reliably applied.

\begin{figure}
	\centering
	\includegraphics[width=\columnwidth]{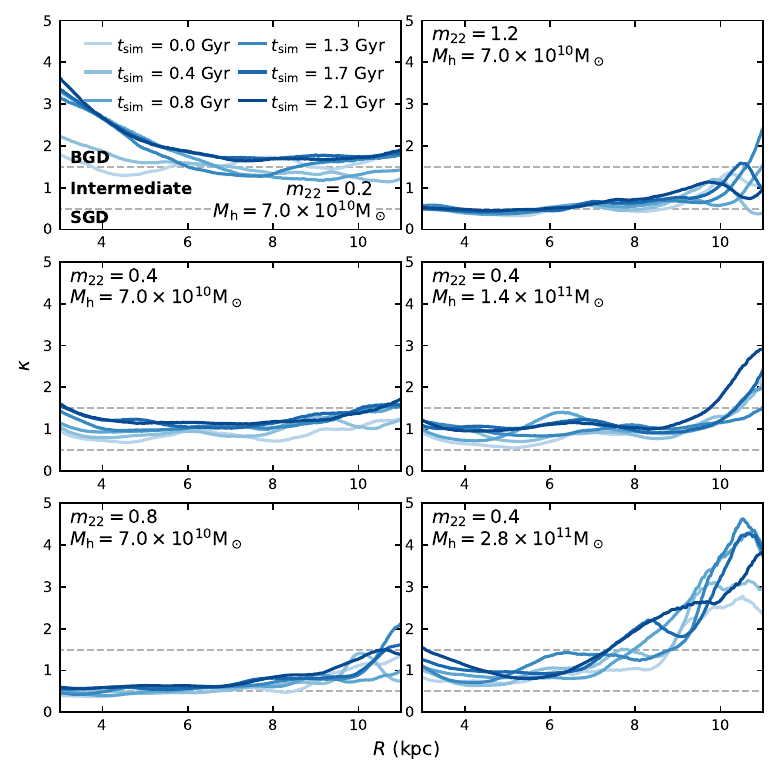}
	\caption{
		$\kappa$ profiles from $\ts = 0$~Gyr (light blue) to $2.1$~Gyr (dark blue). As defined in \eref{eqn:sigma_z_transition}, $\kappa$ provides an effective measure of the relative importance of the disc self-gravity and non-disc background potential such that $\kappa \ll 1$ corresponds to the self-gravity dominated (SGD) limit, while $\kappa \gg1$ gives the background-dominated (BGD) limit. For the case $\ma=0.2$, the disc lies almost entirely within the BGD limit. For the cases $\ma = 0.4$ ($\ma = 0.8$ and $1.2$), the systems are generally in the intermediate regime (SGD limit). Since analytical heating rate estimates exist only in the SGD and BGD limits (see \sref{sec:theo}),  in the intermediate regime $0.5<\kappa<1.5$ we linearly interpolate theoretical heating rates from both limits; the accuracy of such an approach is discussed in \sref{subsec:disc_heating_rate_Fokker_Planck}.
	}
	\label{fig:limit_check}
\end{figure}

The relative contributions of the disc self-potential $\Phi_\text{d}$ and the non-disc background $\Phi_\text{bg}$ to the total gravitational potential $\Phi_\text{tot}$ can be effectively quantified by $\kappa(\sigma_z, \rhobgeff, \Sigma)$ defined in \eref{eqn:sigma_z_transition}, where $\kappa \ll 1$ ($\gg1$) corresponds to the SGD (BGD) limit. \fref{fig:limit_check} shows the time evolution of $\kappa$ from the end of the co-relaxation phase $\ts = 0$~Gyr (light blue) to $2.1$~Gyr (dark blue) for all six cases listed in \tref{tab:SimulationSetup}. We observe that in the three haloes with $\ma = 0.4$, the stellar discs are generally in the intermediate regime $\kappa\simeq1$. For the case $\ma = 0.2$ (0.8 and $1.2$), the inner disc region remains in the BGD (SGD) limit for $\ts = 0$\textendash$2.1$~Gyr.

Provided that the disc surface density $\Sigma$ does not change significantly with time (\fref{fig:surface_dens}), the analytical disc scale height \eref{eqn:Scale_Height_Limits} in the SGD limit $h \propto \sigma_z^2$ naturally accounts for the approximate linear growth in disc scale height shown in \fref{fig:height_t}. Contrastingly in the BGD limit (applicable for the case $\ma=0.2$ at $R = 4$~kpc), \eref{eqn:Scale_Height_Limits} reduces to $h \propto \sigma_z \propto t^{0.5}$, suggesting a gentle decrease in the disc height growth rate $dh/dt$ with time. This mild flattening in $h(\ts)$ can indeed be identified in \fref{fig:height_t} (blue curve in the top-left panel) towards $\ts \simeq 2.1$~Gyr, but a unequivocal corroboration will require additional simulation data extending beyond $\ts \geq 2.1$~Gyr.

Overall, the comparative analysis reveals a general agreement between the simulation measurements and the theoretical estimates. Nonetheless, a number of non-trivial quantitative differences have also been identified in the preceding discussion. In \sref{subsec:disc_heating_rate_Fokker_Planck}, we examine in detail the possible causes of this discrepancy and the validity of the Fokker\textendash Planck approximation assumed in the analytical heating rate estimate of \citetalias{Chiang2023}.

\subsection{Applicability of the Fokker\textendash Planck approximation}
\label{subsec:disc_heating_rate_Fokker_Planck}

\begin{figure}
	\centering
	\includegraphics[width=\columnwidth]{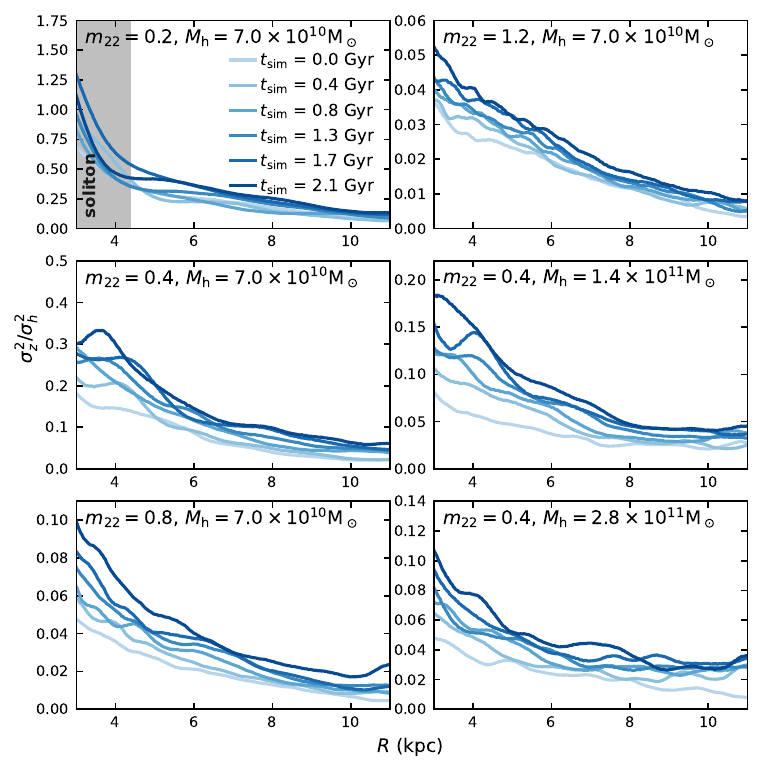}
	\caption{
		Velocity dispersion squared ratios $\sigma_z^2/\sigmah^2$ of the disc star vertical motion to the FDM halo	from $\ts = 0$~Gyr (light blue) to $2.1$~Gyr (dark blue). Except for the soliton-enclosed region (grey shaded) for the case $\ma=0.2$, the stellar discs all remain kinematically cold $\sigma_z^2/\sigmah^2 \lesssim 1$ compared to the respective host haloes.
	}
	\label{fig:sigma_ratio}
\end{figure}
Here we identify major factors that possibly contribute to the heating rate discrepancies between simulations and theory observed in \fref{fig:heating}:
\begin{itemize}
	\item \textbf{The Effect of Soliton:} The significant mismatch at inner radii for the case $\ma=0.2$ is expected and attributed to the proximity to the central soliton core, given that the soliton-halo profile transition occurs at roughly $3.3\rc\simeq4.4$~kpc as shown in \fref{fig:halo_dens}. The Fokker\textendash Planck formalism (\sref{sec:theo}) assumes that all background perturbers can be treated as statistically uncorrelated and \textit{spatially unconfined} quasi-particles, which is appropriate only for modelling a large number of FDM density granules $N_\text{gra} \gg 1$. However, this assumption undoubtedly breaks down at sufficiently small radii $R \leq \mathcal{O}(3.3\rc)$ where the gravitational potential perturbations are dominantly sourced by the soliton. As the ground-state solution to the governing Schr\"{o}dinger\textendash Poisson equations, this coherent soliton core executes confined random-walk like excursions around the host halo centre of mass \citep{Schive:2019rrw, Dutta_Chowdhury_21}. The theory hence overpredicts the disc heating rate by `mistakenly treating' the gravitational interactions with a single soliton as repeated encounters with a group of uncorrelated granules. A similar stellar heating rate overestimate near the central soliton core is also reported in \citet{Dutta_Chowdhury_21}, where the discrepancy was partially alleviated by introducing effective (suppressed) diffusion coefficients within $R\leq2.3\rc$.
	\item \textbf{Validity of Molecular Chaos in the Quasi-particle Treatment:} The roughly factor-two theoretical heating rate overestimate in the outer three radial bins for the case $\ma=0.2$ can partly result from the limited applicability of the molecular chaos assumption \footnote{More precisely, in deriving the analytical expressions of the diffusion coefficient \eref{eqn:FDM_Diffusion_Coeff_1}, \citet{Bar-Or2019ApJ} assumed that the characteristic correlation time of the background perturbers is negligibly small compared to all other dynamical timescales of interest. In the language of classical two-body relaxation, it is equivalent to the assumption that the velocities of colliding background particles are uncorrelated, and consequently the distribution functions of the subject particles and background perturbations are statistically independent (i.e. the assumption of molecular chaos).}. Here the granule effective radius is $R_\text{gra} \simeq 2.2\text{ kpc} \gtrsim h(R, \ts)$ for $\ts = 0$\textendash$2.1$~Gyr (see Figs.~\ref{fig:height_r}~and~\ref{fig:height_t}), and the total number of FDM quasi-particles within $3.3\rc\leq R\leq15$ kpc and $|z|\leq h$ should be less than $N_\text{gra} \lesssim 35$. The small number count $N_\text{gra}$ indicates that these density granules can factually be collisional, and consequently the velocities of colliding quasi-particles are no longer statistically uncorrelated. In addition, the cumulative energy transfer of these correlated potential perturbations can excite non-trivial bulk motion of disc stars, as opposed to all being directed to increasing the disc star random motion $\sigma_z^2$. The analytical framework that treats FDM density granulation as uncorrelated quasi-particles thus expects to be less accurate for sufficiently small $N_\text{gra}$ and overestimate the genuine disc heating rate $d\sigma_z^2/dt$.
	
	Recently, \citet{Zupancic2023arXiv231113352Z} explored the applicability of quasi-particle approximation in lieu of the fully self-consistent Schr\"{o}dinger\textendash Poisson dynamics. Their simulation results, albeit in one spatial dimension, suggest that these two treatments are consistent within the first $\sim$50 dynamical times of evolution, after which the quasi-particle formalism can non-trivially overpredict the stellar heating rates. However, whether this observation holds in three-dimensional systems requires further tailored simulations and is beyond the scope of this work.
	\item \textbf{Coulomb Logarithm:} Another possible cause of the heating rate discrepancy at $\ma=0.2$ lies in the ambiguous nature of the Coulomb logarithm definition $\ln\Lambda$, which can be safely ignored only for a sufficiently large Coulomb factor $\Lambda \equiv b_\text{max}/b_\text{min} \gg 1$ \citep[e.g.][]{Just2005A&A431861J, Binney&Tremaine2008}. For $\ma=0.2$, $\ln\Lambda \simeq 1.5$\textendash$3.0$ within $R \leq 12$~kpc is sufficiently small such that any physical corrections to $\ln\Lambda$ could sizeably impact the predicted heating rate.
	\item \textbf{Heating Behaviour across the SGD and BGD Limits:} Analytical expressions for the granulation-driven ensemble-averaged disc heating rates \eref{eqn:Heating_Eq_SGD_BGD_Limits} exist only if the disc self-gravity either dominates (the SGD limit) or can be ignored relative to the background potential (the BGD limit). For a galactic disc with time-independent $\Sigma(R)$, the theoretical heating rate in the BGD limit is 1.5\textendash2 times higher than that in the SGD limit. In the intermediate regime, where the heating rate cannot be computed directly, we have estimated the analytical heating rate by linearly interpolating the heating rates in both limits as discussed in \sref{sec:theo}. This comparatively uncertain regime applies to the three cases with $\ma=0.4$ (\fref{fig:limit_check}), which could partly contribute to the $\lesssim10$\textendash$50\%$ discrepancy between predicted heating rates and simulation measurements observed in \fref{fig:heating}.
	
	Outside of this intermediate region $0.5\leq \kappa\leq 1.5$, however, comparatively greater heating rate overestimates (underestimates) are observed for the case $\ma=0.2$ (0.8 and 1.2) in the BGD (SGD) limit. In all six disc-halo systems, there appears to a correlation between BGD/SGD limits and overestimate/underestimate of the disc heating rates. Furthermore in the SGD regime, the factor-two simulation\textendash theory discrepancies in the cases $\ma\geq 0.8$ can hardly be accounted for by the aforesaid uncertainties. It thus remains possible that the SGD limit itself has some unknown errors that leads to this non-trivial heating rate underestimation. Future simulations that explore the range $\ma \geq 1.2$ would help clarify this uncertainty.
	\item \textbf{Numerical (Artificial) Heating:} As the granulation-driven disc heating decreases rapidly with increasing $\ma$, other sources of disc heating might become discernible in the simulation data. We assess the level of numerical (artificial) disc heating by conducting a disc-CDM-halo simulation run with $\Mh = 7\times 10^{10}$~M$_\odot$ (see Appendix~\ref{app:CDM} for more details). The heating rates in all four radial bins are below $\lesssim 1$~km$^2$s$^{-2}$Gyr$^{-1}$, which is negligible or sub-dominant compared to the FDM-driven heating rates as seen in \fref{fig:heating}. Hence under the sufficient particle and force resolutions adopted in this work (see \sref{subsec:amr} and Appendix~\ref{app:ConvergenceTest}), we conclude that artificial heating is unimportant and can be safely excluded in our analysis.
	\item \textbf{Evolution in the Disc Kinematic Temperature:} In the Fokker\textendash Planck formalism (\sref{sec:theo}), only perturbers travelling faster than the disc star particles contribute to diffusive heating \cite[e.g.][]{Binney&Tremaine2008}. Since the velocity distribution of FDM quasi-particles is assumed to be Maxwellian, the population of density granules capable of contributing to the disc heating shrinks as a stellar disc becomes kinematically hotter. Namely, the granulation-driven heating becomes ineffective if the stellar disc is kinematically hot relative to the host halo $\sigma_z^2/\sigmah^2 \geq \mathcal{O}(1)$. We examine the time evolution of $\sigma_z^2/\sigmah^2$ from $\ts = 0$~Gyr (light blue) to $2.1$~Gyr (dark blue) for all six disc-halo systems in \fref{fig:sigma_ratio}. Except for the soliton-occupied regions (gray shaded), the stellar discs all remain kinematically cold relative to the respective host haloes. The disc heating rates hence should not be meaningfully impacted by the increase in $\sigma_z^2/\sigmah^2$, consistent with the nearly time-independent growth rate of $\sigma_z$ observed in \fref{fig:sigma_z_sqr} for $\ts = 0$\textendash$2.1$~Gyr in all six cases.
	\item \textbf{Other Physical Disc Heating Sources:} The theoretical underestimate in disc heating rate for the two cases $\ma = 0.8, 1.2$ likely indicates the presence of other physical, non-granulation-driven heating mechanisms such as the outward radial migration \cite[e.g.][]{Sellwood2002MNRAS336785S, Minchev2010ApJ722112M}, transient spiral modes \citep[e.g.][]{Barbanis1967ApJ150461B, Minchev2006MNRAS368623M}, and/or large-scale bending instability \citep[e.g.][]{Rodionov2013MNRAS4342373R}. For these two galactic discs, orbital diffusion proceeds at a rate $10^{-3}$\textendash$10^{-2}\Md$ per Gyr in all 2-kpc-wide radial bins, such that radial migration alone cannot account for the factor-two mismatch in predicted and observed disc heating rates (see also Figs.~\ref{fig:sigma_z}~and~\ref{fig:surface_dens}). Although quantifying these non-FDM disc heating mechanisms is beyond the scope of this work, it is worth noting that the level of theory-simulation discrepancy still remains within a factor of two even when the granulation-driven disc heating rate reaches as low as $d\sigma_z^2/dt \simeq 3$~km$^2$s$^{-2}$Gyr$^{-1}$ for the case $\ma=1.2$ and $\Mh=7\times10$~M$_\odot$. 
\end{itemize}

% section 4
% ----------------------------------------------
\section{Implications for External Disc Galaxies}
\label{sec:implications}
Section~\ref{subsec:disc_scale_height_ins} compares the analytical disc scale height expressions with simulation measurements. Uncertainties in the hydrostatic-equilibrium-based $\sigma_z$ inferences are discussed in \sref{subsec:implications_scale_height}.

\subsection{Instantaneous disc scale height inference}
\label{subsec:disc_scale_height_ins}

Observationally, the fitted vertical structures of galactic disc density profiles are commonly assumed to follow either $\propto \text{sech}^2(z/h)$ \citep[e.g.][]{Comeron2011ApJ72918C} or $\propto e^{-z^2/h^2}$ \citep[e.g.][]{Dutta2009MNRAS398887D}. However, even for locally isothermal discs in hydrostatic equilibrium, these two analytical scaling relations \eref{eqn:Disc_Den_Profiles} are exact only if either the non-disc background potential or the disc self-gravity can be ignored relative to the other component (\citetalias{Chiang2023}). In this subsection, we assess under what necessary conditions do the analytical formulae accurately predict the measured disc scale heights, as well as quantify the level of discrepancy when the conditions are not met (i.e. the total gravitational potential $\Phi_\text{tot}$ is comparably sourced by the disc itself $\Phi_\text{d}$ and the non-disc background $\Phi_\text{bg}$).

As discussed in \sref{subsec:Simulations_vs_Theory}, the dimensionless number $\kappa(\sigma_z, \rhobgeff, \Sigma)$ defined in \eref{eqn:sigma_z_transition} provides a quantifiable measure for the relative contributions of $\Phi_\text{d}$ and $\Phi_\text{bg}$ to $\Phi_\text{tot}$, where $\kappa \ll 1$ ($\gg1$) corresponds to the SGD (BGD) limit. Prior to the disc-halo co-relaxation at $\tr=0$~Gyr, the initial disc configurations (see \sref{subsec:init_disc}) are all kinematically cold with $\kappa$ ranges from $\simeq 0.25$ at $R = 4$~kpc to $0.4$\textendash$0.5$ at $R = 10$~kpc. As each galactic disc thickens with increasing $\sigma_z$ over time, the effect of background potential can gradually dominate over the disc self-gravity in shaping individual disc stars' vertical oscillation motion \eref{eqn:P_Definition}, possibly leading to a transition from the SGD limit to the BGD limit. The post-co-relaxation time evolution of each $\kappa$ profile is shown in \fref{fig:limit_check} over $\ts = 0$~Gyr (light blue; equivalent to $\tr = 1.4$~Gyr) and $2.1$~Gyr (dark blue). By $\ts = 2.1$~Gyr, the galactic discs are generally in the intermediate regime $\kappa\simeq1$, except for the inner disc region of the case $\ma = 0.2$ (0.8 and $1.2$) lying in the BGD (SGD) limit.

\begin{figure}
	\centering
	\includegraphics[width=\columnwidth]{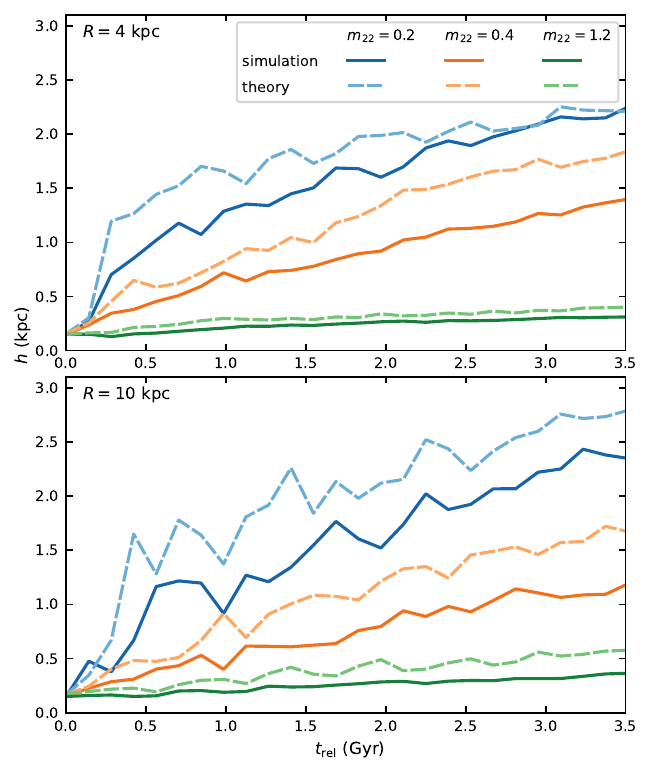}
	\caption{
		Disc scale heights obtained from simulations (solid) or analytical formulae (dashed, \eref{eqn:Scale_Height_Limits}) at $R=4$~kpc (upper panel) and $10$~kpc (lower panel) for the cases $\ma=0.2$ (blue), $0.4$ (orange), and $1.2$ (green) over $\tr = 0$\textendash$3.5$~Gyr. Note that $\ts \equiv \tr - 1.4$~Gyr. Analytical expressions generally overestimate the scale height, by up to a factor of two, and are sufficiently accurate only when the disc is properly in the SGD or BGD limit, as shown in \fref{fig:limit_check}; see \sref{subsec:disc_scale_height_ins} for details.}
	\label{fig:height_compare}
\end{figure}

Focusing on the following three cases with $\ma=0.2, 0.4, 1.2$ (blue, orange, and green respectively) and $\Mh=7.0\times10^{10}\Msun$, \fref{fig:height_compare} compares the disc scale heights in the radial bins centred on $R=4$~kpc (top panel) and $10$~kpc (bottom panel), plotting values from either simulation measurements (solid) or analytical solutions (dashed, \eref{eqn:Scale_Height_Limits}). The latter is defined as the minimum of the two limits $h(R)={\rm min}(h_{\rm SGD}, h_{\rm BGD})$ \footnote{The adoption of theoretical disc heights as the smaller analytical value in the two limits $h(R)={\rm min}(h_{\rm SGD}, h_{\rm BGD})$ is an (overly) optimistic choice that minimises the theory-simulation discrepancy. However even under such a `best-case' scenario, the simulation(observation)\textendash theory mismatch in the disc scale height inferences can still approach a factor of two as shown in \fref{fig:height_compare}, highlighting the need to explicitly verify the validity of \eref{eqn:Scale_Height_Limits} by carefully assessing the relative importance between $\Phi_\text{d}$ and $\Phi_\text{bg}$ when modelling disc vertical structures in external galaxies. If one instead assumes either of the analytical relation \eref{eqn:Scale_Height_Limits} across all $\kappa \geq 0$, the level of scale height overestimation is actually unbounded above and can exceed a factor of two.}, since there exists no closed-form expression for $h$ in the intermediate regime $\kappa\simeq1$. Across the entire simulation time span $\tr = 0$\textendash$3.5$~Gyr, we observe that analytical estimates and simulation results agree well in the SGD limit at the onset of disc-halo co-relaxation $\ts \lesssim 0.25$~Gyr. This general agreement remains for the case $\ma = 1.2$ at $R = 4$~kpc where the SGD limit $\kappa \leq 0.5$ remains applicable until $\tr = 3.5$~Gyr (see the top-right panel of \fref{fig:limit_check} at $\ts = 2.1$~Gyr). At the other extreme where the BGD limit clearly applies, for the case $\ma=0.2$ at $R = 4$~kpc, the theoretical and simulation-inferred scale heights differ by less than $\simeq 15$\% for $\tr \gtrsim 2$~Gyr. In the intermediate regime $\kappa\simeq1$, however, adopting either the $\text{sech}^2(z/h)$ or $e^{-z^2/h^2}$ scale height analytical solutions `undercounts' the total vertical gravity by about half. For the three cases with $\ma =0.4$, \fref{fig:height_compare} shows that the analytical predictions can overestimate the scale height by up to a factor of two. We discuss the relevant implications to observational inferences of disc properties in \sref{subsec:implications_scale_height}.

\subsection[Observational uncertainties of disc scale height and vertical velocity dispersion]{Observational uncertainties of disc scale height and $\sigma_z$}
\label{subsec:implications_scale_height}

In external disc galaxies, radial profiles of stellar/gas disc vertical velocity dispersion $\sigma_z$ for edge-on (or scale height $h$ for face-on) systems cannot be directly measured. With observationally constrained baryonic surface density profile $\Sigma(R)$ and/or effective background density $\rho_\text{bg}^\text{eff}$, it is customary to assume the galactic disc in question to be either self-gravitating (the SGD limit) or dictated by the non-disc background potential (the BGD limit) such that analytical hydrostatic-equilibrium solutions for isothermal discs exist and can be directly applied \citep[e.g.][]{vanderKruit1999A&A352129V, Leroy2008AJ1362782L}. The observationally inaccessible $\sigma_z(R)$ or $h(R)$ can then be directly inferred from either expression of \eref{eqn:Scale_Height_Limits}, depending on which assumption is adopted \citep[e.g.][]{Kregel2002MNRAS334646K, Kasparova2008AstL34152K, Patra2019MNRAS48481P, Das2020ApJ88910D}. However, in the intermediate regime $0.5 < \kappa < 1.5$ as commonly found in our simulations (\fref{fig:limit_check}), both the disc and non-disc background contribute comparably to the local total vertical gravity. By explicitly assuming the SGD (BGD) limit, the total vertical gravity is \textit{always undercounted} by accounting for only the (non-)disc component.

The effect of this vertical gravity undercounting can be substantial in the inferred disc properties, as observed in our self-consistent disc-halo simulations. Analytical solutions \eref{eqn:Scale_Height_Limits} can overestimate the ensemble-averaged disc scale heights by up to a factor of two, as demonstrated in \fref{fig:height_compare}. In the intermediate regime $0.5 < \kappa < 1.5$ where both the SGD and BGD limits fail, `blindingly' applying \eref{eqn:Scale_Height_Limits} can result in $\sigma_z(R)$ underestimated by up to a factor of $\sqrt{2}$ (or $2$) in the SGD (BGD) limit.

This uncertainty applies to all external disc galaxies where the baryonic vertical velocity dispersion is indirectly inferred from $h(R)$ and $\Sigma(R)$. That being said, the caveat is less of a concern when either the SGD limit (e.g. for ultra-thin discs studied by \citet{Matthews2000AJ1201764M} and \citet{Bizyaev2017MNRAS4653784B}) or the BGD limit (e.g. stellar discs that are kinematically hot relative to the host haloes, as for the case $\ma = 0.2$ in Figs.~\ref{fig:limit_check}~and~\ref{fig:height_compare}) is guaranteed to hold. To partially address the uncertainty in this type of hydrostatic-equilibrium-based $\sigma_z$ inferences in observed external galaxies, estimating the value of $\kappa$ could serve as a self-consistency cross-check by requiring $\kappa \lesssim 0.5$ ($\gtrsim 1.5$) if the SGD (BGD) limit is adopted. This source of inference error is expected to be important when $\kappa \simeq 1$.

% section 5
% ----------------------------------------------
\section{Summary And Conclusions}
\label{sec:concl}

In this work, we numerically quantify the stellar disc heating rates caused by FDM halo density granulation by performing the first self-consistent simulations of FDM halo and $N$-body stellar disc, spanning $m_a=0.2$\textendash$1.2\times10^{-22}$~eV (equivalently $\ma = 0.2$\textendash$1.2$) and halo virial masses $\Mh = 0.7$\textendash$2.8\times10^{11}$~M$_\odot$ as listed in \tref{tab:SimulationSetup}. The $\GALIC$-constructed galactic discs weight $\Md = 3.16\times10^9$~M$_\odot$ and are resolved with $\Nd = 0.8$\textendash$1.6\times 10^8$ equal-mass disc particles; prior to initial disc-halo co-relaxation, these discs have the same scale radius $\Rd = 3.0$~kpc and radius-independent scale height $h = 0.15$~kpc. The main results are summarised as follows:

\begin{itemize}
	\item Disc thickening is observed in all disc-halo systems (\fref{fig:disc_edge_on}), at rates $dh/dt \simeq 0.04$\textendash0.4~kpc/Gyr (Figs.~\ref{fig:height_r}~and~\ref{fig:height_t}) increasing with smaller $\ma$ and $\Mh$. The measured disc heating rates $d\sigma^2_z/dt \simeq 4$\textendash$150$~km$^2$/sec$^{2}$ (Figs.~\ref{fig:sigma_z},~\ref{fig:sigma_z_sqr},~and~\ref{fig:heating}) exhibit the same trend. Namely, for a fixed $\Mh$, discs hosted by haloes with smaller $\ma$ have higher heating rates; lighter $\Mh$ with fixed $\ma$ exhibits higher heating rates. In each individual disc-halo system, the ensemble-averaged disc heating rates decrease monotonically with increasing radius (\fref{fig:heating}). Overall, the disc scale height $h$ and vertical velocity dispersion $\sigma^2_z$ increase approximately linearly with time after the halo-disc co-relaxation (Figs.~\ref{fig:height_t}~and~\ref{fig:sigma_z_sqr}).
	\item The FDM granulation-driven disc heating rates quantified in simulations are compared directly with the Fokker\textendash Planck-based analytical estimates of \citetalias{Chiang2023} (\sref{subsec:Simulations_vs_Theory}). The simulation measurements of the disc heating rates agree within a factor of two to the theoretically predicted values for all six cases examined in this work (\fref{fig:heating}), except for the region within the soliton core $r \lesssim 3.3 \rc$. For the three FDM haloes of $\ma = 0.4$, the theory-simulation discrepancy is less than 10\textendash50$\%$ at all radii of interest.
	\item In individual cases, we first note that within the soliton core $r \lesssim 3.3\rc$, the Fokker\textendash Planck formalism is invalid (\sref{subsec:disc_heating_rate_Fokker_Planck}) and expectedly yields a significant overestimation in the innermost radial bin $R = 4$~kpc of the case $\ma = 0.2$ (\fref{fig:heating}). At outer radii, the analytical prediction for $\ma=0.2$ overestimates the disc heating rate within a factor of two. For the three cases of $\ma=0.4$, we observe high level of theory-simulation consistency at all radii of interest. For the two cases $\ma=0.8$ and $1.2$, the theory generally underestimates the genuine disc heating rates (within a factor of two), suggesting that there could be additional heating sources not accounted for in the granulation-driven heating model. We have carefully verified that possible artificial heating due to inadequate numerical and/or improper disc-halo initial conditions is negligible in our FDM simulations by verifying numerical convergence (\aref{app:ConvergenceTest}) and comparing to a CDM simulation (\aref{app:CDM}).
	\item The disc thickening and heating rates are positively correlated to the granule effective mass $\Mgra\propto \ma^{-3}\sigmah^{-3}\rhoh $ as defined in \eref{eqn:M_gra}. This trend is confirmed across all three distinct regimes: $\kappa < 0.5$ (the SGD limit, where the non-disc background is negligible), $0.5 < \kappa < 1.5$ (intermediate regime where the disc and non-disc background contribute comparably to the total vertical gravity), and $\kappa > 1.5$ (the BGD limit, where the disc self-gravity is negligible). However, the heating rate scaling with $\ma$ found in simulations appears to be `shallower' than the analytical prediction $\bigH \propto \ma^{-3}$, \eref{eqn:Heating_Time_Scale}; see \sref{subsec:Simulations_vs_Theory}.
	\item Concerning the usual adoption of analytical disc scale heights \eref{eqn:Scale_Height_Limits} in modelling the observed external disc galaxies, these closed-form solutions exist only in either the SGD or BGD limit and always \textit{undercount} the total vertical gravity, as discussed in \sref{subsec:disc_scale_height_ins}. As a result, the analytical predictions are higher than the true disc scale heights measured in self-consistent simulations (\fref{fig:height_compare}) by up to a factor of two. This implies that hydrostatic-equilibrium-based $\sigma_z$ inferences in observed external disc galaxies can be underestimated by up to a factor of $\sqrt{2}$ (or $2$) in the SGD (BGD) limit; see \sref{subsec:implications_scale_height}.
\end{itemize}

In \citetalias{Chiang2023}, a conservative exclusion bound $\ma \gtrsim 0.4$ was derived from the Galactic disc kinematics by requiring that predicted granulation-driven heating in the solar neighbourhood cannot exceed the observed ensemble-averaged disc velocity dispersion $\sigma_z(R_\odot) \simeq 22$~km~s$^{-1}$ \citep{Sharma2021MNRAS506}. The fact that the theoretical disc heating rates are highly consistent with simulation data at $\ma = 0.4$ further supports the robustness of this FDM particle mass constraint. Relatedly, since the disc heating rates in simulations appear to decline less rapidly than the theoretical prediction $\bigH \propto \ma^{-3}$ with increasing $\ma$, we expect that the reported range $\ma \simeq 0.5$\textendash$0.7$ in \citetalias{Chiang2023} favoured by the observed thick disc kinematics could be corrected upwards to be closer to $\ma \simeq 1.0$. We leave the careful cross-check of this prediction by performing self-consistent $N$-body disc simulations in an MW-sized FDM halo to a future work.

On the other hand, the rough analytical calculations by \citet{Church2019MNRAS4852861C} yield a disc heating rate larger by a factor of $\sim360$ than the estimates of \citetalias{Chiang2023} (see Sec.~4.2.4 therein), under the same disc and FDM halo parameters. To leading order, the discrepancy in $\bigH \propto \Mgra^2$ mainly stems from their overestimate of granulation mass $\Mgra$ by $\mathcal{O}(10)$ and the simplified assumption that a disc is everywhere self-gravity dominated (SGD limit). Regarding the MW thick disc formation, \citet{Church2019MNRAS4852861C} adopted an enormously large one-dimensional halo velocity dispersion $\sigmah = 200$~km~s$^{-1}$~\footnote{This value is larger by a factor of $2.3$ than $\sigmah = 86$~km~s$^{-1}$ measured from a self-consistently constructed MW-sized FDM halo (Su et al., in preparation). This discrepancy could partially stem from the fact that equipartition of energy in an FDM halo was not accounted for by \citet{Church2019MNRAS4852861C}; see footnote~\ref{fn:Halo_Velocity_Dispersion}.}. Due to the strong dependence of granulation-driven disc heating rate $\bigH \propto \ma^{-3}\sigmah^{-6}\rhoh^2$ on FDM parameters, their incorrectly quoted $\sigmah$ significantly reduced the predicted Galactic disc heating rate by a factor of $\simeq 160$, incidentally in the right direction to compensate for their theoretical heating rate overestimate (by a factor of $\sim360$ when adopting the more accurate halo velocity dispersion $\sigmah \simeq 86$~km~s$^{-1}$).

Given the within-factor-two agreement between the heating rate predictions of \citetalias{Chiang2023} and the measurements in all six self-consistent disc-halo simulations performed in this work, we conclude that \citet{Church2019MNRAS4852861C} overestimated the genuine granulation-driven disc heating rate by at least two orders of magnitude. To compare with observed Galactic disc kinematics, they arrived at a similar exclusion bound $\ma \gtrsim 0.6$ by requiring that the predicted disc heating over 12~Gyr cannot exceed a less up-to-date value of $\sigma_z(R_\odot) \simeq 32$~km~s$^{-1}$ \citep{Binney2010MNRAS4012318B}, inferred indirectly from the best-fit analytical distribution functions to the survey data collected by \citet{Gilmore1983MNRAS2021025G}. Although \citet{Church2019MNRAS4852861C} still arrived at a $\ma$ constraint reasonably close to that of \citetalias{Chiang2023}, it should be stressed that their exclusion bound was derived on the bases of rather inaccurate assumptions and estimates.

Lastly, it is worth emphasising that the present work aims to quantify the FDM-granulation-driven disc heating rates in self-consistent FDM-baryon simulations, providing an independent cross-check against the analytical heating rate estimates. As discussed in \sref{subsec:init_disc}, we adopt the kinematically cold, thin discs appropriate in the $\Lambda$CDM cosmology as the initial conditions to perform further FDM halo-stellar disc co-relaxation. It remains to be investigated whether such cold and thin stellar discs could naturally form in the first place in an FDM cosmology. Existing FDM cosmological simulations with baryonic feedback \citep{Mocz:2019pyf, Kulkarni2022ApJ941L18K} still lack the mass resolution and large-sample statistics at sufficiently low redshifts to address this question. That being said, as the granulation-sourced disc heating rates depend primarily on the FDM halo attributes and are comparatively insensitive to the instantaneous disc properties (Figs.~\ref{fig:sigma_z_sqr}, \ref{fig:heating}), we expect the general conclusions of this work to hold irrespective of the adopted disc initial conditions.

Deciphering the thick disc formation processes in the MW and nearby external disc galaxies is indispensable for gaining a more complete picture of galaxy evolution. Having examined analytically and corroborated numerically the granulation-driven disc heating rates in the MW (\citetalias{Chiang2023}) and sub-MW mass haloes in this work, we argue that, in an FDM cosmology, this heating mechanism provides a robust and ubiquitous thick disc formation pathway, comparatively insensitive to the detailed disc morphology and assembly history as required in some other proposed thick disc formation models. Beyond the Local Universe, the recently discovered high-redshift disc galaxies by JWST \citep{Nelson2023ApJ948L18N} also present exciting opportunities to further put to test various proposed thick disc formation models. Future works could perform a detailed analysis of FDM granulation-driven heating over a larger sample of observed disc-halo systems to place a more stringent exclusion bound on $\ma$, or examine the quantitative (dis)agreement between the Gaia phase spiral and the out-of-equilibrium features caused by FDM density granulation in MW-sized haloes.

% acknowledgements
% ----------------------------------------------
\section*{Acknowledgements}
\label{sec:acknowledgements}

We are happy to acknowledge useful conversations with Frank van den Bosch. We also thank the anonymous referee for the constructive and helpful report. We thank National Center for High-performance Computing (NCHC) for the use of supercomputer TAIWANIA 3. We use \texttt{NumPy} \citep{numpy} and \texttt{SciPy} \citep{scipy} for data analysis, and the data visualisation is carried out using \texttt{Matplotlib} \citep{matplotlib} and \texttt{yt} \citep{yt}. HS acknowledges funding support from the Yushan Scholar Program No. NTU-111V1201-5, sponsored by the Ministry of Education, Taiwan. This research is partially supported by the National Science and Technology Council (NSTC) of Taiwan under Grants No. NSTC 111-2628-M-002-005-MY4, No. NSTC 108-2112-M-002-023-MY3, and No. NSTC-110-2112-M-002-018, and the NTU Academic Research-Career Development Project under Grant No. NTU-CDP-111L7779.

% data availability
% ----------------------------------------------
\section*{Data Availability}
\label{sec:data_availability}
The data underlying this article will be shared
on reasonable request to the corresponding author.

% reference
% ----------------------------------------------
\bibliographystyle{mnras}
\bibliography{ref}

\begin{thebibliography}{}
\makeatletter
\relax
\def\mn@urlcharsother{\let\do\@makeother \do\$\do\&\do\#\do\^\do\_\do\%\do\~}
\def\mn@doi{\begingroup\mn@urlcharsother \@ifnextchar [ {\mn@doi@}
  {\mn@doi@[]}}
\def\mn@doi@[#1]#2{\def\@tempa{#1}\ifx\@tempa\@empty \href
  {http://dx.doi.org/#2} {doi:#2}\else \href {http://dx.doi.org/#2} {#1}\fi
  \endgroup}
\def\mn@eprint#1#2{\mn@eprint@#1:#2::\@nil}
\def\mn@eprint@arXiv#1{\href {http://arxiv.org/abs/#1} {{\tt arXiv:#1}}}
\def\mn@eprint@dblp#1{\href {http://dblp.uni-trier.de/rec/bibtex/#1.xml}
  {dblp:#1}}
\def\mn@eprint@#1:#2:#3:#4\@nil{\def\@tempa {#1}\def\@tempb {#2}\def\@tempc
  {#3}\ifx \@tempc \@empty \let \@tempc \@tempb \let \@tempb \@tempa \fi \ifx
  \@tempb \@empty \def\@tempb {arXiv}\fi \@ifundefined
  {mn@eprint@\@tempb}{\@tempb:\@tempc}{\expandafter \expandafter \csname
  mn@eprint@\@tempb\endcsname \expandafter{\@tempc}}}

\bibitem[\protect\citeauthoryear{{Abadi}, {Navarro}, {Steinmetz}  \&
  {Eke}}{{Abadi} et~al.}{2003}]{Abadi2003ApJ59721A}
{Abadi} M.~G.,  {Navarro} J.~F.,  {Steinmetz} M.,   {Eke} V.~R.,  2003, \mn@doi
  [\apj] {10.1086/378316}, \href
  {https://ui.adsabs.harvard.edu/abs/2003ApJ...597...21A} {597, 21}

\bibitem[\protect\citeauthoryear{{Amruth} et~al.,}{{Amruth}
  et~al.}{2023}]{Amruth2023NatAstmp104A}
{Amruth} A.,  et~al., 2023, \mn@doi [Nature Astron.]
  {10.1038/s41550-023-01943-9}, \href
  {https://ui.adsabs.harvard.edu/abs/2023NatAs...7..736A} {7, 736}

\bibitem[\protect\citeauthoryear{{Antoja} et~al.,}{{Antoja}
  et~al.}{2018}]{Antoja2018Natur561360A}
{Antoja} T.,  et~al., 2018, \mn@doi [\nat] {10.1038/s41586-018-0510-7}, \href
  {https://ui.adsabs.harvard.edu/abs/2018Natur.561..360A} {561, 360}

\bibitem[\protect\citeauthoryear{{Banik}, {Weinberg}  \& {van den
  Bosch}}{{Banik} et~al.}{2022}]{Banik2022ApJ935135B}
{Banik} U.,  {Weinberg} M.~D.,   {van den Bosch} F.~C.,  2022, \mn@doi [\apj]
  {10.3847/1538-4357/ac7ff9}, \href
  {https://ui.adsabs.harvard.edu/abs/2022ApJ...935..135B} {935, 135}

\bibitem[\protect\citeauthoryear{{Bar-Or}, {Fouvry}  \& {Tremaine}}{{Bar-Or}
  et~al.}{2019}]{Bar-Or2019ApJ}
{Bar-Or} B.,  {Fouvry} J.-B.,   {Tremaine} S.,  2019, \mn@doi [\apj]
  {10.3847/1538-4357/aaf28c}, \href
  {https://ui.adsabs.harvard.edu/abs/2019ApJ...871...28B} {871, 28}

\bibitem[\protect\citeauthoryear{{Barbanis} \& {Woltjer}}{{Barbanis} \&
  {Woltjer}}{1967}]{Barbanis1967ApJ150461B}
{Barbanis} B.,  {Woltjer} L.,  1967, \mn@doi [\apj] {10.1086/149349}, \href
  {https://ui.adsabs.harvard.edu/abs/1967ApJ...150..461B} {150, 461}

\bibitem[\protect\citeauthoryear{{Bennett} \& {Bovy}}{{Bennett} \&
  {Bovy}}{2019}]{Bennett2019MNRAS4821417B}
{Bennett} M.,  {Bovy} J.,  2019, \mn@doi [\mnras] {10.1093/mnras/sty2813},
  \href {https://ui.adsabs.harvard.edu/abs/2019MNRAS.482.1417B} {482, 1417}

\bibitem[\protect\citeauthoryear{{Bernal}, {Fern{\'a}ndez-Hern{\'a}ndez},
  {Matos}  \& {Rodr{\'\i}guez-Meza}}{{Bernal}
  et~al.}{2018}]{Bernal2018MNRAS4751447B}
{Bernal} T.,  {Fern{\'a}ndez-Hern{\'a}ndez} L.~M.,  {Matos} T.,
  {Rodr{\'\i}guez-Meza} M.~A.,  2018, \mn@doi [\mnras] {10.1093/mnras/stx3208},
  \href {https://ui.adsabs.harvard.edu/abs/2018MNRAS.475.1447B} {475, 1447}

\bibitem[\protect\citeauthoryear{{Binney}}{{Binney}}{2010}]{Binney2010MNRAS4012318B}
{Binney} J.,  2010, \mn@doi [\mnras] {10.1111/j.1365-2966.2009.15845.x}, \href
  {https://ui.adsabs.harvard.edu/abs/2010MNRAS.401.2318B} {401, 2318}

\bibitem[\protect\citeauthoryear{{Binney} \& {Sch{\"o}nrich}}{{Binney} \&
  {Sch{\"o}nrich}}{2018}]{Binney2018MNRAS4811501B}
{Binney} J.,  {Sch{\"o}nrich} R.,  2018, \mn@doi [\mnras]
  {10.1093/mnras/sty2378}, \href
  {https://ui.adsabs.harvard.edu/abs/2018MNRAS.481.1501B} {481, 1501}

\bibitem[\protect\citeauthoryear{{Binney} \& {Tremaine}}{{Binney} \&
  {Tremaine}}{2008}]{Binney&Tremaine2008}
{Binney} J.,  {Tremaine} S.,  2008, {Galactic Dynamics: Second Edition}.
Princeton University Press, Princeton, N.J

\bibitem[\protect\citeauthoryear{{Bizyaev}, {Kautsch}, {Sotnikova},
  {Reshetnikov}  \& {Mosenkov}}{{Bizyaev}
  et~al.}{2017}]{Bizyaev2017MNRAS4653784B}
{Bizyaev} D.~V.,  {Kautsch} S.~J.,  {Sotnikova} N.~Y.,  {Reshetnikov} V.~P.,
  {Mosenkov} A.~V.,  2017, \mn@doi [\mnras] {10.1093/mnras/stw2972}, \href
  {https://ui.adsabs.harvard.edu/abs/2017MNRAS.465.3784B} {465, 3784}

\bibitem[\protect\citeauthoryear{{Bland-Hawthorn} \&
  {Gerhard}}{{Bland-Hawthorn} \& {Gerhard}}{2016}]{Bland-Hawthorn2016ARA&A54}
{Bland-Hawthorn} J.,  {Gerhard} O.,  2016, \mn@doi [Annu. Rev. Astron.
  Astrophys.] {10.1146/annurev-astro-081915-023441}, \href
  {https://ui.adsabs.harvard.edu/abs/2016ARA&A..54..529B} {54, 529}

\bibitem[\protect\citeauthoryear{{Borucki} et~al.,}{{Borucki}
  et~al.}{2010}]{Borucki2010Sci327}
{Borucki} W.~J.,  et~al., 2010, \mn@doi [Science] {10.1126/science.1185402},
  \href {https://ui.adsabs.harvard.edu/abs/2010Sci...327..977B} {327, 977}

\bibitem[\protect\citeauthoryear{{Bournaud}, {Elmegreen}  \&
  {Martig}}{{Bournaud} et~al.}{2009}]{Bournaud2009ApJ707L1B}
{Bournaud} F.,  {Elmegreen} B.~G.,   {Martig} M.,  2009, \mn@doi [\apjl]
  {10.1088/0004-637X/707/1/L1}, \href
  {https://ui.adsabs.harvard.edu/abs/2009ApJ...707L...1B} {707, L1}

\bibitem[\protect\citeauthoryear{{Brook}, {Kawata}, {Gibson}  \&
  {Freeman}}{{Brook} et~al.}{2004}]{Brook2004ApJ612894B}
{Brook} C.~B.,  {Kawata} D.,  {Gibson} B.~K.,   {Freeman} K.~C.,  2004, \mn@doi
  [\apj] {10.1086/422709}, \href
  {https://ui.adsabs.harvard.edu/abs/2004ApJ...612..894B} {612, 894}

\bibitem[\protect\citeauthoryear{{Burstein}}{{Burstein}}{1979}]{Burstein1979ApJ234829B}
{Burstein} D.,  1979, \mn@doi [\apj] {10.1086/157563}, \href
  {https://ui.adsabs.harvard.edu/abs/1979ApJ...234..829B} {234, 829}

\bibitem[\protect\citeauthoryear{Calabrese \& Spergel}{Calabrese \&
  Spergel}{2016}]{Calabrese:2016hmp}
Calabrese E.,  Spergel D.~N.,  2016, \mn@doi [Mon. Not. Roy. Astron. Soc.]
  {10.1093/mnras/stw1256}, 460, 4397

\bibitem[\protect\citeauthoryear{{Carlberg} \& {Sellwood}}{{Carlberg} \&
  {Sellwood}}{1985}]{Carlberg1985ApJ29279C}
{Carlberg} R.~G.,  {Sellwood} J.~A.,  1985, \mn@doi [\apj] {10.1086/163134},
  \href {https://ui.adsabs.harvard.edu/abs/1985ApJ...292...79C} {292, 79}

\bibitem[\protect\citeauthoryear{{Chandrasekhar}}{{Chandrasekhar}}{1942}]{Chandrasekhar1942psdbookC}
{Chandrasekhar} S.,  1942, {Principles of stellar dynamics}

\bibitem[\protect\citeauthoryear{{Chavanis}}{{Chavanis}}{2013}]{Chavanis2013A&A556A93C}
{Chavanis} P.~H.,  2013, \mn@doi [\aap] {10.1051/0004-6361/201220607}, \href
  {https://ui.adsabs.harvard.edu/abs/2013A&A...556A..93C} {556, A93}

\bibitem[\protect\citeauthoryear{{Chen}, {Schive}  \& {Chiueh}}{{Chen}
  et~al.}{2017}]{Chen2017MNRAS4681338C}
{Chen} S.-R.,  {Schive} H.-Y.,   {Chiueh} T.,  2017, \mn@doi [\mnras]
  {10.1093/mnras/stx449}, \href
  {https://ui.adsabs.harvard.edu/abs/2017MNRAS.468.1338C} {468, 1338}

\bibitem[\protect\citeauthoryear{Chiang, Schive  \& Chiueh}{Chiang
  et~al.}{2021}]{Chiang:2021uvt}
Chiang B.~T.,  Schive H.-Y.,   Chiueh T.,  2021, \mn@doi [Phys. Rev. D]
  {10.1103/PhysRevD.103.103019}, 103, 103019

\bibitem[\protect\citeauthoryear{{Chiang}, {Ostriker}  \& {Schive}}{{Chiang}
  et~al.}{2023}]{Chiang2023}
{Chiang} B.~T.,  {Ostriker} J.~P.,   {Schive} H.-Y.,  2023, \mn@doi [\mnras]
  {10.1093/mnras/stac3358}, \href
  {https://ui.adsabs.harvard.edu/abs/2023MNRAS.518.4045C} {518, 4045}

\bibitem[\protect\citeauthoryear{{Church}, {Mocz}  \& {Ostriker}}{{Church}
  et~al.}{2019}]{Church2019MNRAS4852861C}
{Church} B.~V.,  {Mocz} P.,   {Ostriker} J.~P.,  2019, \mn@doi [\mnras]
  {10.1093/mnras/stz534}, \href
  {https://ui.adsabs.harvard.edu/abs/2019MNRAS.485.2861C} {485, 2861}

\bibitem[\protect\citeauthoryear{{Comer{\'o}n} et~al.,}{{Comer{\'o}n}
  et~al.}{2011}]{Comeron2011ApJ72918C}
{Comer{\'o}n} S.,  et~al., 2011, \mn@doi [\apj] {10.1088/0004-637X/729/1/18},
  \href {https://ui.adsabs.harvard.edu/abs/2011ApJ...729...18C} {729, 18}

\bibitem[\protect\citeauthoryear{{Comer{\'o}n}, {Elmegreen}, {Salo},
  {Laurikainen}, {Holwerda}  \& {Knapen}}{{Comer{\'o}n}
  et~al.}{2014}]{Comeron2014A&A571A58C}
{Comer{\'o}n} S.,  {Elmegreen} B.~G.,  {Salo} H.,  {Laurikainen} E.,
  {Holwerda} B.~W.,   {Knapen} J.~H.,  2014, \mn@doi [\aap]
  {10.1051/0004-6361/201424412}, \href
  {https://ui.adsabs.harvard.edu/abs/2014A&A...571A..58C} {571, A58}

\bibitem[\protect\citeauthoryear{{Dalal}, {Bovy}, {Hui}  \& {Li}}{{Dalal}
  et~al.}{2021}]{Dalal2021}
{Dalal} N.,  {Bovy} J.,  {Hui} L.,   {Li} X.,  2021, \mn@doi [\jcap]
  {10.1088/1475-7516/2021/03/076}, \href
  {https://ui.adsabs.harvard.edu/abs/2021JCAP...03..076D} {2021, 076}

\bibitem[\protect\citeauthoryear{{Dalcanton} \& {Bernstein}}{{Dalcanton} \&
  {Bernstein}}{2002}]{Dalcanton2002AJ1241328D}
{Dalcanton} J.~J.,  {Bernstein} R.~A.,  2002, \mn@doi [\aj] {10.1086/342286},
  \href {https://ui.adsabs.harvard.edu/abs/2002AJ....124.1328D} {124, 1328}

\bibitem[\protect\citeauthoryear{{Das}, {McGaugh}, {Ianjamasimanana},
  {Schombert}  \& {Dwarakanath}}{{Das} et~al.}{2020}]{Das2020ApJ88910D}
{Das} M.,  {McGaugh} S.~S.,  {Ianjamasimanana} R.,  {Schombert} J.,
  {Dwarakanath} K.~S.,  2020, \mn@doi [\apj] {10.3847/1538-4357/ab5fcd}, \href
  {https://ui.adsabs.harvard.edu/abs/2020ApJ...889...10D} {889, 10}

\bibitem[\protect\citeauthoryear{{Dootson} \& {Magorrian}}{{Dootson} \&
  {Magorrian}}{2022}]{Dootson2022arXiv220515725D}
{Dootson} D.,  {Magorrian} J.,  2022, \mn@doi [arXiv e-prints]
  {10.48550/arXiv.2205.15725}, \href
  {https://ui.adsabs.harvard.edu/abs/2022arXiv220515725D} {p. arXiv:2205.15725}

\bibitem[\protect\citeauthoryear{{Dutta Chowdhury}, {van den Bosch}, {Robles},
  {van Dokkum}, {Schive}, {Chiueh}  \& {Broadhurst}}{{Dutta Chowdhury}
  et~al.}{2021}]{Dutta_Chowdhury_21}
{Dutta Chowdhury} D.,  {van den Bosch} F.~C.,  {Robles} V.~H.,  {van Dokkum}
  P.,  {Schive} H.-Y.,  {Chiueh} T.,   {Broadhurst} T.,  2021, \mn@doi [\apj]
  {10.3847/1538-4357/ac043f}, \href
  {https://ui.adsabs.harvard.edu/abs/2021ApJ...916...27D} {916, 27}

\bibitem[\protect\citeauthoryear{{Dutta}, {Begum}, {Bharadwaj}  \&
  {Chengalur}}{{Dutta} et~al.}{2009}]{Dutta2009MNRAS398887D}
{Dutta} P.,  {Begum} A.,  {Bharadwaj} S.,   {Chengalur} J.~N.,  2009, \mn@doi
  [\mnras] {10.1111/j.1365-2966.2009.15105.x}, \href
  {https://ui.adsabs.harvard.edu/abs/2009MNRAS.398..887D} {398, 887}

\bibitem[\protect\citeauthoryear{{Gaia Collaboration} et~al.,}{{Gaia
  Collaboration} et~al.}{2016}]{GaiaCollaboration2016A&A595A}
{Gaia Collaboration} et~al., 2016, \mn@doi [Astron. Astrophys.]
  {10.1051/0004-6361/201629272}, \href
  {https://ui.adsabs.harvard.edu/abs/2016A&A...595A...1G} {595, A1}

\bibitem[\protect\citeauthoryear{{Gilmore} \& {Reid}}{{Gilmore} \&
  {Reid}}{1983}]{Gilmore1983MNRAS2021025G}
{Gilmore} G.,  {Reid} N.,  1983, \mn@doi [\mnras] {10.1093/mnras/202.4.1025},
  \href {https://ui.adsabs.harvard.edu/abs/1983MNRAS.202.1025G} {202, 1025}

\bibitem[\protect\citeauthoryear{{Harris} et~al.,}{{Harris}
  et~al.}{2020}]{numpy}
{Harris} C.~R.,  et~al., 2020, \mn@doi [\nat] {10.1038/s41586-020-2649-2},
  \href {https://ui.adsabs.harvard.edu/abs/2020Natur.585..357H} {585, 357}

\bibitem[\protect\citeauthoryear{{Hayashi}, {Ferreira}  \& {Chan}}{{Hayashi}
  et~al.}{2021}]{Hayashi2021ApJ912L3H}
{Hayashi} K.,  {Ferreira} E. G.~M.,   {Chan} H. Y.~J.,  2021, \mn@doi [\apjl]
  {10.3847/2041-8213/abf501}, \href
  {https://ui.adsabs.harvard.edu/abs/2021ApJ...912L...3H} {912, L3}

\bibitem[\protect\citeauthoryear{Hui, Ostriker, Tremaine  \& Witten}{Hui
  et~al.}{2017}]{Hui:2016ltb}
Hui L.,  Ostriker J.~P.,  Tremaine S.,   Witten E.,  2017, \mn@doi [Phys. Rev.
  D] {10.1103/PhysRevD.95.043541}, 95, 043541

\bibitem[\protect\citeauthoryear{{Hunter}}{{Hunter}}{2007}]{matplotlib}
{Hunter} J.~D.,  2007, \mn@doi [Computing in Science and Engineering]
  {10.1109/MCSE.2007.55}, \href
  {https://ui.adsabs.harvard.edu/abs/2007CSE.....9...90H} {9, 90}

\bibitem[\protect\citeauthoryear{{Just} \& {Pe{\~n}arrubia}}{{Just} \&
  {Pe{\~n}arrubia}}{2005}]{Just2005A&A431861J}
{Just} A.,  {Pe{\~n}arrubia} J.,  2005, \mn@doi [\aap]
  {10.1051/0004-6361:20041175}, \href
  {https://ui.adsabs.harvard.edu/abs/2005A&A...431..861J} {431, 861}

\bibitem[\protect\citeauthoryear{{Kasparova} \& {Zasov}}{{Kasparova} \&
  {Zasov}}{2008}]{Kasparova2008AstL34152K}
{Kasparova} A.~V.,  {Zasov} A.~V.,  2008, \mn@doi [Astronomy Letters]
  {10.1007/s11443-008-3003-4}, \href
  {https://ui.adsabs.harvard.edu/abs/2008AstL...34..152K} {34, 152}

\bibitem[\protect\citeauthoryear{{Khoperskov}, {Di Matteo}, {Gerhard}, {Katz},
  {Haywood}, {Combes}, {Berczik}  \& {Gomez}}{{Khoperskov}
  et~al.}{2019}]{Khoperskov2019A&A622L6K}
{Khoperskov} S.,  {Di Matteo} P.,  {Gerhard} O.,  {Katz} D.,  {Haywood} M.,
  {Combes} F.,  {Berczik} P.,   {Gomez} A.,  2019, \mn@doi [\aap]
  {10.1051/0004-6361/201834707}, \href
  {https://ui.adsabs.harvard.edu/abs/2019A&A...622L...6K} {622, L6}

\bibitem[\protect\citeauthoryear{{King}}{{King}}{1966}]{King1966AJ.....71...64K}
{King} I.~R.,  1966, \mn@doi [\aj] {10.1086/109857}, \href
  {https://ui.adsabs.harvard.edu/abs/1966AJ.....71...64K} {71, 64}

\bibitem[\protect\citeauthoryear{{Kregel}, {van der Kruit}  \& {de
  Grijs}}{{Kregel} et~al.}{2002}]{Kregel2002MNRAS334646K}
{Kregel} M.,  {van der Kruit} P.~C.,   {de Grijs} R.,  2002, \mn@doi [\mnras]
  {10.1046/j.1365-8711.2002.05556.x}, \href
  {https://ui.adsabs.harvard.edu/abs/2002MNRAS.334..646K} {334, 646}

\bibitem[\protect\citeauthoryear{{Kroupa}}{{Kroupa}}{2002}]{Kroupa2002MNRAS330707K}
{Kroupa} P.,  2002, \mn@doi [\mnras] {10.1046/j.1365-8711.2002.05128.x}, \href
  {https://ui.adsabs.harvard.edu/abs/2002MNRAS.330..707K} {330, 707}

\bibitem[\protect\citeauthoryear{{Kulkarni}, {Visbal}, {Bryan}  \&
  {Li}}{{Kulkarni} et~al.}{2022}]{Kulkarni2022ApJ941L18K}
{Kulkarni} M.,  {Visbal} E.,  {Bryan} G.~L.,   {Li} X.,  2022, \mn@doi [\apjl]
  {10.3847/2041-8213/aca47c}, \href
  {https://ui.adsabs.harvard.edu/abs/2022ApJ...941L..18K} {941, L18}

\bibitem[\protect\citeauthoryear{{Lacey}}{{Lacey}}{1984}]{Lacey1984MNRAS208687L}
{Lacey} C.~G.,  1984, \mn@doi [\mnras] {10.1093/mnras/208.4.687}, \href
  {https://ui.adsabs.harvard.edu/abs/1984MNRAS.208..687L} {208, 687}

\bibitem[\protect\citeauthoryear{{Lacey} \& {Ostriker}}{{Lacey} \&
  {Ostriker}}{1985}]{Lacey1985ApJ}
{Lacey} C.~G.,  {Ostriker} J.~P.,  1985, \mn@doi [Astrophys. J.]
  {10.1086/163729}, \href
  {https://ui.adsabs.harvard.edu/abs/1985ApJ...299..633L} {299, 633}

\bibitem[\protect\citeauthoryear{{Leroy}, {Walter}, {Brinks}, {Bigiel}, {de
  Blok}, {Madore}  \& {Thornley}}{{Leroy} et~al.}{2008}]{Leroy2008AJ1362782L}
{Leroy} A.~K.,  {Walter} F.,  {Brinks} E.,  {Bigiel} F.,  {de Blok} W.~J.~G.,
  {Madore} B.,   {Thornley} M.~D.,  2008, \mn@doi [\aj]
  {10.1088/0004-6256/136/6/2782}, \href
  {https://ui.adsabs.harvard.edu/abs/2008AJ....136.2782L} {136, 2782}

\bibitem[\protect\citeauthoryear{{Leung}, {Broadhurst}, {Lim}, {Diego},
  {Chiueh}, {Schive}  \& {Windhorst}}{{Leung}
  et~al.}{2018}]{Leung2018ApJ862156L}
{Leung} E.,  {Broadhurst} T.,  {Lim} J.,  {Diego} J.~M.,  {Chiueh} T.,
  {Schive} H.-Y.,   {Windhorst} R.,  2018, \mn@doi [\apj]
  {10.3847/1538-4357/aacdad}, \href
  {https://ui.adsabs.harvard.edu/abs/2018ApJ...862..156L} {862, 156}

\bibitem[\protect\citeauthoryear{{Lin}, {Schive}, {Wong}  \& {Chiueh}}{{Lin}
  et~al.}{2018}]{Lin2018}
{Lin} S.-C.,  {Schive} H.-Y.,  {Wong} S.-K.,   {Chiueh} T.,  2018, \mn@doi
  [\prd] {10.1103/PhysRevD.97.103523}, \href
  {https://ui.adsabs.harvard.edu/abs/2018PhRvD..97j3523L} {97, 103523}

\bibitem[\protect\citeauthoryear{{Loebman}, {Ro{\v{s}}kar}, {Debattista},
  {Ivezi{\'c}}, {Quinn}  \& {Wadsley}}{{Loebman}
  et~al.}{2011}]{Loebman2011ApJ7378L}
{Loebman} S.~R.,  {Ro{\v{s}}kar} R.,  {Debattista} V.~P.,  {Ivezi{\'c}}
  {\v{Z}}.,  {Quinn} T.~R.,   {Wadsley} J.,  2011, \mn@doi [\apj]
  {10.1088/0004-637X/737/1/8}, \href
  {https://ui.adsabs.harvard.edu/abs/2011ApJ...737....8L} {737, 8}

\bibitem[\protect\citeauthoryear{{Mackereth} et~al.,}{{Mackereth}
  et~al.}{2019}]{Mackereth2019MNRAS}
{Mackereth} J.~T.,  et~al., 2019, \mn@doi [Mon. Not. Roy. Astron. Soc.]
  {10.1093/mnras/stz1521}, \href
  {https://ui.adsabs.harvard.edu/abs/2019MNRAS.489..176M} {489, 176}

\bibitem[\protect\citeauthoryear{{Majewski} et~al.,}{{Majewski}
  et~al.}{2017}]{Majewski2017AJ154}
{Majewski} S.~R.,  et~al., 2017, \mn@doi [Astron. J.]
  {10.3847/1538-3881/aa784d}, \href
  {https://ui.adsabs.harvard.edu/abs/2017AJ....154...94M} {154, 94}

\bibitem[\protect\citeauthoryear{Marsh \& Pop}{Marsh \&
  Pop}{2015}]{Marsh:2015wka}
Marsh D. J.~E.,  Pop A.-R.,  2015, \mn@doi [Mon. Not. Roy. Astron. Soc.]
  {10.1093/mnras/stv1050}, 451, 2479

\bibitem[\protect\citeauthoryear{{Martell} et~al.,}{{Martell}
  et~al.}{2017}]{Martell2017MNRAS465}
{Martell} S.~L.,  et~al., 2017, \mn@doi [Mon. Not. Roy. Astron. Soc.]
  {10.1093/mnras/stw2835}, \href
  {https://ui.adsabs.harvard.edu/abs/2017MNRAS.465.3203M} {465, 3203}

\bibitem[\protect\citeauthoryear{{Martin}, {Ibata}, {Bellazzini}, {Irwin},
  {Lewis}  \& {Dehnen}}{{Martin} et~al.}{2004}]{Martin2004MNRAS34812M}
{Martin} N.~F.,  {Ibata} R.~A.,  {Bellazzini} M.,  {Irwin} M.~J.,  {Lewis}
  G.~F.,   {Dehnen} W.,  2004, \mn@doi [\mnras]
  {10.1111/j.1365-2966.2004.07331.x}, \href
  {https://ui.adsabs.harvard.edu/abs/2004MNRAS.348...12M} {348, 12}

\bibitem[\protect\citeauthoryear{{Matthews}}{{Matthews}}{2000}]{Matthews2000AJ1201764M}
{Matthews} L.~D.,  2000, \mn@doi [\aj] {10.1086/301555}, \href
  {https://ui.adsabs.harvard.edu/abs/2000AJ....120.1764M} {120, 1764}

\bibitem[\protect\citeauthoryear{{Michtchenko}, {Barros}, {P{\'e}rez-Villegas}
  \& {L{\'e}pine}}{{Michtchenko} et~al.}{2019}]{Michtchenko2019ApJ87636M}
{Michtchenko} T.~A.,  {Barros} D.~A.,  {P{\'e}rez-Villegas} A.,   {L{\'e}pine}
  J. R.~D.,  2019, \mn@doi [\apj] {10.3847/1538-4357/ab11cd}, \href
  {https://ui.adsabs.harvard.edu/abs/2019ApJ...876...36M} {876, 36}

\bibitem[\protect\citeauthoryear{{Miglio} et~al.,}{{Miglio}
  et~al.}{2021}]{Miglio2021A&A645A}
{Miglio} A.,  et~al., 2021, \mn@doi [Astron. Astrophys.]
  {10.1051/0004-6361/202038307}, \href
  {https://ui.adsabs.harvard.edu/abs/2021A&A...645A..85M} {645, A85}

\bibitem[\protect\citeauthoryear{{Minchev} \& {Famaey}}{{Minchev} \&
  {Famaey}}{2010}]{Minchev2010ApJ722112M}
{Minchev} I.,  {Famaey} B.,  2010, \mn@doi [\apj]
  {10.1088/0004-637X/722/1/112}, \href
  {https://ui.adsabs.harvard.edu/abs/2010ApJ...722..112M} {722, 112}

\bibitem[\protect\citeauthoryear{{Minchev} \& {Quillen}}{{Minchev} \&
  {Quillen}}{2006}]{Minchev2006MNRAS368623M}
{Minchev} I.,  {Quillen} A.~C.,  2006, \mn@doi [\mnras]
  {10.1111/j.1365-2966.2006.10129.x}, \href
  {https://ui.adsabs.harvard.edu/abs/2006MNRAS.368..623M} {368, 623}

\bibitem[\protect\citeauthoryear{Mocz et~al.}{Mocz et~al.}{2019}]{Mocz:2019pyf}
Mocz P.,  et~al., 2019, \mn@doi [Phys. Rev. Lett.]
  {10.1103/PhysRevLett.123.141301}, 123, 141301

\bibitem[\protect\citeauthoryear{{Nadler} et~al.,}{{Nadler}
  et~al.}{2021}]{Nadler2021PhRvL126i1101N}
{Nadler} E.~O.,  et~al., 2021, \mn@doi [\prl] {10.1103/PhysRevLett.126.091101},
  \href {https://ui.adsabs.harvard.edu/abs/2021PhRvL.126i1101N} {126, 091101}

\bibitem[\protect\citeauthoryear{{Navarro}, {Frenk}  \& {White}}{{Navarro}
  et~al.}{1996}]{NFW1996ApJ}
{Navarro} J.~F.,  {Frenk} C.~S.,   {White} S. D.~M.,  1996, \mn@doi [\apj]
  {10.1086/177173}, \href
  {https://ui.adsabs.harvard.edu/abs/1996ApJ...462..563N} {462, 563}

\bibitem[\protect\citeauthoryear{{Nelson} et~al.,}{{Nelson}
  et~al.}{2023}]{Nelson2023ApJ948L18N}
{Nelson} E.~J.,  et~al., 2023, \mn@doi [\apjl] {10.3847/2041-8213/acc1e1},
  \href {https://ui.adsabs.harvard.edu/abs/2023ApJ...948L..18N} {948, L18}

\bibitem[\protect\citeauthoryear{{Patra}}{{Patra}}{2019}]{Patra2019MNRAS48481P}
{Patra} N.~N.,  2019, \mn@doi [\mnras] {10.1093/mnras/sty3493}, \href
  {https://ui.adsabs.harvard.edu/abs/2019MNRAS.484...81P} {484, 81}

\bibitem[\protect\citeauthoryear{{Powell}, {Vegetti}, {McKean}, {White},
  {Ferreira}, {May}  \& {Spingola}}{{Powell}
  et~al.}{2023}]{Powell2023arXiv230210941P}
{Powell} D.~M.,  {Vegetti} S.,  {McKean} J.~P.,  {White} S. D.~M.,  {Ferreira}
  E. G.~M.,  {May} S.,   {Spingola} C.,  2023, \mn@doi [arXiv e-prints]
  {10.48550/arXiv.2302.10941}, \href
  {https://ui.adsabs.harvard.edu/abs/2023arXiv230210941P} {p. arXiv:2302.10941}

\bibitem[\protect\citeauthoryear{{Quinn}, {Hernquist}  \& {Fullagar}}{{Quinn}
  et~al.}{1993}]{Quinn1993ApJ40374Q}
{Quinn} P.~J.,  {Hernquist} L.,   {Fullagar} D.~P.,  1993, \mn@doi [\apj]
  {10.1086/172184}, \href
  {https://ui.adsabs.harvard.edu/abs/1993ApJ...403...74Q} {403, 74}

\bibitem[\protect\citeauthoryear{{Rodionov} \& {Sotnikova}}{{Rodionov} \&
  {Sotnikova}}{2013}]{Rodionov2013MNRAS4342373R}
{Rodionov} S.~A.,  {Sotnikova} N.~Y.,  2013, \mn@doi [\mnras]
  {10.1093/mnras/stt1183}, \href
  {https://ui.adsabs.harvard.edu/abs/2013MNRAS.434.2373R} {434, 2373}

\bibitem[\protect\citeauthoryear{{Rogers} \& {Peiris}}{{Rogers} \&
  {Peiris}}{2021}]{Rogers2021PhRvL126g1302R}
{Rogers} K.~K.,  {Peiris} H.~V.,  2021, \mn@doi [\prl]
  {10.1103/PhysRevLett.126.071302}, \href
  {https://ui.adsabs.harvard.edu/abs/2021PhRvL.126g1302R} {126, 071302}

\bibitem[\protect\citeauthoryear{{Sanders} \& {Das}}{{Sanders} \&
  {Das}}{2018}]{Sanders2018MNRAS4814093S}
{Sanders} J.~L.,  {Das} P.,  2018, \mn@doi [\mnras] {10.1093/mnras/sty2490},
  \href {https://ui.adsabs.harvard.edu/abs/2018MNRAS.481.4093S} {481, 4093}

\bibitem[\protect\citeauthoryear{{Schive}, {Chiueh}  \& {Broadhurst}}{{Schive}
  et~al.}{2014a}]{Schive2014a}
{Schive} H.-Y.,  {Chiueh} T.,   {Broadhurst} T.,  2014a, \mn@doi [Nature
  Physics] {10.1038/nphys2996}, \href
  {http://adsabs.harvard.edu/abs/2014NatPh..10..496S} {10, 496}

\bibitem[\protect\citeauthoryear{{Schive}, {Liao}, {Woo}, {Wong}, {Chiueh},
  {Broadhurst}  \& {Hwang}}{{Schive} et~al.}{2014b}]{Schive2014PhRvL}
{Schive} H.-Y.,  {Liao} M.-H.,  {Woo} T.-P.,  {Wong} S.-K.,  {Chiueh} T.,
  {Broadhurst} T.,   {Hwang} W. Y.~P.,  2014b, \mn@doi [\prl]
  {10.1103/PhysRevLett.113.261302}, \href
  {https://ui.adsabs.harvard.edu/abs/2014PhRvL.113z1302S} {113, 261302}

\bibitem[\protect\citeauthoryear{{Schive}, {ZuHone}, {Goldbaum}, {Turk},
  {Gaspari}  \& {Cheng}}{{Schive} et~al.}{2018}]{Schive2018}
{Schive} H.-Y.,  {ZuHone} J.~A.,  {Goldbaum} N.~J.,  {Turk} M.~J.,  {Gaspari}
  M.,   {Cheng} C.-Y.,  2018, \mn@doi [\mnras] {10.1093/mnras/sty2586}, \href
  {https://ui.adsabs.harvard.edu/abs/2018MNRAS.481.4815S} {481, 4815}

\bibitem[\protect\citeauthoryear{Schive, Chiueh  \& Broadhurst}{Schive
  et~al.}{2020}]{Schive:2019rrw}
Schive H.-Y.,  Chiueh T.,   Broadhurst T.,  2020, \mn@doi [Phys. Rev. Lett.]
  {10.1103/PhysRevLett.124.201301}, 124, 201301

\bibitem[\protect\citeauthoryear{{Sch{\"o}nrich} \& {Binney}}{{Sch{\"o}nrich}
  \& {Binney}}{2009}]{Schonrich2009MNRAS396203S}
{Sch{\"o}nrich} R.,  {Binney} J.,  2009, \mn@doi [\mnras]
  {10.1111/j.1365-2966.2009.14750.x}, \href
  {https://ui.adsabs.harvard.edu/abs/2009MNRAS.396..203S} {396, 203}

\bibitem[\protect\citeauthoryear{{Sch{\"o}nrich} \& {Dehnen}}{{Sch{\"o}nrich}
  \& {Dehnen}}{2018}]{Schonrich2018MNRAS4783809S}
{Sch{\"o}nrich} R.,  {Dehnen} W.,  2018, \mn@doi [\mnras]
  {10.1093/mnras/sty1256}, \href
  {https://ui.adsabs.harvard.edu/abs/2018MNRAS.478.3809S} {478, 3809}

\bibitem[\protect\citeauthoryear{{Schutz}}{{Schutz}}{2020}]{Schutz2020PhRvD101l3026S}
{Schutz} K.,  2020, \mn@doi [\prd] {10.1103/PhysRevD.101.123026}, \href
  {https://ui.adsabs.harvard.edu/abs/2020PhRvD.101l3026S} {101, 123026}

\bibitem[\protect\citeauthoryear{{Sellwood} \& {Binney}}{{Sellwood} \&
  {Binney}}{2002}]{Sellwood2002MNRAS336785S}
{Sellwood} J.~A.,  {Binney} J.~J.,  2002, \mn@doi [\mnras]
  {10.1046/j.1365-8711.2002.05806.x}, \href
  {https://ui.adsabs.harvard.edu/abs/2002MNRAS.336..785S} {336, 785}

\bibitem[\protect\citeauthoryear{{Sharma} et~al.,}{{Sharma}
  et~al.}{2021}]{Sharma2021MNRAS506}
{Sharma} S.,  et~al., 2021, \mn@doi [Mon. Not. Roy. Astron. Soc.]
  {10.1093/mnras/stab1086}, \href
  {https://ui.adsabs.harvard.edu/abs/2021MNRAS.506.1761S} {506, 1761}

\bibitem[\protect\citeauthoryear{{Spitzer} \& {Schwarzschild}}{{Spitzer} \&
  {Schwarzschild}}{1951}]{Spitzer1951ApJ114385S}
{Spitzer} Lyman J.,  {Schwarzschild} M.,  1951, \mn@doi [\apj]
  {10.1086/145478}, \href
  {https://ui.adsabs.harvard.edu/abs/1951ApJ...114..385S} {114, 385}

\bibitem[\protect\citeauthoryear{{Steinmetz} et~al.,}{{Steinmetz}
  et~al.}{2006}]{RAVE2006AJ1321645S}
{Steinmetz} M.,  et~al., 2006, \mn@doi [\aj] {10.1086/506564}, \href
  {https://ui.adsabs.harvard.edu/abs/2006AJ....132.1645S} {132, 1645}

\bibitem[\protect\citeauthoryear{{Ting} \& {Rix}}{{Ting} \&
  {Rix}}{2019}]{Ting2019ApJ878}
{Ting} Y.-S.,  {Rix} H.-W.,  2019, \mn@doi [Astrophys. J.]
  {10.3847/1538-4357/ab1ea5}, \href
  {https://ui.adsabs.harvard.edu/abs/2019ApJ...878...21T} {878, 21}

\bibitem[\protect\citeauthoryear{{Toomre}}{{Toomre}}{1964}]{Toomre1964ApJ1391217T}
{Toomre} A.,  1964, \mn@doi [\apj] {10.1086/147861}, \href
  {https://ui.adsabs.harvard.edu/abs/1964ApJ...139.1217T} {139, 1217}

\bibitem[\protect\citeauthoryear{{Toth} \& {Ostriker}}{{Toth} \&
  {Ostriker}}{1992}]{Toth1992ApJ3895T}
{Toth} G.,  {Ostriker} J.~P.,  1992, \mn@doi [\apj] {10.1086/171185}, \href
  {https://ui.adsabs.harvard.edu/abs/1992ApJ...389....5T} {389, 5}

\bibitem[\protect\citeauthoryear{{Tremaine}, {Frankel}  \& {Bovy}}{{Tremaine}
  et~al.}{2023}]{Tremaine2023MNRAS521114T}
{Tremaine} S.,  {Frankel} N.,   {Bovy} J.,  2023, \mn@doi [\mnras]
  {10.1093/mnras/stad577}, \href
  {https://ui.adsabs.harvard.edu/abs/2023MNRAS.521..114T} {521, 114}

\bibitem[\protect\citeauthoryear{{Tsikoudi}}{{Tsikoudi}}{1979}]{Tsikoudi1979ApJ234842T}
{Tsikoudi} V.,  1979, \mn@doi [\apj] {10.1086/157565}, \href
  {https://ui.adsabs.harvard.edu/abs/1979ApJ...234..842T} {234, 842}

\bibitem[\protect\citeauthoryear{{Turk}, {Smith}, {Oishi}, {Skory}, {Skillman},
  {Abel}  \& {Norman}}{{Turk} et~al.}{2011}]{yt}
{Turk} M.~J.,  {Smith} B.~D.,  {Oishi} J.~S.,  {Skory} S.,  {Skillman} S.~W.,
  {Abel} T.,   {Norman} M.~L.,  2011, \mn@doi [Astrophys. J. Suppl.]
  {10.1088/0067-0049/192/1/9}, \href
  {http://adsabs.harvard.edu/abs/2011ApJS..192....9T} {192, 9}

\bibitem[\protect\citeauthoryear{Veltmaat, Niemeyer  \& Schwabe}{Veltmaat
  et~al.}{2018}]{Veltmaat:2018dfz}
Veltmaat J.,  Niemeyer J.~C.,   Schwabe B.,  2018, \mn@doi [Phys. Rev. D]
  {10.1103/PhysRevD.98.043509}, 98, 043509

\bibitem[\protect\citeauthoryear{{Virtanen} et~al.,}{{Virtanen}
  et~al.}{2020}]{scipy}
{Virtanen} P.,  et~al., 2020, \mn@doi [Nature Methods]
  {10.1038/s41592-019-0686-2}, \href
  {https://ui.adsabs.harvard.edu/abs/2020NatMe..17..261V} {17, 261}

\bibitem[\protect\citeauthoryear{{Wasserman} et~al.,}{{Wasserman}
  et~al.}{2019}]{Wasserman2019ApJ885}
{Wasserman} A.,  et~al., 2019, \mn@doi [Astrophys. J.]
  {10.3847/1538-4357/ab3eb9}, \href
  {https://ui.adsabs.harvard.edu/abs/2019ApJ...885..155W} {885, 155}

\bibitem[\protect\citeauthoryear{{Widrow}, {Gardner}, {Yanny}, {Dodelson}  \&
  {Chen}}{{Widrow} et~al.}{2012}]{Widrow2012ApJ750L41W}
{Widrow} L.~M.,  {Gardner} S.,  {Yanny} B.,  {Dodelson} S.,   {Chen} H.-Y.,
  2012, \mn@doi [\apjl] {10.1088/2041-8205/750/2/L41}, \href
  {https://ui.adsabs.harvard.edu/abs/2012ApJ...750L..41W} {750, L41}

\bibitem[\protect\citeauthoryear{{Yoachim} \& {Dalcanton}}{{Yoachim} \&
  {Dalcanton}}{2005}]{Yoachim2005ApJ624701Y}
{Yoachim} P.,  {Dalcanton} J.~J.,  2005, \mn@doi [\apj] {10.1086/428922}, \href
  {https://ui.adsabs.harvard.edu/abs/2005ApJ...624..701Y} {624, 701}

\bibitem[\protect\citeauthoryear{{Yoachim} \& {Dalcanton}}{{Yoachim} \&
  {Dalcanton}}{2006}]{Yoachim2006AJ131226Y}
{Yoachim} P.,  {Dalcanton} J.~J.,  2006, \mn@doi [\aj] {10.1086/497970}, \href
  {https://ui.adsabs.harvard.edu/abs/2006AJ....131..226Y} {131, 226}

\bibitem[\protect\citeauthoryear{{Yoachim} \& {Dalcanton}}{{Yoachim} \&
  {Dalcanton}}{2008}]{Yoachim2008ApJ683707Y}
{Yoachim} P.,  {Dalcanton} J.~J.,  2008, \mn@doi [\apj] {10.1086/590246}, \href
  {https://ui.adsabs.harvard.edu/abs/2008ApJ...683..707Y} {683, 707}

\bibitem[\protect\citeauthoryear{{Yurin} \& {Springel}}{{Yurin} \&
  {Springel}}{2014}]{Yurin2014}
{Yurin} D.,  {Springel} V.,  2014, \mn@doi [\mnras] {10.1093/mnras/stu1421},
  \href {https://ui.adsabs.harvard.edu/abs/2014MNRAS.444...62Y} {444, 62}

\bibitem[\protect\citeauthoryear{Zhao, Zhao, Chu, Jing  \& Deng}{Zhao
  et~al.}{2012}]{Zhao2012LAMOST}
Zhao G.,  Zhao Y.-H.,  Chu Y.-Q.,  Jing Y.-P.,   Deng L.-C.,  2012, \mn@doi
  [Res. Astron. Astrophys.] {10.1088/1674-4527/12/7/002}, 12, 723

\bibitem[\protect\citeauthoryear{{Zupancic} \& {Widrow}}{{Zupancic} \&
  {Widrow}}{2023}]{Zupancic2023arXiv231113352Z}
{Zupancic} B.,  {Widrow} L.~M.,  2023, \mn@doi [\mnras]
  {10.1093/mnras/stad3620}, \href
  {https://ui.adsabs.harvard.edu/abs/2023MNRAS.tmp.3484Z} {}

\bibitem[\protect\citeauthoryear{{van der Kruit} \& {Searle}}{{van der Kruit}
  \& {Searle}}{1981}]{vanderKruit1981A&A95105V}
{van der Kruit} P.~C.,  {Searle} L.,  1981, \aap, \href
  {https://ui.adsabs.harvard.edu/abs/1981A&A....95..105V} {95, 105}

\bibitem[\protect\citeauthoryear{{van der Kruit} \& {de Grijs}}{{van der Kruit}
  \& {de Grijs}}{1999}]{vanderKruit1999A&A352129V}
{van der Kruit} P.~C.,  {de Grijs} R.,  1999, \mn@doi [\aap]
  {10.48550/arXiv.astro-ph/9910288}, \href
  {https://ui.adsabs.harvard.edu/abs/1999A&A...352..129V} {352, 129}

\makeatother
\end{thebibliography}

% appendix
% ----------------------------------------------
\appendix
\section{Relaxation of halo-disc initial conditions}
\label{app:RelaxationOfInitialConditions}

\begin{figure}
	\centering
	\includegraphics[width=\columnwidth]{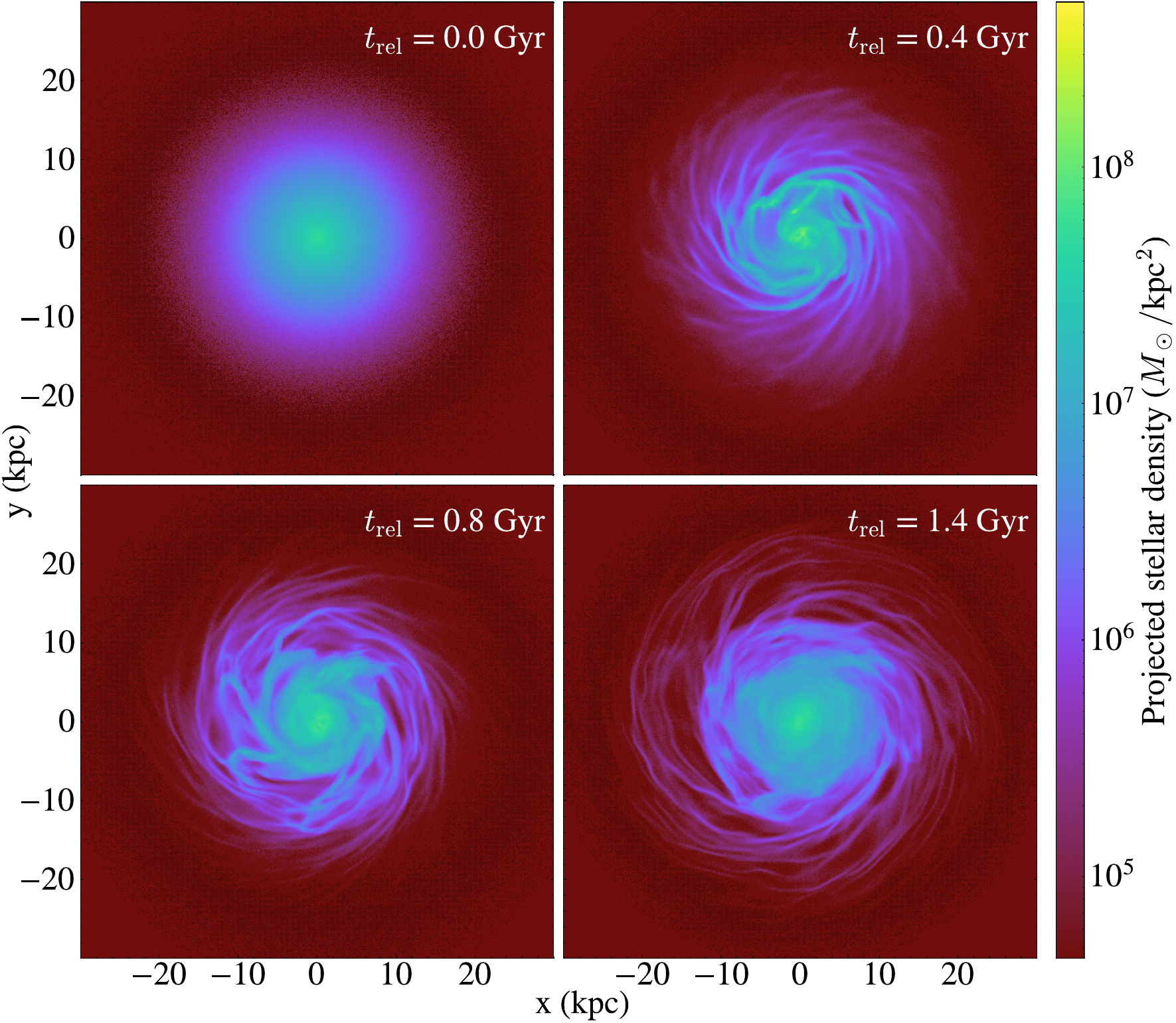}
	\caption{
		Face-on density projections of the stellar disc hosted by the halo with $\ma=0.4$ and $\Mh=7\times10^{10}\Msun$ during the initial co-relaxation process. At $\tr=0$ (upper left panel), the initial disc configuration constructed by $\GALIC$ appears smooth and free of inhomogeneous substructures. 
		At the onset of the co-evolution (upper right), the development of spiral arm structures and local density clumps rapidly destabilises the disc. As the disc-halo system evolves further in time (lower left), these transient substructures gradually disappear from the inner halo region. At the end of the co-relaxation phase $\tr=1.4$~Gyr (lower right), disc clumps have largely dissolved within $R\lesssim10$~kpc. The entire disc remains Toomre stable $Q(R\leq 15\text{~kpc})>1$ and the disc rotation curve becomes stable in time beyond this point (see Figs.~\ref{fig:rotation}~and~\ref{fig:rotation_diff}).
	}
	\label{fig:disc_relax}
\end{figure}
Each initial $N$-body stellar disc constructed with $\GALIC$ assumes a shell-averaged, substructure-free, and static background halo density profile. However, during the initial stage of disc co-evolution with a dynamical FDM halo, the gravitational potential perturbations sourced by locally fluctuating density granulation can temporarily destabilise and drive irreversible evolution of disc star orbits. During this period, the rapid increase in $\sigma_z(R)$ arises the initially sub-unity Toomre $Q(R)$ parameter across the disc. Empirically, such halo-disc systems generally reach quasi-equilibrium after $\lesssim 0.5T_{\rm ff} \simeq 1.0$~Gyr. \fref{fig:disc_relax} shows the face-on density projections of the stellar disc with $\ma=0.4$ and $\Mh=7\times10^{10}\Msun$ during this phase of initial co-relaxation $\tr = 0$\textendash$1.4$~Gyr. The initially smooth stellar disc (upper left panel) quickly develops pronounced spiral density waves across all radii by $\tr = 0.4$~Gyr (upper right). By $\tr = 0.8$~Gyr (lower left), the central disc region becomes less clumpy. This trend continues inside-out until the end of the initial co-relaxation phase at $\tr = 1.4$~Gyr (lower right), after which the inner disc appears smooth and comparatively more coherent while some spiral arm structures remain visible at large radii. We observe that $Q(R\leq15\text{~kpc}) > 1$ holds true beyond this point.

\begin{figure}
	\centering
	\includegraphics[width=\columnwidth]{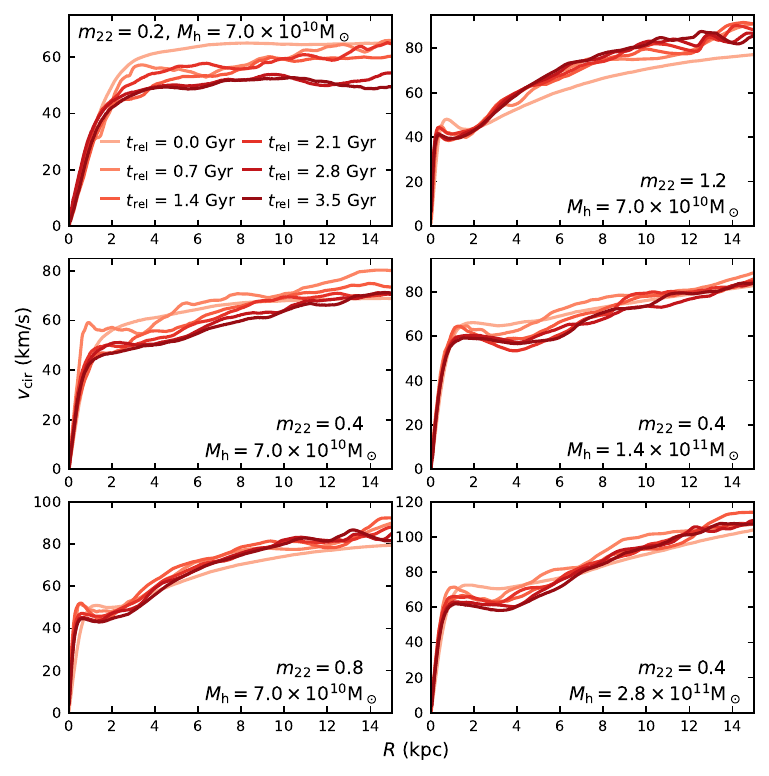}
	\caption{
		Disc rotation curve profiles from $\tr = 0$~Gyr (light red) to $3.5$~Gyr (dark red). During the initial disc-halo co-relaxation $\tr = 0$\textendash$1.4$~Gyr, transient disc substructures cause these rotation curves to undergo instability-driven dynamical evolution. In contrast for $\tr\geq 1.4$~Gyr, each relaxed disc configuration and the corresponding rotation curve remain comparatively stable with time.
	}
	\label{fig:rotation}
\end{figure}

\begin{figure}
	\centering
	\includegraphics[width=\columnwidth]{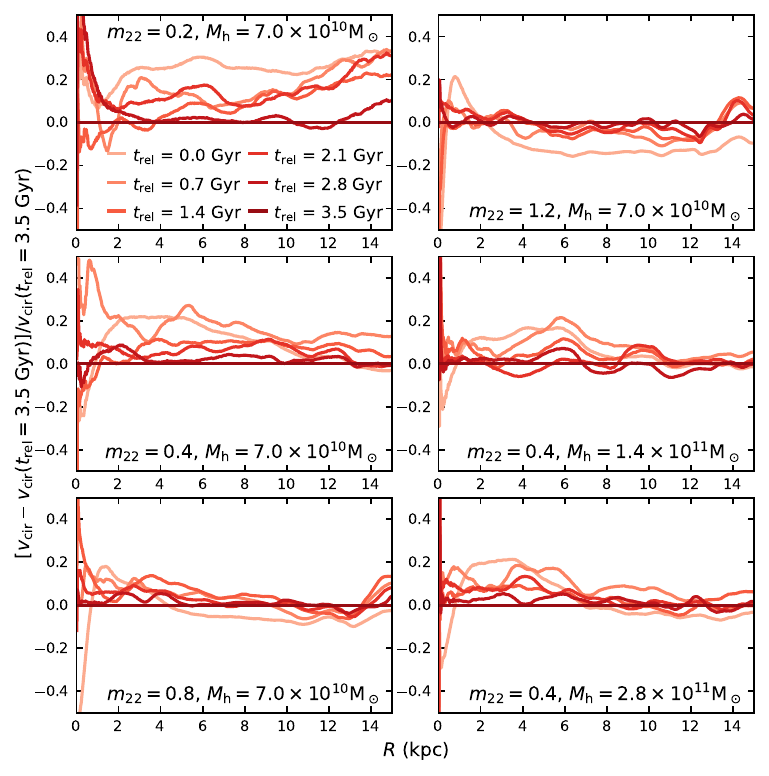}
	\caption{
		The relative difference of the disc rotation curve profiles between a given time slice $\tr = 0$\textendash$3.5$~Gyr (light to dark red) and the final snapshot at $3.5$~Gyr. During the halo-disc co-relaxation phase $\tr \leq 1.4$~Gyr, comparative large relative velocity differences $\gtrsim20\%$ are ubiquitously present within $R\leq$15~kpc. Post co-relaxation $\tr \geq 1.4$~Gyr, the rotation curve profiles largely stabilise, with relative local velocity differences less than $\lesssim10\%$.
	}
	\label{fig:rotation_diff}
\end{figure}
\begin{figure}
\centering
\includegraphics[width=\columnwidth]{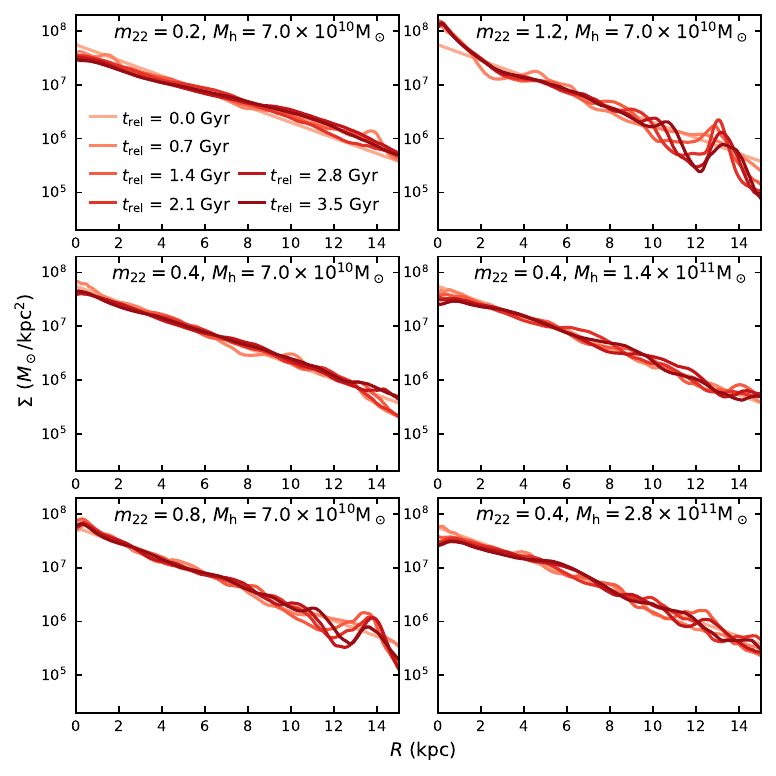}
\caption{
Disc surface density profiles from $\tr = 0$~Gyr (light red) to $3.5$~Gyr (dark red). Each profile still roughly follows the exponential parameterisation $\Sigma(R)\propto e^{-R/\Rd}$ throughout the entire simulation. Sizable profile fluctuations at large radii $R\gtrsim 10$~kpc, even at late times, are caused by the persistent spiral arm structures in the outer disc regions (e.g. see lower right panel of \fref{fig:disc_relax}).
}
\label{fig:surface_dens}
\end{figure}

A more quantitative view of this initial co-relaxation phase is to examine the time evolution of azimuthally averaged disc rotation curve profiles in \fref{fig:rotation} across $\tr = 0$\textendash$3.5$~Gyr (from light to dark red). Disc rotation curves exhibit noticeable variation during the first 0.7~Gyr of initial co-relaxation, while after 1.4~Gyr the profiles show little evolution with time. The relative difference of the rotation curve profile of each snapshot and the final snapshot is presented in \fref{fig:rotation_diff}. A similar trend is also observed in the disc surface density evolution from the initially exponential profile $\Sigma(R)\propto e^{-R/\Rd}$ (see \sref{subsec:init_disc} for details). As shown in \fref{fig:surface_dens}, within the cylindrical radius $R\lesssim2.5$~kpc, the disc surface density profiles change noticeably at the onset of co-relaxation and become comparatively stable after $\tr \geq 1.4$~Gyr. For $4\text{ kpc}\lesssim R\lesssim 10$~kpc, the density profiles remain approximately exponential throughout the simulation. However in the outer disc regions, the persistent spiral arms and density ripples (e.g. see \fref{fig:disc_relax}) lead to continuous fluctuations in the $\Sigma(R\gtrsim10\text{ kpc})$ profiles even at late times. Lastly, since the disc surface densities beyond $R\gtrsim10\text{ kpc}$ are generally too low to be statistically reliable given the adopted particle resolution, we thus exclude the simulation-inferred granulation-driven heating rates in radial bins beyond $R \geq 10$~kpc, as shown in Figs.~\ref{fig:sigma_z_sqr}~and~\ref{fig:heating}.

\begin{figure}
	\centering
	\includegraphics[width=\columnwidth]{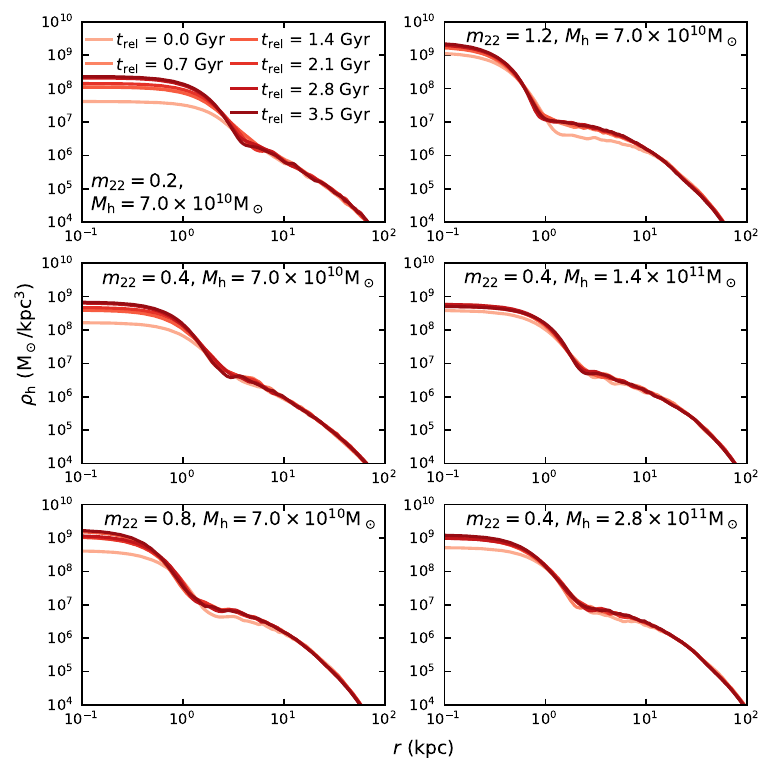}
	\caption{
		The shell-averaged density profiles of six simulated FDM haloes at time slices $\tr=0$~Gyr (light red) to 3.5~Gyr (dark red). The initial introduction of the disc component deepens the central gravitational potential, which in turn causes adiabatic contraction in the FDM central soliton core. The dynamical relaxation of FDM haloes proceeds much more rapidly than the stellar components (cf. Figs.~\ref{fig:rotation}~and~\ref{fig:rotation_diff}), and the FDM density profiles are comparatively stable after $\tr=0.7$~Gyr.
	}
	\label{fig:halo_dens_evo}
\end{figure}

\begin{figure}
	\centering
	\includegraphics[width=\columnwidth]{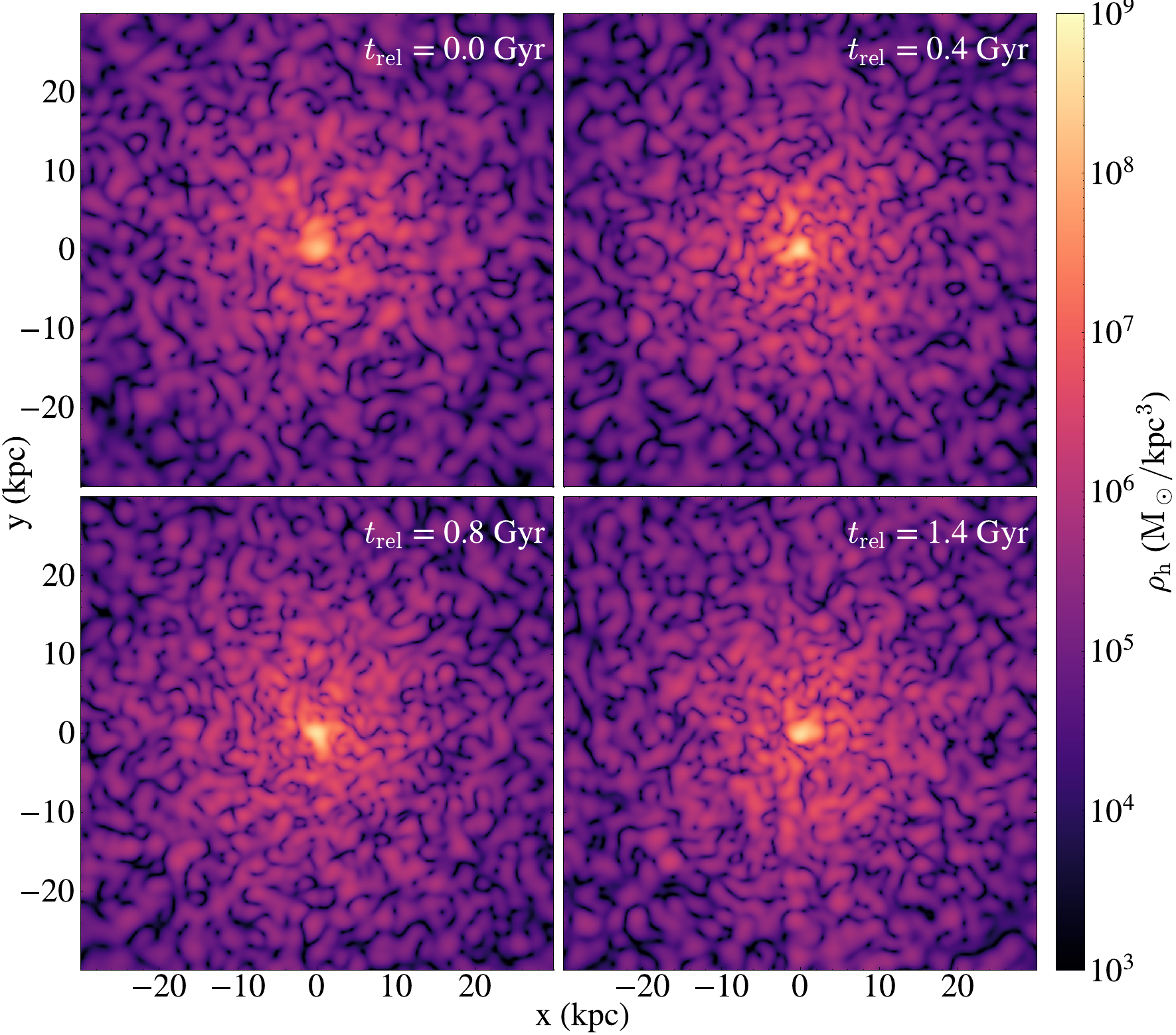}
	\caption{
		The density slices of the $\ma=0.4$, $\Mh=7.0\times10^{10}\Msun$ halo during the co-relaxation process. Relative to the initial granular structures at $\tr=0$ (upper left), the typical granulation sizes shrink during the co-relaxation due to the introduction of the disc component. Beyond $\tr\gtrsim0.4$~Gyr, the physical scales of FDM granulation do not evolve significantly with time, again indicating that the FDM haloes relax more rapidly than the corresponding disc components (cf. \fref{fig:disc_relax}).
	}
	\label{fig:halo_relax}
\end{figure}

The presence of an additional stellar disc also slightly affects each FDM halo and the associated granular structures. The initial introduction of the disc component causes adiabatic construction in the inner halo, causing both the central soliton core to compactify and the typical halo granulation size to shrink. The relaxation of an FDM halo generally occurs much more rapidly than the stellar disc, as shown in \fref{fig:halo_dens_evo}. The halo density profiles become comparatively stable after $\tr=0.7$~Gyr, with small temporal fluctuations contained largely within the central soliton region. \fref{fig:halo_relax} shows the density slices of the $\ma=0.4$, $\Mh=7.0\times10^{10}\Msun$ halo during the initial disc-halo co-relaxation. The typical granulation sizes shrink noticeably during $\tr=0$\textendash0.4~Gyr. After $\tr=0.4$~Gyr, no visible changes in the granulation sizes are observed, indicating that the FDM halo has reached a quasi-equilibrium state. It should be emphasised that although the presence of an additional disc component does modify the initial FDM granulation structures, these changes in the FDM host halo post-relaxation have been self-consistently (at least in the linear regime) accounted for in the analytical heating rate estimates via $\sigmah$ and $\rhoh$ in Eqs.~(\ref{eqn:Heating_Eq_SGD_BGD_Limits}, \ref{eqn:Heating_Eq_SGD_BGD_Limits_2}).

\section{Numerical Convergence tests}
\label{app:ConvergenceTest}

\begin{figure}
	\centering
	\includegraphics[width=\columnwidth]{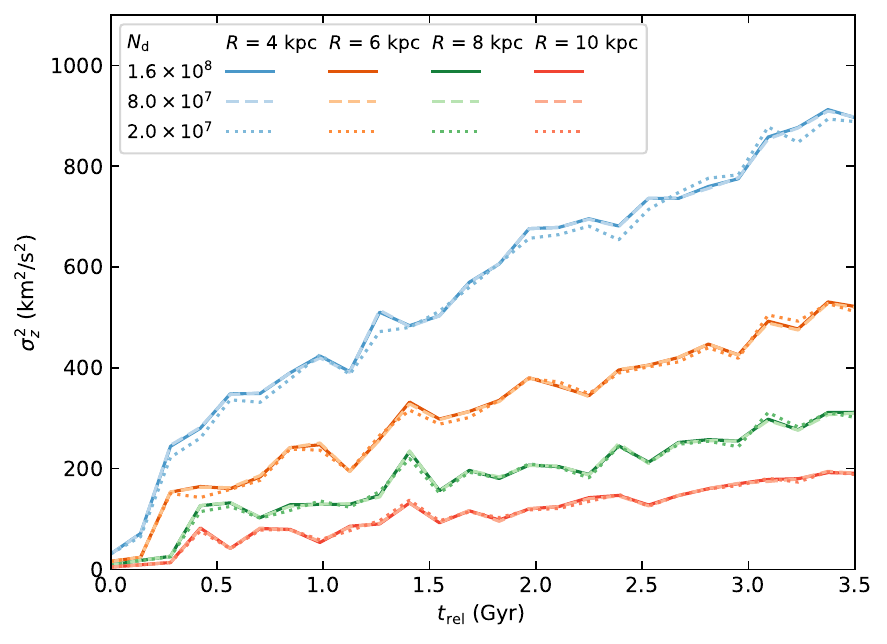}
	\caption{Convergence of ensemble-averaged disc vertical velocity dispersion squared $\sigma_z^2(R,\tr)$ in 2-kpc-wide radial bins centred on $R = 4$~kpc (blue), 6~(orange), 8~(green), and 10~kpc (red) over $\tr = 0$\textendash$3.5$~Gyr, where the stellar discs are sampled by a total number of $\Nd = 2\times10^7$~(dotted), $8\times10^7$~(dashed), and $1.6\times10^8$~(solid) equal-mass particles. The host FDM halo has $\ma=0.2$ and $\Mh=7\times10^{10}\Msun$, and the maximum spatial resolution is 0.062 kpc (see \tref{tab:SimulationSetup}). Numerical convergence is achieved among all three runs with varying particle resolutions $\Nd \geq 2\times10^7$.
	}
	\label{fig:conv_evolve}
\end{figure}

To further validate the robustness of our simulation results, we have carried out simulation runs with varying particle and spatial resolutions for each of the six cases listed in \tref{tab:SimulationSetup}. The level of numerical convergence is assessed by comparing the time evolution of measured $\sigma_z^2$ across resolution-varying runs. \fref{fig:conv_evolve} shows the time evolution of ensemble-averaged disc vertical velocity dispersion squared $\sigma_z^2(\tr)$ profiles under different particle resolutions $\Nd=2.0\times10^7$ (dotted), $8.0\times10^7$ (dashed), and $1.6\times10^8$ (solid), evolved in the FDM host halo with $\ma=0.2$ and $\Mh=7\times10^{10}\Msun$ as an example. During the initial $0.5$~Gyr co-relaxation as discussed in Appendix~\ref{app:RelaxationOfInitialConditions}, instability-driven rapid growth in $\sigma_z^2$ is observed in all radial bins across the three resolution-varying runs. The post-co-relaxation heating curves $\tr \geq 1.4$~Gyr from the highest to the lowest particle resolution runs agree within $\pm 3\,{\rm km}^2{\rm s}^{-2}{\rm Gyr}^{-1}$ ($\lesssim 6\%$ at outer radii and $\lesssim 1\%$ at inner radii). This high level of consistency implies that the numerical convergence is achieved for $\Nd\geq2.0\times10^7$.

\begin{figure*}
	\centering
	\includegraphics[width=\textwidth]{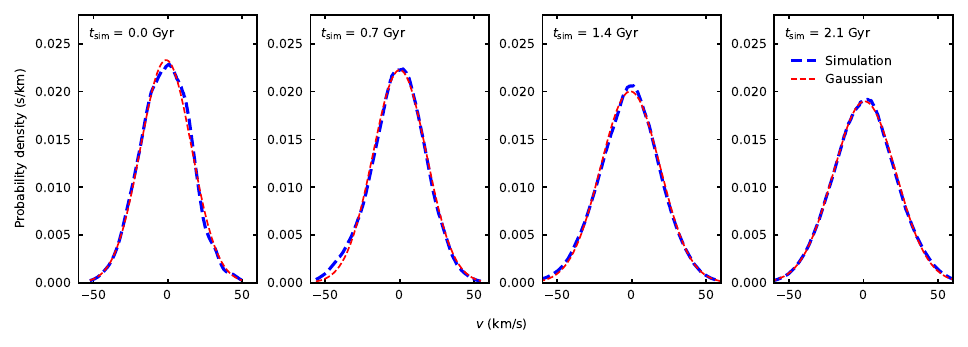}
	\caption{
		Ensemble-averaged disc vertical velocity probability density distributions within $R = 5$\textendash7~kpc at $\ts = 0.0, 0.7, 1.4, 2.1$~Gyr (left to right panels; $\ts \equiv \tr -1.4$~Gyr) in the host halo with $\ma=0.2$ and $\Mh=7\times10^{10}\Msun$. Here the disc is resolved with $\Nd = 1.6\times 10^8$ equal-mass particles, and the simulation data (blue) properly preserve the Gaussian distributions (red). Note that the initial disc velocity distribution is Gaussian and symmetric with respect to $v_z = 0$~km~s$^{-1}$ (see \sref{subsec:init_disc}). Under the local approximation of the Fokker\textendash Planck formalism, the ensemble-averaged granulation-driven disc heating should be symmetric with respect to the disc mid-plane. Furthermore, this second-order diffusive heating broadens the width, while preserving the Gaussian shape, of the disc velocity distribution. This heating behaviour is consistent with simulations with adequate particle resolution.
	}
	\label{fig:vel_distribution}
\end{figure*}

\begin{figure}
	\centering
	\includegraphics[width=\columnwidth]{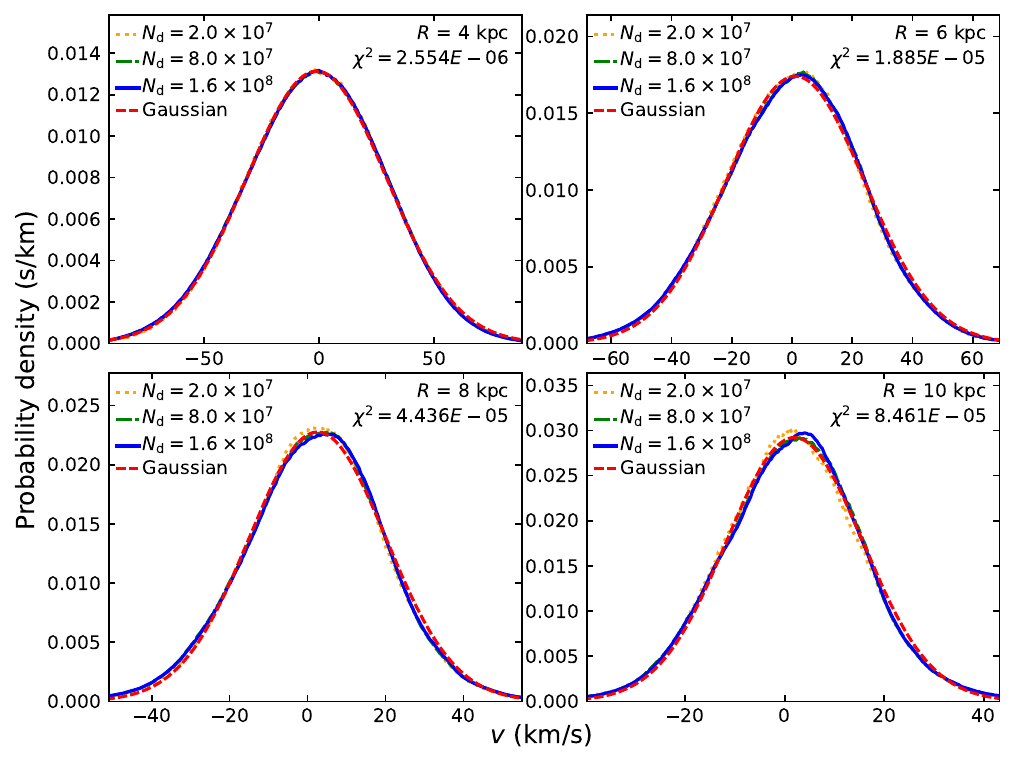}
	\caption{
		Ensemble-averaged vertical velocity probability density distributions in $2$-kpc-wide radial bins centred on $R = 4$~kpc, 6, 8, and 10~kpc at $\tr = 3.5$~Gyr, for stellar discs sampled with $\Nd = 2\times 10^7$ (yellow), $8\times 10^7$ (green), and $1.6\times 10^8$ (blue) identical particles. All three runs well preserve the Gaussian distributions (red), indicating that discs sampled with $\Nd\geq 2\times 10^7$ have sufficient particle resolution to yield numerically convergent heating behaviour.
    }
	\label{fig:conv_distri}
\end{figure}

How the distribution of disc particle vertical velocities evolves with time provides an additional cross-check for numerical convergence. For an initially Gaussian velocity distribution centred at $v_z = 0$~km/s (see \sref{subsec:init_disc}), the net contribution of FDM-granulation-driven stellar heating is diffusive and purely second-order. Hence the local velocity distribution should still remain Gaussian while undergoing continuous broadening with time, provided that the adopted particle resolution is sufficient to resolve the disc dynamics. \fref{fig:vel_distribution} shows the time evolution of disc velocity probability density distribution in a $2$-kpc-wide radial bin centred on $R = 6$~kpc (i.e. including disc particles within $R=5$\textendash7~kpc); the disc is resolved with $\Nd = 1.6\times 10^8$ equal-mass particles and evolved in the FDM halo with $\ma=0.2$ and $\Mh=7\times10^{10}\Msun$. For the disc particle with velocities within $3\sigma_z$, the simulation data (blue) are consistent with the best-fit Gaussian distributions (red), as the full width at half maximum increases with time due to the continuous granulation-driven diffusive heating. We have verified that this consistency condition is satisfied at all radii of interest under this particle resolution.

To examine the impact of particle resolution on the evolved vertical velocity probability density distribution, we compare in \fref{fig:conv_distri} the probability density distributions in $2$-kpc-wide radial bins centred on $R = 4$~kpc, 6, 8, and 10~kpc at $\tr = 3.5$~Gyr, for stellar discs resolved with $\Nd = 2\times 10^7$ (yellow), $8\times 10^7$ (green), and $1.6\times 10^8$ (blue; identical to \fref{fig:vel_distribution}) equal-mass particles. The FDM host halo has $\ma=0.2$ and $\Mh=7\times10^{10}\Msun$. The Gaussian probability density distributions for all three runs are well preserved until the end of each simulation. Contrastingly in simulation runs with inadequate particle resolution $\Nd$ (not shown here), we have observed skewed distributions that noticeably deviate from the best-fit Gaussian distributions (red). The presence of such distorted non-Gaussian features in the disc vertical velocity distribution allows one to promptly identify insufficient particle resolution adopted to simulate disc dynamics.

Lastly, at a fixed particle resolution, we have also performed simulations with different maximum spatial resolutions. For the FDM halo with $\ma=0.4$ and $\Mh=7\times10^{10}\Msun$ and the stellar disc resolved with $\Nd=8.0\times10^7$ equal-mass particles, the two simulation runs with respective maximum spatial resolutions of $120$~pc and $60$~pc (adopted in the production run) yield ensemble-averaged disc vertical heating rates consistent within $\leq 10\%$ at all radii of interest. We thus conclude that a maximum spatial resolution of $60$~pc is sufficient in this halo-disc setup.

\section{Heating from CDM halo}
\label{app:CDM}

\begin{figure}
	\centering
	\includegraphics[width=\columnwidth]{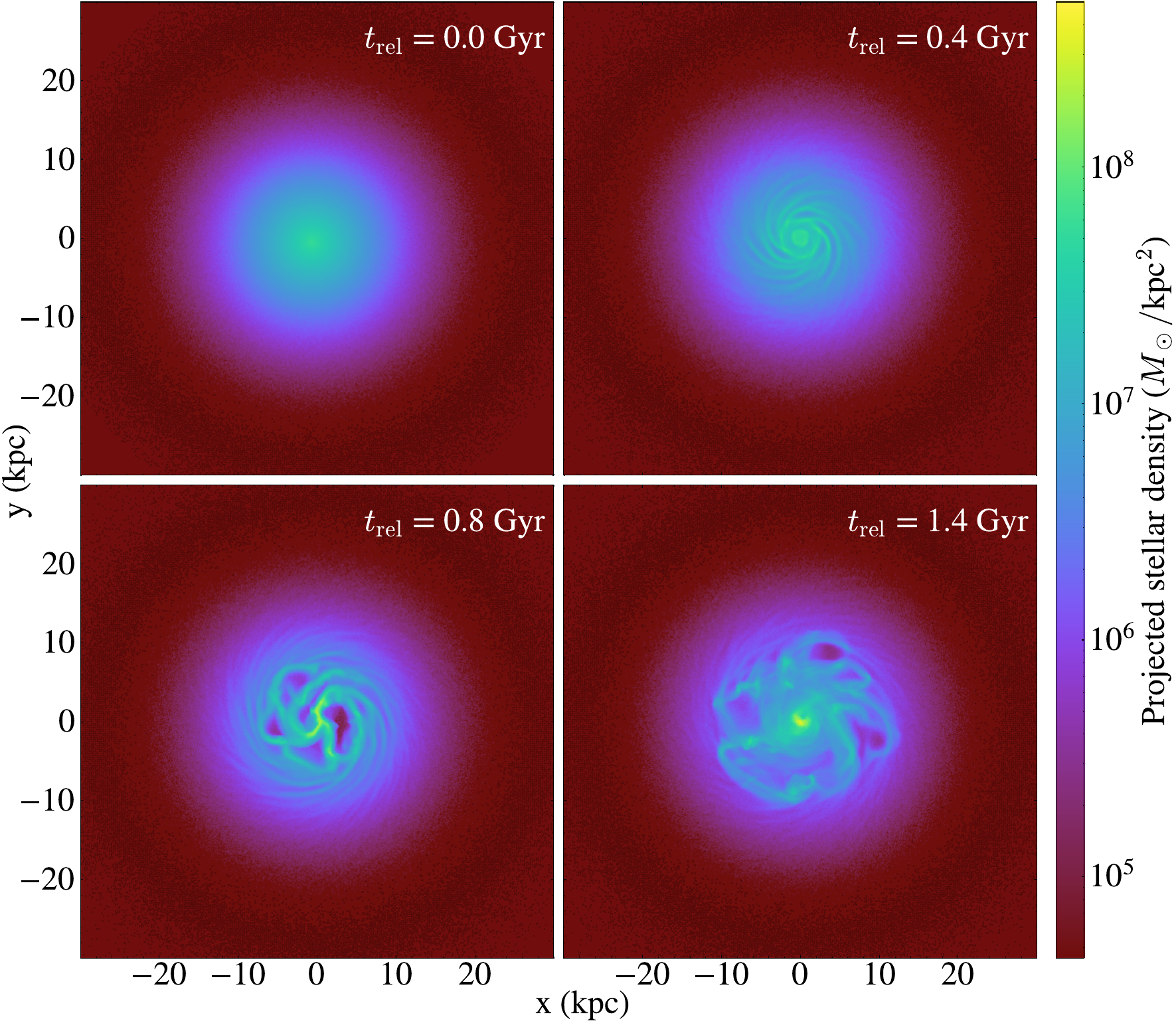}
	\caption{
		Face-on density projections of the stellar disc hosted by a CDM halo with $\Mh=7\times10^{10}\Msun$ and $c_{\rm h}=8$ during the initial co-relaxation process. The disc configuration remains largely smooth for $\tr\lesssim 0.4$~Gyr, compared to the counterparts relaxed in FDM haloes (e.g. \fref{fig:disc_relax}). Perturbed by quadrupolar oscillations in the background CDM halo potential (due to the presence of the stellar disc that breaks the spherical symmetry), pronounced density waves and clumps gradually develop and peak at $\tr\simeq0.8$~Gyr. After $\tr=1.4$~Gyr, a new disc-halo quasi-equilibrium is established and the disc vertical velocity dispersion profiles become stable with time (see \fref{fig:cdm}).
	}
	\label{fig:disc_relax_cdm}
\end{figure}

To better disentangle different perturbative sources present during the initial disc-halo co-relaxation (\sref{subsec:init_disc}) and further assess the degree of possible artificial heating in our FDM halo-disc simulations (\sref{subsec:disc_heating_rate_Fokker_Planck}), we perform an additional simulation replacing the FDM halo with CDM using the same code $\gamer$. The $\GALIC$-generated spherical $N$-body halo has a virial mass of $\Mh=7\times10^{10}\Msun$ and an NFW concentration parameter of $c_{\rm h}=8$ (see \sref{subsec:init_disc} for more details). The stellar disc initial condition is identical to that in all FDM simulation runs; the density profile follows \eref{eq:disc} with $\Md=3.16\times 10^{9}$~$\Msun$, $\Rd=3.0$~$\kpc$, and $h=0.15$~$\kpc$. The CDM halo and the stellar disc are respectively sampled with $N_{\rm h}=1.6\times10^{8}$ and $\Nd=8.0\times10^{7}$ equal-mass particles. The highest spatial resolution is $0.05$~kpc.

\begin{figure}
	\centering
	\includegraphics[width=\columnwidth]{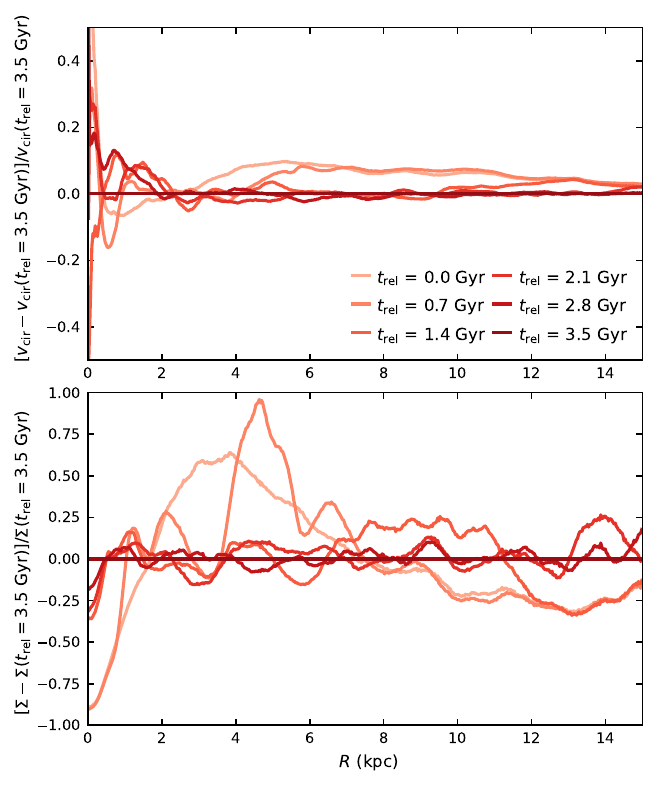}
	\caption{
		The relative difference of disc rotation curve (upper) and the surface density (lower) profiles in the CDM host halo during $\tr=0$\textendash$3.5$~Gyr between the $\tr=3.5$~Gyr snapshot. Both profiles exhibit negligible evolution after $\tr\geq1.4$~Gyr.
		}
	\label{fig:cdm_profile_diff}
\end{figure}

\fref{fig:disc_relax_cdm} shows the face-on density projections of the stellar disc hosted by a spherical CDM halo during the initial co-relaxation $\tr = 0$\textendash1.4~Gyr. The system is initially stable since both the $\GALIC$-constructed CDM halo and stellar disc are constructed as an approximate equilibrium solution to the collisionless Boltzmann equation. The disc spiral density waves driven by quadrupolar oscillations in the halo potential (due to the presence of the stellar disc that breaks the spherical symmetry) gradually develop in the innermost region and peak around $\tr\simeq0.8$~Gyr. The steady increase in the disc vertical velocity dispersion during this period raises the Toomre $Q$ parameter across the entire disc. By $\tr=1.4$~Gyr, we observe that $Q(R) > 1$ for $R \leq 15$~kpc and a new disc-halo quasi-equilibrium is established. \fref{fig:cdm_profile_diff} shows the relative difference of disc rotation curve (upper) and surface density (lower) profiles from $\tr=0$~Gyr to $3.5$~Gyr with the $\tr=3.5$~Gyr snapshot being the reference. Beyond $\tr\geq1.4$~Gyr, low relative differences indicate that both profiles remain approximately constant with time. Following the same setup as the FDM halo-disc simulations (\fref{fig:flowchart}), we define this time slice as the end of disc-halo co-relaxation $\ts=\tr-1.4\text{~Gyr}=0\text{~Gyr}$.

\begin{figure}
	\centering
	\includegraphics[width=\columnwidth]{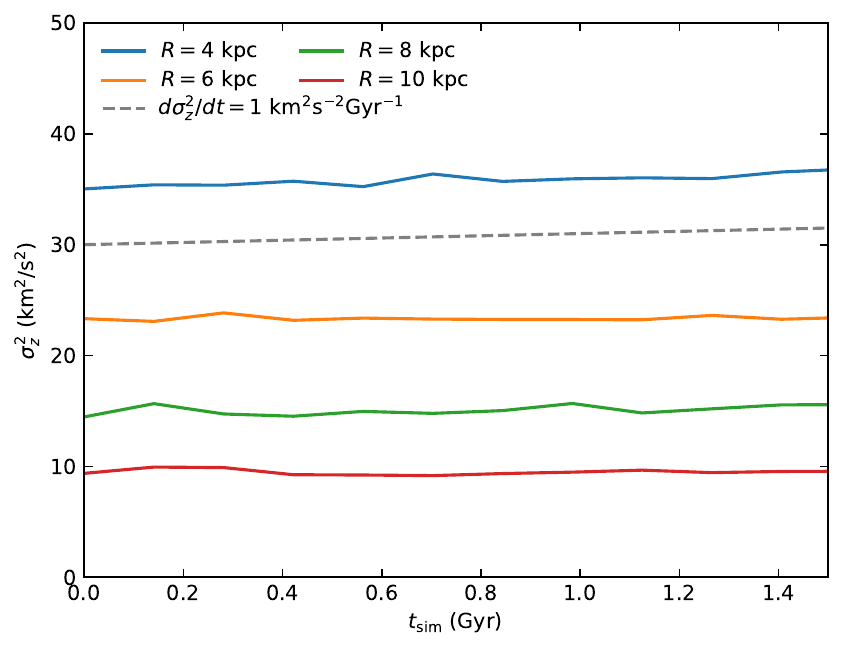}
	\caption{
		Ensemble-averaged $\sigma_z^2$ in radial bins centred on $R = 4$~kpc~(blue), 6~(orange), 8~(green), and 10~kpc~(red) over $\ts = 0$\textendash1.5~Gyr in a $\GALIC$-constructed CDM halo with $\Mh=7\times10^{10}\Msun$ and $c_{\rm h}=8$. The measured heating rates $d\sigma_z^2/dt \lesssim1$~km$^2$s$^{-2}$Gyr$^{-1}$ at all radii are negligible compared to the granulation-driven disc heating in all six FDM simulations carried out in this work (cf. Figs.~\ref{fig:sigma_z_sqr}~and~\ref{fig:heating}). We thus conclude that numerical heating is unimportant in our simulation setup.
	}
	\label{fig:cdm}
\end{figure}

Next we quantify the ensemble-averaged disc vertical heating rates over $\ts = 0$\textendash1.5~Gyr, identical to the procedure outlined in \sref{subsec:Simulations_vs_Theory}. We observe in \fref{fig:cdm} that disc vertical heating rates in the CDM halo $d\sigma_z^2/dt\lesssim1\,{\rm km}^2{\rm s}^{-2}{\rm Gyr}^{-1}$ are negligibly small compared with the values observed in FDM simulations, as shown in Figs.~\ref{fig:sigma_z_sqr}~and~\ref{fig:heating}. This result further confirms that the comparatively low heating rates observed in the two cases $\ma = 0.8$ and $1.2$ (e.g. see \fref{fig:heating}) cannot not be meaningfully attributed to numerical noises, and the observed theory-simulation discrepancy is physical in origin.

As a concluding remark, the perturbations sourced by the CDM and FDM haloes are physically distinct. The former stems from the inconsistency in the initial condition, where the initial CDM halo is constructed with the disc and its density profile undergoes little changes during the simulation. The resulting disc heating saturates after $\tr = 1.4$~Gyr and is generally contained within the central $R\lesssim10$~kpc. In contrast, stellar discs in FDM haloes exhibit comparatively richer spiral arm structures (\fref{fig:disc_relax}). The gravitational perturbations post-relaxation are dominantly sourced by the FDM granulation and show sustained disc heating at all radii (e.g. \fref{fig:disc_edge_on}).

\bsp
\label{lastpage}
\end{document}